\documentclass[a4paper,11pt]{article}
\pdfoutput=1 

\usepackage{jheppub} 
\usepackage{amsmath,amssymb,amsthm,amscd,graphicx}
\usepackage[bbgreekl]{mathbbol}
\usepackage[utf8]{inputenc}
\input epsf.sty

\theoremstyle{definition}

\newcommand{\CA}{{\cal A}}

\newcommand{\CC}{{\cal C}}

\newcommand{\CH}{{\cal H}}

\newcommand{\CL}{{\cal L}}
\newcommand{\CM}{{\cal M}}
\newcommand{\CN}{{\cal N}}
\newcommand{\CO}{{\cal O}}

\newcommand{\bba}{{\mathbb{\bbalpha}}}
\newcommand{\bbb}{{\mathbb{\bbbeta}}}

\def\IZ{{\mathbb Z}}

\newcommand{\re}{{\rm e}}
\newcommand{\ri}{{\rm i}}
\newcommand{\rd}{{\rm d}}
\renewcommand{\d}{\partial}

\newcommand{\be}{\begin{equation}}
\newcommand{\ee}{\end{equation}}
\newcommand{\ba}{\begin{aligned}}
\newcommand{\ea}{\end{aligned}}
\newcommand{\ben}{\begin{eqnarray}\displaystyle}
\newcommand{\een}{\end{eqnarray}}

\newcommand{\mb}{\mathsf{b}}

\newdimen\tableauside\tableauside=1.0ex
\newdimen\tableaurule\tableaurule=0.4pt
\newdimen\tableaustep
\def\phantomhrule#1{\hbox{\vbox to0pt{\hrule height\tableaurule width#1\vss}}}
\def\phantomvrule#1{\vbox{\hbox to0pt{\vrule width\tableaurule height#1\hss}}}
\def\sqr{\vbox{%
  \phantomhrule\tableaustep
  \hbox{\phantomvrule\tableaustep\kern\tableaustep\phantomvrule\tableaustep}%
  \hbox{\vbox{\phantomhrule\tableauside}\kern-\tableaurule}}}
\def\squares#1{\hbox{\count0=#1\noindent\loop\sqr
  \advance\count0 by-1 \ifnum\count0>0\repeat}}
\def\tableau#1{\vcenter{\offinterlineskip
  \tableaustep=\tableauside\advance\tableaustep by-\tableaurule
  \kern\normallineskip\hbox
    {\kern\normallineskip\vbox
      {\gettableau#1 0 }%
     \kern\normallineskip\kern\tableaurule}%
  \kern\normallineskip\kern\tableaurule}}
\def\gettableau#1{\ifnum#1=0\let\next=\null\else
\squares{#1}\let\next=\gettableau\fi\next}

\def\Z{{\mathbb{Z}}}
\def\Q{{\mathbb{Q}}}
\def\R{{\mathbb{R}}}
\def\C{{\mathbb{C}}}

\def\Tr{{\mathrm{Tr}}}

\def\fig8{{\mathbf{4_1}}}

\def\tilde{\widetilde}
\renewcommand{\bar}{\overline}
\renewcommand{\hat}{\widehat}

\renewcommand{\d}{\partial}

\title{\boldmath Resurgence in complex Chern-Simons theory}

\author{Sergei Gukov,$^{a}$}

\affiliation{$^a$Max-Planck-Institut f\"ur Mathematik, Vivatsgasse 7, D-53111 Bonn, Germany\\
\phantom{$^a$}Walter Burke Institute for Theoretical Physics, California Institute of Technology, Pasadena, CA 91125, USA}

\emailAdd{gukov@theory.caltech.edu}

\author{Marcos Mari\~no,$^{b}$}

\affiliation{$^b$D\'epartement de Physique Th\'eorique et Section de Math\'ematiques\\
\phantom{$^b$}Universit\'e de Gen\`eve, Gen\`eve, CH-1211 Switzerland}

\emailAdd{marcos.marino@unige.ch}

\author{Pavel Putrov$^{\; c}$}

\affiliation{$^c$School of Natural Sciences, Institute for Advanced Study, Princeton, NJ 08540, USA}

\emailAdd{putrov@ias.edu}

\abstract{
We study resurgence properties of partition function of $SU(2)$ Chern-Simons theory (WRT invariant) on closed three-manifolds. We check explicitly that in various examples Borel transforms of asymptotic expansions posses expected analytic properties. In examples that we study we observe that contribution of irreducible flat connections to the path integral can be recovered from asymptotic expansions around abelian flat connections.
We also discuss connection to Floer instanton moduli spaces, disk instantons in 2d sigma models, and length spectra of ``complex geodesics'' on
the A-polynomial curve.
}

\preprint{CALT 2016-011}
\begin{document}

\maketitle
\flushbottom


\section{Introduction}

Resurgent analysis seems to work best for quantum field theories without renormalon effects
where one has the complete control over perturbative expansion but limited understanding
of non-perturbative physics
(for comprehensive introduction and illustration in quantum theories see {\it e.g.} \cite{mm-lectures,Dorigoni:2014hea})\footnote{Recently there has been also progress in applying the tools of resurgence to conventional QFTs, see for example \cite{Dunne:2012ae,Cherman:2013yfa,Aniceto:2014hoa} and also \cite{Dunne:2015eaa} for
a review and references.}.
The Chern-Simons gauge theory is a perfect example of such theory, where one has the luxury
perturbative calculations that can produce the explicit form of perturbative coefficients to all-loop order,
sometimes in several different way.
These exact perturbative calculations, around real classical solutions and also in the analytically continued theory,
are waiting to be Borel resummed.
Yet, they mostly managed to escape attention of resurgent analysis, and one of our main goals is to fill this gap.

In general, one might expect that the Feynman path integral of a quantum system --- be it quantum mechanics or QFT ---
is simply a sum of perturbative contributions $e^{- \frac{1}{\hbar} S_{\alpha}} Z_{\alpha} (\hbar)$ from different
saddle points labeled by $\alpha$.
However, this sum, where each term enters with weight 1, does not correctly capture the analytic continuation of
the original Feynman path integral when the perturbative expansion parameter $\hbar$ becomes large and/or acquires a complex phase.
For such values of the ``coupling constant'' $\hbar$, the exact partition function can be obtained using
the resurgent analysis of Jean \'Ecalle \cite{Ecalle},
\be
Z(\hbar) \; = \; \sum_{\alpha} n_{\alpha} e^{- \frac{1}{\hbar} S_{\alpha} } Z_\alpha^{\text{pert}} (\hbar)
\label{ZEcallea}
\ee
where the coefficients $n_{\alpha}$ are called the {\it transseries parameters}.
Each term in this sum is a contribution of the complexified version of the path integral
along the convergent integration cycle $\Gamma_{\alpha}$ obtained by the steepest descent from the critical point $\alpha$,
\be
Z (\hbar) \; = \; \sum_{\alpha} n_{\alpha} \int_{\Gamma_{\alpha}} D A e^{ - \frac{1}{\hbar} S(A)}\,.
\label{ZEcalleb}
\ee
The transseries parameters $n_{\alpha}$ are constant away from the Stokes rays,
where their values may jump in order to correctly capture the dependence
of the exact partition function on $\hbar$ \cite{berry1990hyperasymptotics,berry1991hyperasymptotics}.

In our context of the Chern-Simons path integral, there are several alternatives to the coupling constant $\hbar$
related to each other via\footnote{For example, for categorification purposes, it might be useful to study asymptotic expansions in $(1-q)$ instead of $\hbar$. This will be explored elsewhere. }
\be
q = e^{\hbar} = e^{2 \pi i /k} = e^{2\pi i \tau}\,.
\label{qvskvstau}
\ee


\subsection{An application to mock modular forms and wall crossing}

In the process of Borel resummation of perturbative Chern-Simons expansions,
we encounter a curious application to mock modular forms producing new contour integral expressions for the latter:
\be
f(q) \; = \; \frac{1}{\sqrt{\tau}} \int_{\gamma} d \xi \, B(\xi) \, e^{- \xi / \tau}
\label{mockBcontint}
\ee
where $q = e^{2\pi i \tau}$ as in \eqref{qvskvstau}.
The contour integrals over mid-dimensional cycles, such as one-dimensional integrals in the Borel plane \eqref{mockBcontint}
or their higher-dimensional cousins over Lefschetz cycles \eqref{ZEcalleb}, play a central role in the Borel-\'Ecalle theory
(which, in turn, is rooted in work of Stokes \cite{Stokes}).
Therefore, the appearance of such contour integrals in resurgent analysis is not only natural, but unavoidable.
However, modular properties of the resulting function $f(q)$ are completely unexpected.
{}From the viewpoint of mock modular forms, the situation is, roughly speaking, reversed: the modular properties
are usually introduced as part of the definition of $f(q)$, while meaningful contour integral presentations are
something one has to work for.

We hope that exploring mock modular forms through the ``looking glass'' of resurgent analysis can shed new light
on their somewhat mysterious nature. We shall see some of the promising hints later in the paper,
{\it e.g.} a relation between the structure of singularities in the Borel plane and orbits of the group of Atkin-Lehner involutions.
Moreover, in the previous instances where mock modular forms emerged in string theory, they ``count'' BPS states
(of black holes or more general systems) and it was proposed by Sen \cite{Sen:2007vb} that changing the integration contour
in integral presentations like \eqref{ZEcalleb} or \eqref{mockBcontint}
corresponds to wall crossing phenomena.
We hope that our work here, especially when combined with \cite{Gukov:2016gkn}, can considerably expand the list
of physical systems where the precise map between the integration contours and chambers can be described explicitly,
much as in the famous DVV dyon counting \cite{Dijkgraaf:1996it} where this map is fairly well
understood \cite{Cheng:2007ch,Cheng:2008fc}.\footnote{See {\it e.g.} \cite{Dabholkar:2012nd,Chung:2014qpa}
for further concrete efforts to identify integration contours in various systems.}

Similar to their role in black hole microstate counting, mock modular forms in our story appear as generating
functions of BPS degeneracies or, put differently, as ``characters'' of graded vector spaces~\cite{Gukov:2016gkn}:
\be
f(q) \; = \; \Tr_{\CH_{\text{BPS}}} (-1)^i q^{j}
\; , \qquad
\CH_{\text{BPS}} \; = \; \bigoplus_{i,j} \CH_{\text{BPS}}^{i,j}\,.
\ee
This can serve as an ``explanation'' for why this class of mock modular forms have integer coefficients
(or rational coefficients if we work with their space over $\Q$) and brings us to another application.


\subsection{An application to categorification of WRT invariants of $M_3$}

The content of the present paper can be also viewed as the first step in the following sequence of ``upgrades'':
$$
Z_a^{\text{pert}} (\hbar)
\; \xrightarrow[~~\text{(Borel sum)}~~]{~~\text{resurgence}~~} \;
Z_a (q = e^{\hbar})
\; \xrightarrow[~~\text{transform}~~]{~~\text{ modular}~~} \;
\hat Z_a (q)
\; \xrightarrow[~~\text{(refinement)}~~]{~~\text{categorification}~~} \;
\hat Z_a (q,t)
$$
where the starting point --- think of it as a ``version 1.0'' --- is the perturbative series around an {\it abelian} flat connection (labeled by~$a$)
and the end result --- ``version 4.0'' --- is the Poincar\'e polynomial of the doubly-graded homology $H^{i,j} (M_3; a)$:
\be
\hat Z_a (q,t) \; = \; \sum_{i,j} q^i t^j  \dim H^{i,j} (M_3; a)\,.
\label{Zaqt}
\ee
The last two steps in the above sequence were already described in detail in \cite{Gukov:2016gkn}.
Namely, the second step can be formulated entirely in terms of the classical geometry of the moduli space
of flat connections $\CM_{\text{flat}} (G_{\C}, M_3)$ and the $SL(2,\Z)$ modular group action
that in practice can be evaluated very concretely with the help of 4-manifolds bounded by $M_3$.
The third step in the above sequence involves interpreting $\hat Z_a (q)$
as a vortex partition function ({\it i.e.} partition function on $D^2 \times_q S^1$) in the 3d $\CN=2$ theory $T[M_3]$
and then passing to its BPS spectrum ({\it i.e.} $Q$-cohomology).
In other words, skipping the first step, it was proposed in \cite{Gukov:2016gkn} that Witten-Reshetikhin-Turaev (WRT) invariant of a 3-manifold $M_3$
can be categorified when expressed as a linear combination of ``homological blocks'' $Z_a$,
\be
Z_{\text{CS}} (M_3) \; = \; \sum_{a \in \text{abelian}} e^{2\pi i k S_a} Z_a (M_3)
\label{ZviaZa}
\ee
which are labeled by abelian flat connections, that for $G=SU(2)$ take values in $H_1 (M_3) / \Z_2$.
Since the index `$a$' labels abelian representations in $\CM_{\text{flat}} (G,M_3)$,
it plays a role analogous to the choice of the Spin$^c$ structure in the Heegaard Floer homology~\cite{MR2249248}
or in monopole Floer homology~\cite{MR2388043}.
This point, in fact, is important for connecting these classical theories~\cite{Gukov:2016gkn}
to categorification of WRT invariants of $M_3$.

In this paper, we give a new interpretation to homological blocks by interpreting each $Z_a$ as a Borel resummation
of the corresponding perturbative series $Z_a^{\text{pert}}$.
It is remarkable to see how contributions of non-abelian flat connections get ``attached'' to a (formal) power series $Z_a^{\text{pert}}$
in order to form a $q$-series~$Z_a$ convergent inside the unit disk $|q|<1$.
In the classical limit $\hbar = \frac{2\pi i}{k} \to 0$, the $q$-series $Z_a (q)$
has the same asymptotic expansion as the perturbative series $Z^{\text{pert}}_a$
computed by summing Feynman diagrams around an abelian flat connection labeled by $a$.


\section{Borel resummation in Chern-Simons theory}
\label{sec:BorelCS}

Consider path integral for Chern-Simons theory on $M_3$:
\begin{equation}
 Z_\text{CS}(M_3)=\int\limits_{\CA_{SU(2)}} DA e^{2\pi i k S(A)},
\label{ZCSinit}
\end{equation}
where $\CA_{SU(2)}$ is the space of gauge connections on $M_3$ modulo gauge equivalence.
In order to study resurgence properties of the path integral,
one needs to extend its definition to complex values of the coupling constant $k$.
The idea \cite{Pham} is to introduce the steepest descent integration cycles and apply the standard toolbox of
Picard-Lefschetz theory and Stokes phenomena to the Feynman path integral as if it were a finite dimensional one;
see {\it e.g.} \cite{mm-lectures,MR2501849,MR2470516,Argyres:2012ka,Kashani-Poor:2016edc,costin2011resurgence}
for some of the relevant implementations of this idea and especially \cite{Kontsevich,Witten:2010cx} that we will follow.
Namely, we consider CS partition function for complex value of $k$ as an integral over a middle-dimensional contour in a larger complex space:
\begin{equation}
 Z_\text{CS}(M_3)=\int\limits_{\Gamma_{SU(2)}\subset \tilde{\CA}_{SL(2,\C)}} DA e^{2\pi i k S(A)},
 \label{ZCScont}
\end{equation}
where $\tilde{\CA}_{SL(2,\C)}$ is the universal cover of the space of $SL(2,\C)$ flat connections modulo gauge equivalence. The contour $\Gamma_{SU(2)}$ is chosen such that for integer $k$ the value of (\ref{ZCScont}) coincides with the value of (\ref{ZCSinit})\footnote{Note that $\Gamma_{SU(2)}$ is not just a lift of $\CA_{SU(2)}$ to $\tilde{\CA}_{SL(2,\C)}$ since the latter is not closed.}.
It is also useful to consider a quotient by the so-called {\it based gauged transformations},
that is gauge transformations that are fixed to be $1$ at a particular reference point on $M_3$.
This group acts freely on the space of connections and, therefore, the quotient space is expected
to be smooth, so that one can hope to apply the familiar intuition about resurgence properties
of integrals over contours in finite-dimensional smooth complex spaces.
However, one should still use caution since in principle there might be some new features related
to the fact that we are actually dealing with infinite dimensional spaces.

The critical submanifolds of the complex CS action are connected components of the moduli space of flat connections
\begin{equation}
 \CM_\bba\subset \text{Hom}(\pi_1(M_3),SL(2,\C))\,/SL(2,\C)\times \Z \equiv \CM_\text{flat}(M_3,SL(2,\C))
 \label{conn-flat} \times \Z
\end{equation}
with an extra label corresponding to a value of CS action in the universal cover.
In what follows we will use different notations $\bba$ or $\alpha$ depending on whether we are talking about an element
\begin{equation}
	 \bba\in\pi_0(\CM_\text{flat}(M_3,SL(2,\C)))\times \Z
\end{equation}
or its equivalence class,
\begin{equation}
	\alpha \in\pi_0(\CM_\text{flat}(M_3,SL(2,\C)))
\end{equation}
for which we ``remember'' the value of the CS action only modulo 1. Sometimes we will write
\begin{equation}
	\bba=(\alpha,S_\bba),\qquad S_\bba \in \Z+\text{CS}(\alpha)
\end{equation}
where $S_\bba$ is the value of CS action in $\C$ while $\text{CS}$ is the Chern-Simons invariant, which is defined only modulo $1$. The value of $S_\bba$ determines a lift of $\alpha$ to $\bba$.

Note that, since in the path integral we want to quotient only by the based gauge transformations,
the connected components of critical submanifolds are actually
\begin{equation}
 \tilde\CM_\bba\subset \text{Hom}(\pi_1(M_3),SL(2,\C)) \times \Z
\end{equation}
which are lifts of (\ref{conn-flat}) and form orbits of $SL(2,\C)$ action. For each $\bba$ one can define a Lefschetz thimble $\Gamma_\bba$ as the union of steepest descent trajectories originating from $\tilde\CM_\bba$. It is a middle-dimensional cycle in $\tilde \CA_{SL(2,\C)}$.

If we assume that qualitatively the picture is the same as for finite dimensional integrals,
we can decompose the contour $\Gamma_{SU(2)}$ into Lefschetz thimbles corresponding to connected components:
\begin{equation}
\Gamma_{SU(2)} = \sum_{\bba\in \pi_0(\CM_\text{flat}(M_3,SL(2,\C)))\times \Z} n_{\mathbb{\bbalpha},\theta}\Gamma_{\mathbb{\bbalpha},\theta}
 \label{SU2-thimbles}
\end{equation}
where $\theta$ is an argument of $k=|k|e^{i\theta}$ which is assumed to be of general value,
so that there are no steepest descent flows (with respect to $2\pi ikS(A)$) between different critical submanifolds.

One can define integrals over particular Lefschetz thimbles:
\begin{equation}
 I_{\bba,\theta}=\int\limits_{\Gamma_{\bba,\theta}} DA e^{2\pi i k S(A)}.
\end{equation}
As explained in detail in Appendix \ref{app:PL},
they have asymptotic expansion for large $k$ of the following form\footnote{We use the standard physical normalization such that:
\begin{equation}
Z_\text{CS}(S^2\times S^1)=1,\qquad Z_\text{CS}(S^3)=\sqrt\frac{2}{k}\,\sin\frac{\pi}{k} \,.
\end{equation}
Note, that the choice of normalization is quite important here since it affects significantly the form of the Borel transform.
}:
\begin{equation}
 I_{\bba,\theta}=e^{2\pi ikS_\bba}Z_{\alpha}^\text{pert},\qquad Z_{\alpha}^\text{pert}\in k^{(d_{\alpha}-3)/2}\C[[1/k]]
\label{Iintegralat}
\end{equation}
where $S_\bba$ is the corresponding value of the CS functional and
\begin{equation}
d_{\alpha}=\dim_\C \tilde\CM_\bba.
\end{equation}
Note that $Z_{\alpha}^\text{pert}$ depends only on $\alpha$, and not on its lift to the universal cover, $\bba$. The exact value of $I_{\bba,\theta}$ can be restored via directional Borel summation, where the Laplace transform is performed along the ray $e^{-i\theta}\R_+$. Later in the text we will sometimes suppress explicit dependence on $\theta$.

This gives the transseries expansion \eqref{ZEcalleb} of the original partition function:
\begin{equation}
 Z_\text{CS}(M_3;k)= \sum_{\bba\in \pi_0(\CM_\text{flat}(M_3,SL(2,\C)))\times \Z} n_{\bba,\theta}I_{\bba,\theta}\sim \sum_\bba n_{\bba,\theta} e^{2\pi ikS_\bba}Z_{\alpha}^\text{pert}(k)
 \label{ZCS-transseries}
\end{equation}
from which the total partition function can be recovered by applying directional Borel summation.

As we change $\theta$, in general,
the quantities $I_\bba$ may experience a jump that is usually referred to as the Stokes phenomenon. Jumps can happen when \cite{mm-lectures}:
\begin{equation}
 \theta=\theta_{\bba\bbb}\equiv \arg (S_\bba-S_\bbb)/i
\end{equation}
so that there are possible steepest descent flows from $\tilde{\CM}_\bba$ to $\tilde{\CM}_\bbb$, and have the following form\footnote{In a simple
case when the critical points of $S$ are isolated, there is a Picard-Lefschetz / Cecotti-Vafa wall crossing formula \cite{AGZV,Cecotti:1992rm}:
\be
m^\bba_\bbb = \# (\Gamma_{\bba, \theta_{\bba\bbb}+\epsilon} \cap \Gamma_{\bbb, \theta_{\bba\bbb}-\epsilon+\pi}).
\ee
However, in the case of Chern-Simons theory, the critical points are usually not isolated and this formula does not apply.}:
\begin{equation}
I_{\bba,\theta_{\bba\bbb}+\epsilon}=I_{\bba,\theta_{\bba\bbb}-\epsilon}+m_\bba^\bbb I_{\bbb,\theta_{\bba\bbb}-\epsilon},\qquad m_\bba^\bbb\in \Z
\label{Stokes-theta}
\end{equation}
for a small $\epsilon$. The coefficient $n_{\bba,\theta}$ in the transseries expansion jump accordingly so that the sum changes continuously with $\theta$.

Note that when $k\in \Z$, the weight $e^{2\pi ikS_\bba}$ depends only on the class $\alpha\in \pi_0(\CM_\text{flat}(M_3,SL(2,\C)))$ and the sum (\ref{ZCS-transseries}) collapses to a much smaller sum over $\alpha$.  The left-hand side of (\ref{ZCS-transseries}) then should reproduce the usual CS partition function, also known as the WRT invariant. In principle, analytic continuation of Chern-Simons partition function is not unique since one can choose different contours $\Gamma_{SU(2)}$ in the universal cover which give the same result for integer values of $k$. Only the following sums are fixed:
\begin{equation}
 n_{\alpha} \; \equiv\sum_{\bba\,\text{(fixed $\alpha$)}} n_{\bba,0}\qquad\in \Z
\end{equation}
where we put $\theta=0$. They give us reduction of transseries (\ref{ZCS-transseries}) for general complex $k$ to the following transseries for $k\in \Z_+$:
\begin{equation}
 Z_\text{CS}(M_3;k) \; = \sum_{\alpha\in \pi_0(\CM_\text{flat}(M_3,SL(2,\C)))} n_{\alpha}
 e^{2\pi ik\text{CS}(\alpha)}Z_{\alpha}^\text{pert}(k)\,.
 \label{ZCS-int-k-trans}
\end{equation}
The CS partition function is expected to have the following asymptotics for integer $k\rightarrow +\infty$:
\begin{equation}
 Z_\text{CS}(k)\; \approx \sum_{\alpha\in \pi_0(\CM_\text{flat}(M_3,SU(2)))} e^{2\pi ik\text{CS}(\alpha)}Z_{\alpha}^\text{pert}(k)\,.
\end{equation}
It follows that, necessarily,
\begin{equation}
n_{\alpha}=1,\qquad \forall \alpha\in  \pi_0(\CM_\text{flat}(M_3,SU(2)))
 \label{trans-cond-su2}
\end{equation}
and
\begin{equation}
n_{\alpha}=0,\qquad \forall \alpha\in  \pi_0(\CM_\text{flat}(M_3,SL(2,\C))),\;\text{Im}\, \text{CS}(\alpha) \leq 0.
 \label{trans-cond-im}
\end{equation}
where the sums are performed over the elements of the $\Z$ factor in (\ref{conn-flat}) and we put $\theta=0$.
In the case of $SU(2)$ gauge group that we consider in this paper,
there are three basic types of flat connections that can be characterized by their stabilizers of $SU(2)$ action on $\text{Hom}(\pi_1(M_3),SU(2))$:
\begin{itemize}

\item
$SU(2)$ -- central;

\item
$U(1)$ -- abelian;

\item
$\{ \pm 1 \}$ -- irreducible.

\end{itemize}

From a practical point of view,
it is often useful to consider central and abelian connections on the same footing. Therefore in what follows we use terminology in which
\begin{equation}
\{\text{central}\} \subset \{\text{abelian}\}\,.
\end{equation}

Consider in more detail a simple case when $\CM_\text{flat}(M_3,SL(2,\C))$ is a discrete set. More general case is reviewed in Appendix \ref{app:PL}. The dimensions $d_{\alpha}\equiv \dim_\C \tilde\CM_\bba$ corresponding to central, other abelian and irreducible flat connections then have values 0, 2 and 3 respectively. It follows that perturbative expansions have the following forms:
\begin{equation}
Z_{\alpha}^\text{pert}  = \left\{
\begin{array}{cl}
 \sum_{n=0}^\infty a^{\alpha}_n\,k^{-n-1/2}\, &{\alpha}\in \, \text{abelian ($a^\alpha_0=0$ for central)},\\
 \sum_{n=0}^\infty a^{\alpha}_n\,k^{-n} \, &{\alpha}\in \, \text{irreducible}.\\
\end{array}\right.
\label{ZaexpansSU2}
\end{equation}
In the case of degenerate critical point/manifold, the ``offset'' of integer powers of $k$ by $-\frac{1}{2}$
or $-\frac{3}{2}$ that we saw here is replaced by the {\it spectrum} of the singularity \cite{AGZV}
(see \cite{Elliot:2015pra} for a physics-oriented introduction and an application).
In particular, the corresponding generalization of \eqref{Iintegralat} and \eqref{ZaexpansSU2} looks like:
\be
I_{\bba,\theta} \; = \; e^{2\pi ikS_\bba}
\, \sum_{s=0}^{\mu_{\alpha}-1} \, \sum_{r \, \in \, \text{Spectrum} (\alpha)} \frac{(\log k)^s}{k^{r - 1/2}} \, Z_{\alpha,s,r}^\text{pert} \,,\qquad
Z_{\alpha,s,r}^\text{pert} \in k^{\frac{d_{\alpha}-3}{2}} \C[[k^{-1}]]
\ee
where $\mu_{\alpha}$ is the Milnor number of the vanishing cycle.
In the non-degenerate case, $\mu_{\alpha} = 1$ and the action $S$ has quadratic expansion (of type $A_1$)
around a critical set, {\it cf.} \eqref{actionA1}, so that its spectrum comprises only one rational number $r = \frac{1}{2}$.

The Borel transform of \eqref{ZaexpansSU2} has the following expansion in the vicinity of $\xi=\xi_{\bba}\equiv -2\pi iS_\bba$:
\begin{equation}
 B_\text{pert}^{\bba}(\xi)\equiv BZ_{\alpha}^\text{pert}(\xi-\xi_\bba)
 =\left\{
 \begin{array}{cl}
  \sum_{n=0}^\infty \frac{a^{\alpha}_n}{\Gamma(n+1/2)}\,(\xi-\xi_\bba)^{n-1/2}\, &\bba\in \, \text{abelian},\\
  &\\
  \sum_{n=1}^\infty \frac{a^{\alpha}_n}{\Gamma(n)}\,(\xi-\xi_\bba)^{n-1} \, &\bba\in \, \text{irreducible}.\\
 \end{array}\right.
\end{equation}
\begin{figure}[ht]
\centering
 \includegraphics[scale=2.0]{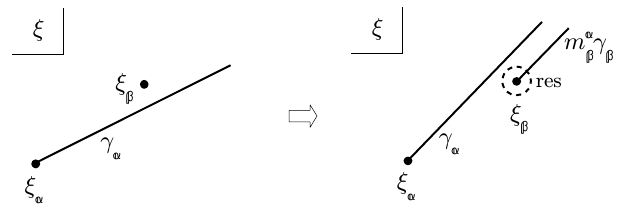}
\caption{Graphical representatation of a Stokes phenomenon in the Borel plane. The dashed circle depicts contribution of non-integral terms in (\ref{Borel-stokes-0}) which are given by residue of $B^{\bba}(\xi)e^{-k\xi}$ at $\xi=\xi_{\bbb}$.}
\label{fig:PL-stokes-0}
\end{figure}
If finite dimensional intuition does not fail, they are expected to have finite radii
of convergence and can be analytically continued to (multivalued) functions $B^\bba(\xi)$
on $\C$ with possible singularities/branch points at $\{\xi_\bba\}_\bba$.
The integral over the Lefschetz thimbles is then given by (see Appendix \ref{app:PL} for details):
\begin{equation}
I^{\bba}(\xi)=e^{2\pi ikS_\bba}
 \left\{
 \begin{array}{cl}
  \int_{\gamma_\bba} d\xi e^{-k(\xi-\xi_{\bba})} B^{\bba}(\xi)\, &\qquad\bba\in \, \text{abelian},\\
  &\\
 a_0^{\alpha}+\int_{\gamma_\bba}d\xi e^{-k(\xi-\xi_{\bba})} B^{\bba}(\xi) \, &\qquad\bba\in \, \text{irreducible}.\\
 \end{array}\right.
 \label{Borel-stokes-0}
\end{equation}
where
\begin{equation}
\gamma_\bba = \{\xi_\bba+e^{-i\theta}\R_+\}
\end{equation}
is the ray of steepest descent in the Borel plane starting at $\xi_\bba$. The behavior of $B^{\bba}(\xi)$ near $\xi=\xi_{\bbb}$ has the following form:
\begin{equation}
B^{\bba}(\xi)\;=\;\text{regular}\;+\;\\
 m^{\bba}_{\bbb}\,\left[\frac{1}{2\pi i}\,\frac{a_0^{\beta}}{\xi-\xi_{\bbb}}+\frac{\log(\xi-\xi_{\bbb})}{2\pi i}
 \,B^\bbb(\xi)\right],\qquad {\bbb} \in\text{irreducible} 
 \label{Borel-general-pole}
\end{equation}
which is, of course, in agreement with the Stokes phenomenon (\ref{Stokes-theta})
that happens when $\gamma_{\bba}$ passes through $\xi_{\bbb}$ and schematically depicted in Figure~\ref{fig:PL-stokes-0}.
As was argued in \cite{Witten:2010cx}, one expects that
\begin{equation}
 m^{\bba}_{\bbb}=0,\; {\bba} \in\text{irreducible},\; {\bbb} \in\text{abelian}
\end{equation}
which indeed will be the case in our examples\footnote{
Similarly one expects that
\begin{equation}
 m^{\bba}_{\bbb}=0,\; {\bba} \in\text{abelian},\; {\bbb} \in\text{central}.
\end{equation}
Later in the text we will see that, in the case of Seifert rational homology spheres, a stronger statement is actually true:
\begin{equation}
 m^{\bba}_{\bbb}=0,\; {\bba} \in\text{abelian},\; {\bbb} \in\text{abelian}.
\end{equation}
In the ``abelianization'' approach of \cite{Blau:2006gh}, this can be argued from the fact that different abelian connections correspond to different topological sectors of $U(1)$ bundle over the base of Seifert fibration which is obtained via ``diagonalization'' procedure.
}.


\subsection{Decomposition into abelian flat connections}

We would like to represent analytically continued CS partition function (that reproduces exactly WRT invariant for $k\in \Z$) as a sum over abelian elements of $\CM_\text{flat}(M_3,SU(2))$ which has the following form:
\begin{equation}
 Z_\text{CS}(M_3;k)=\sum_{a\in\text{abelian}\subset \CM_\text{flat}(M_3,SU(2))}e^{2\pi i kS_\mathbb{a}}Z_{a}(k)
 \label{abelian-resurgence}
\end{equation}
where $\mathbb{a}$ is some fixed representative of $a$  and
so that
\begin{equation}
Z_{a}(k)\sim Z^\text{pert}_{a}(k)
\end{equation}
at the perturbative level. It is therefore natural to assume that it has the following form
\begin{equation}
e^{2\pi i k\text{CS}(a)}Z_{a}(k)=I_{\mathbb{a},0}+\sum_{\bbb\notin \text{abelian}} n_{a,\bbb}I_{\bbb,0}
\end{equation}
where we took $\theta=0$. From (\ref{ZCS-transseries}) it follows that
\begin{equation}
\sum_{{a}\in\text{abelian}} n_{{a},\bbb}=n_{\bbb,0}\,.
\end{equation}
There is an ambiguity in the choice of $Z_{{a}}(k)$ since only
\begin{equation}
 n_{\beta}= \sum_{\bbb\,(\text{fixed }\beta)}\sum_{{a}\in\text{abelian}} n_{{a},\bbb}
\end{equation}
is fixed.
However, in principle, one can impose additional constraints coming from the requirement that
\begin{equation}
Z_{{a}}(k)=\frac{1}{i\sqrt{2k}}\sum_{b\in\text{abelian}}{S_{{a}b}}\hat{Z}_{b}(q),\qquad \hat{Z}_{{a}}(q)\in q^{\Delta_a} \Z[[q]]
\end{equation}
where $S$ is a $k$-independent matrix whose explicit form appeared in \cite{Gukov:2016gkn}.

Note that, as long as there is non-trivial Stokes phenomenon from abelian to irreducible flat connections, that is
 \begin{equation}
 	\forall {\mathbb{a}}\in\text{abelian},\; \exists \bbb \notin\text{abelian}\;\text{s.t. }
 	m^\bbb_{\mathbb{a}} \neq 0
 	\label{good-resurgence}
 \end{equation}
one, in principle, can always recover the full partition function from asymptotic expansions around abelian flat connections. In particular, there exist contours (or linear combination of) $\tilde\gamma_{\mathbb{a}}$ which start at ${\mathbb{a}}$ and the following choice of $Z_{a}(k)$ in (\ref{abelian-resurgence})
 \begin{equation}
 	Z_{a}(k)=\int_{\tilde\gamma_{\mathbb{a}}} B^{\mathbb{a}}(\xi) e^{-k(\xi-\xi_{\mathbb{a}})}\,d\xi\,.
 \end{equation}
However, apart from the obvious technical difficulties, there is also a problem that $n_\beta$ for complex flat connections are in general unknown.

The property (\ref{good-resurgence}) appears to hold for various examples that we consider in the paper.
In particular, there is a nice simple class of Seifert 3-manifolds with three exceptional fibers,
where all singularities in the Borel plane are localized along $i\R_+$ and one can choose\footnote{There is evidence that the similar statement holds for more general Seifert manifolds.} (\textit{cf.} \cite{costin2011resurgence}):
\begin{equation}
e^{2\pi i kS_{\mathbb{a}}}Z_{{a}}(k)=\frac{1}{2}\left(I_{{\mathbb{a}},\pi}+I_{{\mathbb{a}},0}\right)=
I_{{\mathbb{a}},0}+\frac{1}{2}\sum_{\bbb}m_{\mathbb{a}}^\bbb I_{\bbb,0}
\label{Borel-half-sum}
\end{equation}
where, as before,  ${\mathbb{a}}$ is a particular representative of ${a}$, and we have applied Stokes phenomenon at $\theta =\pi/2$.
Via resurgent analysis, we will check explicitly that
\begin{equation}
n_{\beta} \; = \; \frac{1}{2}\sum_{\bbb\,(\text{fixed }\beta)}\sum_{{a}\in\text{abelian}} m_{\mathbb{a}}^\bbb
\end{equation}
have correct values, that is (\ref{trans-cond-su2})-(\ref{trans-cond-im}) are satisfied.

In the setting of ``abelianization'' of \cite{Blau:2006gh}, which can be applied to CS theory on a Seifert manifold, the integrals $I_{{\mathbb{a}},0}$ and $I_{{\mathbb{a}},\pi}$ can be also defined as follows.
The ``abelianization'' procedure fixes a maximal torus $T=U(1)\subset SU(2)$, where holonomy around fiber takes values.
Then, $I_{{\mathbb{a}},0}$ and $I_{{\mathbb{a}},\pi}$ are realized as the path integrals over (aprropriate lifts to universal cover of) contours where the components of the connection along the base of Seifert fibration are taken to be in the $SL(2,\R)$ subgroup of $SL(2,\C)$ that has the same $T=U(1)$ as a subgroup. The label ${a}$ in (\ref{Borel-half-sum}) corresponds to a choice of the first Chern class of the $T$-bundle over the base $\Sigma$ of the Seifert fibration, $M_3 \to \Sigma$.


\section{Non-abelian flat connections as transseries}

In this section, we study resurgence for mock modular forms and for their composites,
quantum invariants of certain Seifert 3-manifolds.\footnote{As we mentioned around \eqref{Zaqt},
in our story, these mock modular forms are the generating functions
of BPS spectra (a.k.a. $Q$-cohomology) of 3-manifolds, see~\cite{Gukov:2016gkn} for more details.
It would be a great opportunity missed by Nature if there is no relation to
other instances in string theory where mock modular forms appear as generating functions of BPS
spectra~\cite{Eguchi:2008gc,Troost:2010ud,Alexandrov:2012au,Dabholkar:2012nd,Cheng:2013kpa,Harvey:2013mda,Cheng:2014zpa,Cheng:2014owa,Haghighat:2015ega},
some of which also involve compactifications on Calabi-Yau 3-folds and share other similarities with BPS spectra of 3-manifolds~\cite{Gukov:2016gkn}.
As a part of developing the dictionary (or, perhaps, a concrete string duality) one might ask if the resurgent analysis
here has any interpretation in other string backgrounds where a connection with mock modular forms was found.}
We will be able to write the exact form of the Borel transform and see explicitly a number of peculiar phenomena.
Thus, it is remarkable to see how contributions of non-abelian flat connections get ``attached'' to the Borel resummation
of the perturbative series around an abelian flat connection.


\subsection{Warm-up: Ramanujan's mock theta-functions}

In his famous last letter to Hardy, Ramanujan wrote a list of 17 mock modular forms --- which back
then he called {\it mock theta-functions} --- of ``order 3'', ``order 5'', and ``order 7''.
For example, one of his order-5 mock theta-functions is the following $q$-series:
\be
\chi_0 (q) \; = \; \sum_{n=0}^{\infty} \frac{q^n}{(q^{n+1})_n}
\label{Ramanujanchi0}
\ee
where $(x)_n \equiv (x;q)_n = (1-x)(1-xq) \ldots (1 - xq^{n-1})$ is the standard shorthand notation for the $q$-Pochhammer symbol.
In fact, Ramanujan wrote five pairs of such functions which transform as vector-valued modular forms under $SL(2,\Z)$,
and $\chi_0 (q)$ is a member of one of the pairs.
Soon, we shall see $\chi_0 (q)$ as a quantum group invariant of the Poincar\'e sphere $\Sigma (2,3,5)$,
but for now we wish to ignore its interpretation and simply try out the power of resurgent analysis on mock modular forms.

For this, it is convenient to introduce the following basis of nearly modular (or, mock modular) theta-functions:
\be
\tilde \Psi^{(a)}_p (q)  :=  \sum_{n=0}^\infty \psi^{(a)}_{2p}(n) q^{\frac{n^2}{4p}} \qquad \in q^\frac{a^2}{4p}\,\Z[[q]]
\label{Psis}
\ee
which are Eichler integrals of weight-$\tfrac{3}{2}$ vector-valued modular forms.
Here,
\begin{equation}
 \psi^{(a)}_{2p}(n)  =  \left\{
\begin{array}{cl}
 \pm 1, & n\equiv \pm a\mod 2p\,, \\
0, & \text{otherwise}.
\end{array}\right.
\end{equation}
Both Ramanujan's mock theta-functions as well as 3-manifold invariants
can be expressed as linear combinations of $\tilde \Psi^{(a)}_p (q)$ with various values of $p$ and $a \in \Z / p\Z$.

Therefore, without loss of generality, we can study resurgence of the functions $\tilde \Psi^{(a)}_p (q)$
by writing the ``perturbative'' expansion in $\hbar$ --- or, equivalently, in $\frac{1}{k}$, where $q = e^{\hbar} = e^{2 \pi i / k}$ ---
and then comparing the original function $\tilde \Psi^{(a)}_p (q)$ to the Borel resummation.
We will see shortly that, with functions $\tilde \Psi^{(a)}_p (q)$, we are in a very special lucky situation,
where the Borel transform can be performed exactly. In particular, it will allow to analyze the singularities on the Borel plane.
For now, however, we proceed slowly and gain some intuition by the direct approach,
which may be the only tool available for the analysis of perturbative expansions of more general 3-manifolds.
The reader may find it helpful to work with a concrete example of the function \eqref{Psis}, {\it e.g.}
\be
Z(q) \; = \; \tilde \Psi^{(1)}_6 (q)  \; = \; \sum_{n=0}^\infty \psi^{(1)}_{12}(n)q^{\frac{n^2}{24}}\qquad \in q^\frac{1}{24}\,\Z[[q]]
\label{Psis16}
\ee
which, roughly speaking, contains half of the terms in the $q$-series expansion of the Dedekind eta-function.\footnote{The function
$\tilde\Psi^{(1)}_6 (q)$ also enters \cite{hikami2006quantum} quantum group invariants of Seifert manifolds
$\Sigma (2,3,3)$, $\Sigma (3,3,6)$, and $\Sigma (3,4,5)$, but we promised to ignore this for now.}

\begin{figure}[h]
\centering
\begin{minipage}{0.45\textwidth}
\centering
\includegraphics[width=0.8\textwidth, height=0.2\textheight]{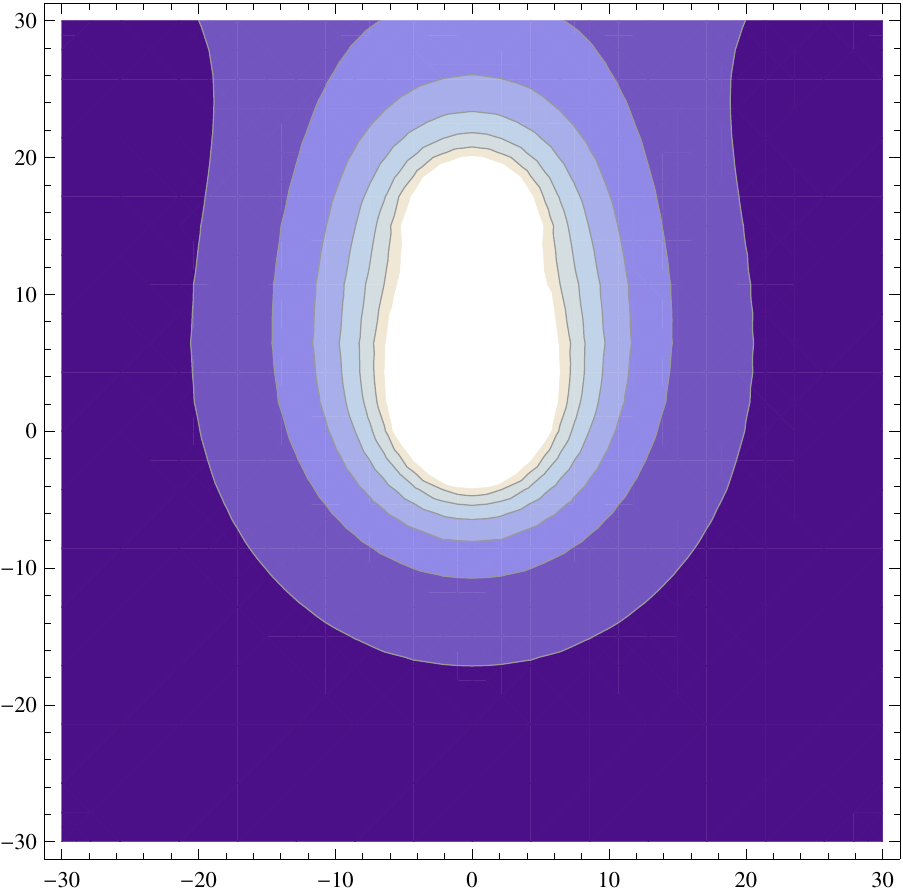}
\caption{The Borel plane.}
\label{fig:res2}
\end{minipage}
\begin{minipage}{0.45\textwidth}
\centering
\includegraphics[width=0.8\textwidth, height=0.2\textheight]{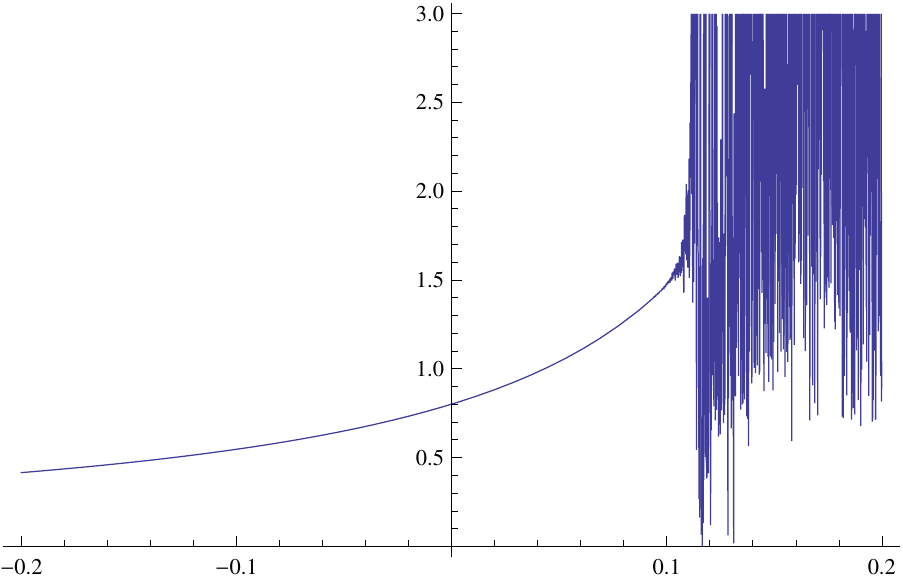}
\caption{A numerical approximation to $|\tilde{BZ}_{\text{pert}} (\xi)|$ for $\xi \in i \R$.}
\label{fig:res3}
\end{minipage}
\end{figure}

Given the exact function $Z(q)$, it is easy to write its perturbative expansion in $\hbar$ or, equivalently,
in $1/k$, where $\hbar = \frac{2\pi i}{k}$:
\be
Z_{\text{pert}} = \sum_{n=0}^{\infty} \frac{a_n}{k^n} \quad \text{as} \quad k \to \infty\,.
\label{Zpertank}
\ee
In our concrete example of \eqref{Psis16}, we have
\be
a_n = - \frac{(12 \pi i)^n}{(2n+1) n!} \left( B_{2n+1} (\tfrac{1}{12}) - B_{2n+1} (\tfrac{11}{12}) \right)
\ee
where $B_{2n+1}$ is Bernoulli polynomial. From this perturbative series one can construct its Borel transform,
\be
B Z_{\text{pert}} (\xi) \; = \; \sum_{n=1}^{\infty} \frac{a_n}{(n-1)!} \xi^{n-1}
\label{Borel16}
\ee
whose analytic structure is shown in Figure~\ref{fig:res2}.
In order to construct the Borel resummation of $Z_{\text{pert}}$ we need
an analytic continuation $\tilde{BZ}_{\text{pert}} (\xi)$ of the series \eqref{Borel16}.
In practice, we can take a finite-order approximation to \eqref{Borel16},
\be
B Z^{(2N+1)}_{\text{pert}} (\xi) \; = \; \sum_{n=1}^{2N+1} \frac{a_n}{(n-1)!} \xi^{n-1}
\ee
and introduce the diagonal Pad\'e approximants to $B Z^{(2N+1)}_{\text{pert}} (\xi)$, {\it i.e.} rational functions
\be
\tilde{BZ}^{(2N+1)}_{\text{pert}} (\xi) = \frac{\sum_{i=0}^N c_i \xi^i}{1 + \sum_{j=1}^N d_j \xi^j}
\ee
whose power series expansion at $\xi \approx 0$ agrees with that of \eqref{Borel16} up to order $2N+1$.
By improving the approximation (that is, increasing the values of $N$) it is easy to see that
the original Borel transform \eqref{Borel16} has no singularities, except on the imaginary axis:
\be
\xi \in i \R_+\,.
\ee
For this reason, in Figure~\ref{fig:res3} we focus specifically on the values of $|\tilde{BZ}_{\text{pert}} (\xi)|$
along the imaginary axis.
Both Figures \ref{fig:res2} and \ref{fig:res3} may appear to suggest that singularities
of the Borel transform \eqref{Borel16} start only at some non-zero imaginary value of $\xi$,
\be
\text{Im} (\xi) \ge \xi_0 \quad \text{where} \quad \xi_0 > 0 \,.
\ee
Shortly, by constructing the exact Borel transform, we will confirm that this is indeed the case.
At this point, however, it is not clear (even with the improved precision to higher values of $N$)
whether the singularities along the imaginary axis $\xi \in i \R_+$ are poles or a brunch cut.
A good news, though, is that the diagonal Pad\'e approximants $\tilde{BZ}^{(2N+1)}_{\text{pert}} (\xi)$
are all non-singular at $\xi = 0$.

Therefore, our next and the final step is to construct the generalized Borel sum of $Z_{\text{pert}}$:
\be
S_{\theta} Z_{\text{pert}} (k) = a_0 + \int_{e^{-i\theta}\R_+} d \xi \, \tilde{BZ}_{\text{pert}} (\xi) \, e^{- k \xi}
\label{genBorelint}
\ee
where for the analytic continuation of \eqref{Borel16} we take its Pad\'e approximation $\tilde{BZ}^{(2N+1)}_{\text{pert}} (\xi)$.
Since we concluded that the latter are non-singular along the real axis, we can simply take the standard Borel sum
$SZ_{\text{pert}} (z)$ with $\theta =0$ and compare it with the exact function \eqref{Psis16}.

Comparing $Z(q) = \tilde \Psi^{(1)}_6 (q)$ with its Borel resummation $SZ_{\text{pert}} (k)$
we first notice that the former is manifestly real for real values of $q \in (-1,1)$, where the series \eqref{Psis16} converges.
Comparing this to the behavior of $SZ_{\text{pert}} (z)$ with $q = e^{2 \pi i/k}$,
we find that the latter is {\it not} real-valued at $k \in - i \R_+$,
but their real parts coincide (at least to the accuracy we were able to achieve):
\be
\text{Re}\, SZ_{\text{pert}} (k) \vert_{k \in -i \R_+} \; = \; \tilde \Psi^{(1)}_6 (q = e^{2 \pi i/k})\,.
\ee
The exact analysis we are going to perform next will tell us that the correct prescription does involve
the generalized Borel sum \eqref{genBorelint}; namely, the functions $\tilde \Psi^{(a)}_p (q)$
(including our example with $p=6$ and $a=1$)
are given by the average of two generalized Borel sums (\textit{cf.} \cite{costin2011resurgence}):
\be
Z(q) \; = \; \frac{1}{2} \Big[ S_{0} Z_{\text{pert}} (k) + S_{\pi} Z_{\text{pert}} (k) \Big]\,.
\label{Zaveragei}
\ee
However, to produce the exact Borel transform we need a little bit of magic.


\subsection{The unreasonable effectiveness of physics}

To paraphrase a famous quote by Eugene Wigner, complex Chern-Simons theory and its formulation
as the vortex partition function of $T[M_3]$ turns out to be unreasonably effective in the resurgent analysis of mock
modular forms, even though at the first sight the two subjects seem to be completely unrelated.

Indeed, while our original problem was purely mathematical and centered around resurgent analysis
of the remarkable function $\tilde \Psi^{(a)}_p (q)$, it turns out that making a small modification
motivated from physics can dramatically simplify the analysis and produce a simple closed form expression
for the Borel transform $\tilde{BZ}_{\text{pert}} (\xi)$.
The modification consists of multiplying $Z_{\text{pert}}$ by $\frac{1}{\sqrt{k}}$:
\be
Z_{\text{pert}}^{\text{CS}} \; = \; \sum_{n=0}^{\infty} \frac{a_n}{k^{n+1/2}}\,.
\label{Zpertofkhalf}
\ee
We label the resulting series by the superscript ``CS'' since it represents the way $\tilde \Psi^{(a)}_p (q)$
enters homological blocks and the partition function of Chern-Simons theory.
Then, the corresponding Borel transform involves $\Gamma (n+\tfrac{1}{2})$ instead of the standard $(n-1)!$ that was used in \eqref{Borel16}:
\be
B Z^{\text{CS}}_{\text{pert}} (\xi) \; = \; \sum_{n=1}^{\infty} \frac{a_n}{\Gamma (n+\tfrac{1}{2})} \xi^{n-\frac{1}{2}}\,.
\ee
Moreover, using the explicit expression for the gamma function at half-integer values
\be
\Gamma \left( n + \tfrac{1}{2} \right) \; = \; \frac{\sqrt{\pi}}{4^n} \cdot \frac{(2n)!}{n!}
\ee
we learn that
\be
B Z^{\text{CS}}_{\text{pert}} (\xi) \; = \; \frac{1}{\sqrt{\xi}}
\sum_{n=0}^{\infty} a_n \frac{4^n}{\sqrt{\pi}} \cdot \frac{n!}{(2n)!} \xi^{n}
\; = \;
\frac{1}{\sqrt{\pi \xi}} \sum_{n=0}^{\infty} c_n \frac{n!}{(2n)!}
\left( \frac{2 \pi i}{p} \right)^n \xi^{n}
\label{BZac}
\ee
where in the last expression we introduced a new set of perturbative coefficients $c_n$ related to $a_n$ in a simple way.
Now comes the magic moment.

The reason we introduced the new coefficients $c_n$ is that, precisely in this form, the right-hand side of \eqref{BZac}
can be matched with the generating function
\be
\frac{\sinh ((p-a)z)}{\sinh (pz)} \; = \; \sum_{n=0}^{\infty} c_n \frac{n!}{(2n)!} z^{2n}
\label{sinhgen}
\ee
that conveniently packages all ``perturbative'' coefficients of $\tilde \Psi^{(a)}_p (q)$.
Specifically, mock modularity implies \cite{hikami2004quantum} (see also \cite{lawrence1999modular}):
\be
\tilde \Psi_p^{(a)} (q) \; = \; - \sqrt{\frac{k}{i}} \sum_{b=1}^{p-1} M_{ab}
\tilde\Psi_p^{(b)}(e^{-2\pi ik})
\; + \; \sum_{n=0}^{\infty} \frac{c_n}{k^n} \left( \frac{\pi i}{2p} \right)^n
\label{Psi-modular}
\ee
where
\begin{equation}
	M_{ab}=\sqrt\frac{2}{p}\,\sin\frac{\pi ab}{p}\,.
\end{equation}
If we take $k$ to be integer (\ref{Psi-modular}) reduces to
\be
\tilde \Psi_p^{(a)} (q) \; = \; - \sqrt{\frac{k}{i}} \sum_{b=1}^{p-1} M_{ab}
\left( 1 - \frac{b}{p} \right) e^{- \frac{b^2}{2p} \pi i k}
\; + \; \sum_{n=0}^{\infty} \frac{c_n}{k^n} \left( \frac{\pi i}{2p} \right)^n\,.
\label{Psipertnonpert}
\ee
The first parts in (\ref{Psi-modular}) or (\ref{Psipertnonpert}) are ``non-perturbative'' in $\hbar = \frac{2 \pi i}{k}$,
and the second part contains the same perturbative coefficients $c_n$ as in our original $\frac{1}{k}$-expansion \eqref{Zpertofkhalf}
and, more importantly, as in the Borel transform \eqref{BZac}.
Comparing the right-hand side of \eqref{BZac} with the generating function \eqref{sinhgen}
it is easy to see that the two coincide if we identify
\be
z^2 \; = \; \frac{2\pi i}{p} \xi \,.
\label{zviaxi}
\ee
As a result, we obtain the exact form of the Borel transform:
\be
\tilde{BZ}^{\text{CS}}_{\text{pert}} (\xi) \; = \;
\frac{1}{\sqrt{\pi \xi}}
\frac{\sinh \left( (p-a)\sqrt{\frac{2\pi i}{p} \xi} \right)}{\sinh \left( p\sqrt{\frac{2\pi i}{p} \xi} \right)} \,,
\label{exactBorelk}
\ee
in fact, not only for our concrete example, but for any vector-valued mock modular form $\tilde \Psi_p^{(a)} (q)$.
For any $p$ and $a$, it has a cut starting at $\xi=0$ and simple poles along the positive imaginary axis of $\xi$,
just as anticipated from our earlier numerical analysis.

It is easy to see that the contribution of poles should give the first, ``non-perturbative'' part in \eqref{Psi-modular}. We will now show what choice of integration contour ({\it i.e.} resummation prescription) in the Borel plane reproduces exact $\tilde\Psi^{(a)}_p(q)$. In the process we will give a simpler way to obtain (\ref{exactBorelk}), without using mock modular property.

Suppose we are interested in finding an integral representation of mock modular forms of the following type:
\begin{equation}
 \frac{1}{\sqrt{k}}\,\tilde{\Psi}^{(a)}_p(q)
=
\int_\gamma d\xi B(\xi) e^{-k\xi}
\end{equation}
where $\gamma$ is a certain contour (or a formal linear combination of such that coefficients sum to 1).
Remember that the mock modular form in the left-hand side is an analytic function in the unit disk $|q|<1$ defined by the series
\begin{equation}
 \tilde{\Psi}^{(a)}_p(q)=\sum_{n=0}^\infty \psi^{(a)}_{2p}(n)\,q^\frac{n^2}{4p}
\end{equation}
convergent in this domain. Therefore in what follows we assume $\mathrm{Im}k<0\Leftrightarrow |q|<1$.

\begin{figure}[ht]
\centering
 \includegraphics[scale=2.1]{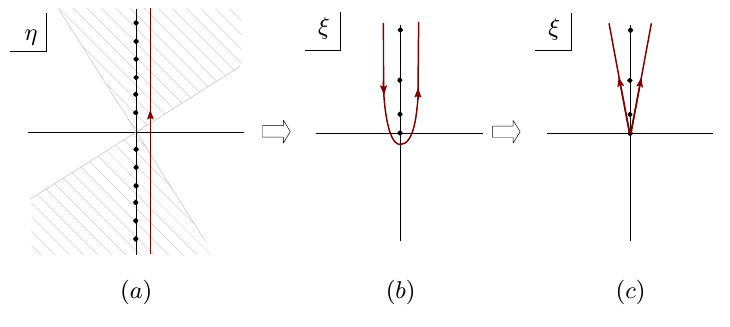}
\caption{The integration contours that give $\tilde \Psi^{(a)}_p (q)$. $(a)$ Integration contour in $\eta$-plane. The directions in which integrand decays are shown as a hatched domain in the case of generic value $k$ such that $\mathrm{Im}\,k<0$. $(b)$ The image of the contour in $\xi$-plane under the coordinate change. $(c)$ Equivalent contour in $\xi$-plane (we used the fact that the integrand changes sign when $\xi$ goes around the origin.  $(all)$ The black dots depict singularities of the integrands. }
\label{fig:Psi-contour}
\end{figure}

First, let us note that
\begin{equation}
 \frac{\sinh(p-a)\eta}{\sinh p\eta}=
\sum_{n=0}^\infty \psi^{(a)}_{2p}(n)\,e^{-n\eta}
\end{equation}
where the infinite sum in the right-hand side is uniformly convergent in the domain $\mathrm{Re}\eta\geq\epsilon>0$ for any $\epsilon$. Therefore we can integrate both sides of this equality over the line $\{\mathrm{Re}\eta=\epsilon\}$ in the $\eta$-plane together with $e^{-\frac{kp\eta^2}{2\pi i}}$ factor, which decays quickly enough at $\eta=\pm i\infty$ so that the integral is absolutely convergent (see Figure~\ref{fig:Psi-contour}$a$):
\begin{equation}
 \int_{i\R+\epsilon}d\eta\, \frac{\sinh(p-a)\eta}{\sinh p\eta}\,e^{-\frac{kp\eta^2}{2\pi i}}=\int_{i\R+\epsilon}d\eta
\sum_{n=0}^\infty \psi^{(a)}_{2p}(n)\,e^{-n\eta}\,e^{-\frac{kp\eta^2}{2\pi i}}\,.
\end{equation}
Gaussian integration on the right-hand side gives us $\tilde{\Psi}^{(a)}_p(q)$, so we get the following formula:
\begin{equation}
 \frac{1}{\sqrt{k}}\tilde{\Psi}^{(a)}_p(q)=
\sqrt\frac{p}{2\pi^2 i}\int_{i\R+\epsilon}d\eta\,  \frac{\sinh(p-a)\eta}{\sinh p\eta}\,e^{-\frac{kp\eta^2}{2\pi i}}\,.
\end{equation}
Then, making the change of variables
\begin{equation}
 \frac{p\eta^2}{2\pi i}=\xi
\end{equation}
and manipulating contours as shown in Figure~\ref{fig:Psi-contour}, we arrive at the following expression
\begin{equation}
  \frac{1}{\sqrt{k}}\tilde{\Psi}^{(a)}_p(q)=
\frac{1}{2}\left(\int_{ie^{+i\delta}\R_+}+\int_{ie^{-i\delta}\R_+}\right)
\,\frac{d\xi}{\sqrt{\pi\xi}}\,
\frac{\sinh(p-a)\sqrt\frac{2\pi i\xi}{p}}{\sinh p\sqrt\frac{2\pi i\xi}{p}}
\,e^{-k\xi}
\label{Psi-Laplace}
\end{equation}
for some small $\delta$. We see that the integrand coincides with (\ref{exactBorelk}) so that the exact function $\tilde \Psi_p^{(a)} (q)$ can be recovered by taking the average of the generalized Borel sums, ({\it cf.} \eqref{Zaveragei} and (\ref{Borel-half-sum})):
\be
Z(q) \; = \; \frac{1}{2} \Big[ \, S_{\frac{\pi}{2}-\delta} Z_{\text{pert}} (k) \; + \; S_{\frac{\pi}{2}+\delta} Z_{\text{pert}} (k) \, \Big]\,.
\label{Zaverageii}
\ee


Note, if instead of $\frac{1}{k}$-expansion we use the expansion in $\hbar = \frac{2\pi i}{k}$,
\be
Z_{\text{pert}}^{\text{CS}} \; = \; \sum_{n=0}^{\infty} a_n^{(\hbar)} \hbar^{n+1/2}
= \sum_{n=0}^{\infty} a^{(\hbar)}_n \left( \frac{2\pi i}{k} \right)^{n+1/2}
= \sum_{n=0}^{\infty} \frac{a_n}{k^{n+1/2}}
\ee
then $a^{(\hbar)}_n = a_n \cdot ( 2\pi i)^{-n-1/2}$ and the Borel transform looks like
\be
BZ^{(\hbar)}_{\text{pert}} (\xi)
\; = \;
\frac{1}{\sqrt{2 i \pi^2 \xi}} \sum_{n=0}^{\infty} c_n \frac{n!}{(2n)!}
\frac{\xi^n}{p^n}\,.
\ee
As in the previous analysis, comparing with \eqref{sinhgen} at $z = \sqrt{\frac{\xi}{p}}$ gives the exact form of the Borel transform:
\be
\tilde{BZ}^{(\hbar)}_{\text{pert}} (\xi) \; = \;
\frac{1}{\sqrt{2 i \pi^2 \xi}}
\frac{\sinh \left( (p-a)\sqrt{\frac{\xi}{p}} \right)}{\sinh \left( p\sqrt{\frac{\xi}{p}} \right)}\,.
\label{exactBorelhbar}
\ee
Whether we prefer a $\frac{1}{k}$-expansion or $\hbar$-expansion, \eqref{exactBorelk} or \eqref{exactBorelhbar}
completely describe the structure of the singularities in the Borel $\xi$-plane.
The crucial steps that led to these expressions were $(i)$ passing from \eqref{Zpertank} to \eqref{Zpertofkhalf}
and $(ii)$ the special fact about mock theta-functions $\tilde \Psi_p^{(a)} (q)$ that their perturbative coefficients
can be packaged into a nice generating function \eqref{sinhgen}.

Now, thanks to  \eqref{exactBorelk} and \eqref{exactBorelhbar}, we can easily describe the resurgence for
any linear combination of the functions $\tilde \Psi_p^{(a)} (q)$.
To this end, it is convenient to adopt the shorthand notation~\cite{hikami2011decomposition}:
\be
\tilde\Psi^{n_a (a) + n_b (b) + \cdots}_p (q) \; := \;
n_a \tilde\Psi^{(a)}_p (q)
+ n_b \tilde\Psi^{(b)}_p (q)
+ \ldots
\label{linearPsitilde}
\ee
Note, under this operation the generating functions \eqref{sinhgen} are additive.
In particular, homological blocks of Seifert 3-manifolds with three singular fibers are all of the form \eqref{linearPsitilde}.


\subsection{General $q$-series}

Note, that the relation (\ref{Psi-Laplace}) can be naively generalized to the case of a more general $q$-series of the form
\begin{equation}
	\Psi(q) \; = \; \sum_{n\geq0}c_mq^\frac{m}{p}
\end{equation}
for some $p\in \Z_+$. Namely,
\begin{equation}
  \frac{1}{\sqrt{k}}{\Psi}(q) \;=\;
\frac{1}{2}\left(\int_{ie^{+i\epsilon}\R_+}+\int_{ie^{-i\epsilon}\R_+}\right)
\frac{d\xi}{\sqrt{\pi\xi}}\,\,
f(\xi)
\,e^{-k\xi}
\label{Psi-Laplacexx}
\end{equation}
where the function $f(\xi)$ has the following expansion:
\begin{equation}
	f(\xi) \; = \; \sum_{m\geq 0}c_m\,e^{-2\sqrt\frac{\pi i m\xi}{p}}\,.
\end{equation}


\subsection{Homological blocks as Borel sums}

Now, we are ready to discuss 3-manifold invariants.
Our general strategy is very simple: starting with a perturbative invariant $Z_{\text{pert}} (M_3)$ around a given flat connection,
we wish to study what the resurgent analysis gives us and, in particular, what kind of invariant is produced by the Borel resummation
of this perturbative series.

If the flat connection is abelian, we shall see that the Borel resummation of $Z_{\text{pert}} (M_3)$ gives precisely
the homological blocks of \cite{Gukov:2016gkn}, attaching suitable contributions of non-abelian flat connections as transseries.
This will explain, from the viewpoint of resurgent analysis, the surprising special role of abelian flat connections that seem to label
basic building blocks $Z_a$ and their linear combinations $\hat Z_a$ obtained by $SL(2,\Z)$ action on
abelian representations $a \in \CM_{\text{flat}} (M_3, G_{\C})$ and suitable for categorification~\cite{Gukov:2016gkn}.
In the rank-1 case and for rational homology spheres, the label $a$ simply takes values in $H_1 (M_3) / \Z_2$.

For simplicity, we limit ourselves here to Seifert 3-manifolds with three singular fiber.
Although some other types of 3-manifolds will be discussed later in the paper, it would be very interesting to extend
this analysis further, to more general 3-manifolds as well as to higher-rank groups.
With our assumptions about $M_3$, its Chern-Simons partition function can be expressed as a linear combination
of vector-valued mock modular forms \eqref{linearPsitilde} or, rather, its ``corrected version'' \eqref{Zpertofkhalf}
with $\frac{1}{\sqrt{k}}$ factor that was essential for the Borel resummation.
Therefore, for this class of 3-manifolds, the resurgent analysis of $Z_{\text{CS}}(M_3)$ simply boild down to that of \eqref{linearPsitilde},
which is why we spent the beginning of this section on mock modular forms.

For example, for the Poincar\'e sphere $M_3=\Sigma (2,3,5)$, the partition function\footnote{\label{WRTfootnote}Note, the WRT invariant $\tau_k (M_3)$ often
used in the mathematics literature is related to the standard normalization of the Chern-Simons partition function via
$$
Z_{\text{CS}}(M_3) = \frac{q-1}{i \sqrt{2k} q^{\frac{1}{2}}} \tau_k (M_3) .
$$}
is given by~\cite{lawrence1999modular}:
\be
Z_{\text{CS}} (\Sigma (2,3,5)) \; = \;
\frac{1}{i \sqrt{2k} q^{\frac{181}{120}}}
\left( q^{\frac{1}{120}} - \frac{1}{2} \tilde\Psi^{(1) + (11) + (19) + (29)}_{30} (q) \right)\,.
\label{Z235}
\ee
The expression in parenthesis on the right-hand side is equal to $q^{\frac{1}{120}} \left( \chi_0 (1/q) - 1 \right)$,
where $\chi_0 (q)$ is precisely the Ramanujan's mock theta-function \eqref{Ramanujanchi0} of order five \cite{zagier2001vassiliev,lawrence1999modular,MR2191375}.
As usual, our goal is to write the perturbative expansion of \eqref{Z235} in $\hbar$ or in $1/k$ and then study its Borel resummation.
According to \eqref{sinhgen}, the Borel transform of $\tilde\Psi^{(1) + (11) + (19) + (29)}_{30} (q)$
is governed by
\footnote{The CS partition function (\ref{Z235}), of course, differs from it by a shift and multiplication by a power of $q$. However since $q^\Delta=e^\frac{2\pi i\,\Delta}{k}$ is an entire function of $1/k$, it does not affect singularity structure of the Borel transform. See Appendix \ref{app:PL} for more details.}
\be
\frac{\sinh (29z)}{\sinh (30z)}
+ \frac{\sinh (19z)}{\sinh (30z)}
+ \frac{\sinh (11z)}{\sinh (30z)}
+ \frac{\sinh (z)}{\sinh (30z)}
\; = \;
\frac{2 \cosh (5z) \cosh (9z) }{\cosh (15z)}\,.
\ee
It has poles at $z = \frac{n \pi i}{p}$ with certain positive integer values of $n$,
whose residues contribute to the part of asymptotic around the suddle with the Chern-Simons invariant
\be
\text{CS} \; = \; -\frac{n^2}{4p} \mod 1\,.
\label{CSnp}
\ee
Indeed, the position of the poles in the Borel $\xi$-plane is precisely the classical ``instanton'' action,
and from \eqref{zviaxi} it follows that a pole at $z = \frac{n \pi i}{p}$ has classical action
\be
\xi \; = \; \frac{p}{2\pi i} \, z^2 \; = \; 2\pi i \, \frac{n^2}{4p}\,.
\ee
In our present example, there are two groups of poles (which repeat modulo $2p=60$):
\begin{itemize}

\item
$n=1$, 11, 19, 29, 31, 41, 49, 59 have $\text{CS} = -\frac{1}{120}$ and residues $\{ -1,-1,-1,-1,1,1,1,1\}$, respectively, up to an overall factor $\frac{i}{10} \sqrt{\frac{1}{6} (5 - \sqrt{5})}$;

\item
$n=7$, 13, 17, 23, 37, 43, 47, 53 have $\text{CS} = -\frac{49}{120}$ and residues $\{ -1,-1,-1,-1,1,1,1,1\}$, respectively, up to an overall factor $\frac{i}{10} \sqrt{\frac{1}{6} (5 + \sqrt{5})}$.

\end{itemize}
This is in total agreement with the expected structure of singularities (\ref{Borel-general-pole}). Note that logarithmic terms in this case vanish, which corresponds to the fact that contribution of irreducible flat connections in finite-loop exact for Seifert manifolds \cite{lawrence1999witten}.

The values of $n$ listed here may appear somewhat random at first, but they actually have significance in the Atkin-Lehner theory
and its variant for mock modular forms, see {\it e.g.} \cite{Dabholkar:2012nd,MirandaNT}.
Namely, for $p=30$ there are only two orbits of the group of Atkin-Lehner involutions on the group of residue classes modulo $2p$,
which perfectly match the two groups of poles in the Borel plane.
The correspondence between the analytic structure on the Borel plane and number theory certainly deserves further study.

When the Chern-Simons level is integer, $k \in \Z$, the poles with $n = \pm a$ mod $2p$ and fixed $a$
correspond to the same value of the Chern-Simons invariant modulo 1 \eqref{CSnp}:
\be
e^{2 \pi i k \text{CS}} = e^{-2\pi i k \frac{a^2}{4p}}\,.
\label{CSviakap}
\ee
\begin{figure}[ht]
\centering
 \includegraphics[scale=2.1]{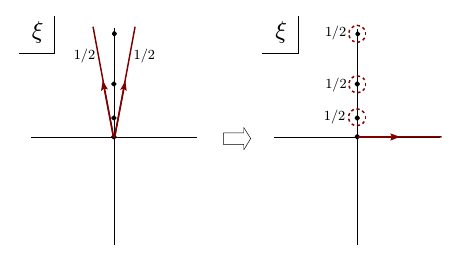}
\caption{The deformation of the contour for generalized Borel resummation (\ref{Zaverageii})  (cf. \ref{Borel-half-sum}) into contours corresponding to Lefschetz thimbles for $k>0$. Note that logarithmic terms in (\ref{Borel-general-pole}) vanish in this case.}
\label{fig:Psi-contour-thimbles}
\end{figure}
Moreover, the residues of these poles determine the weights of these contributions to the integral in
the generalized Borel resummation (\ref{Zaverageii}) (see Fig. \ref{fig:Psi-contour-thimbles}) and, therefore, to the Chern-Simons partition function.
Thus, the first group of poles with $a=1$, 11, 19, 29 contributes
\begin{multline}
- \frac{i}{20} \sqrt{\frac{1}{6} (5 - \sqrt{5})}
\left(\sum_{n=\pm 1 \, \text{mod} \,60}^{\infty} \pm 1 \; + \sum_{n=\pm 11 \, \text{mod} \, 60}^{\infty} \pm 1 \;\right.\\ \left.
+ \sum_{n=\pm 19 \, \text{mod} \, 60}^{\infty} \pm 1 \; + \sum_{n=\pm 29 \, \text{mod} \, 60}^{\infty} \pm 1 \right)
e^{-\frac{\pi i k}{60}}\,.
\end{multline}
Each sum needs to be regularized and gives ({\it e.g.} via zeta-function regularization\footnote{
\begin{equation}
	\left.\sum_{n=\pm a \mod 2p}\pm t^n\right|_{t=1}=
	\left.\left(\frac{t^a}{1-t^{2p}}-\frac{t^{2p-a}}{1-t^{2p}}\right)\right|_{t=1}=1-\frac{a}{p}\,.
\end{equation}
}, which is equivalent to using the modular tranform (\ref{Psi-modular})):
\be
\sum_{n \; = \;\pm a \, \text{mod} \,2p}^{\infty} \pm 1 \; = \; 1 - \frac{a}{p}
\label{ressumregap}
\ee
so that the total contribution of such poles is
\be
- \frac{i}{20} \sqrt{\frac{1}{6} (5 - \sqrt{5})} \; e^{-\frac{\pi i k}{60}}
\sum_{a=1, 11, 19, 29} (1 - \tfrac{a}{p})=
- \frac{i}{10} \sqrt{\frac{1}{6} (5 - \sqrt{5})} \; e^{-\frac{\pi i k}{60}}\,.
\label{poles235i}
\ee
Similarly, the second group of poles with $a=7$, 13, 17, 23 contributes to the Chern-Simons partition function
\be
- \frac{i}{20} \sqrt{\frac{1}{6} (5 + \sqrt{5})} \; e^{-\frac{49 \pi i k}{60}}
\sum_{a=7, 13, 17, 23} (1 - \tfrac{a}{p})=- \frac{i}{10} \sqrt{\frac{1}{6} (5 + \sqrt{5})} \; e^{-\frac{49 \pi i k}{60}}\,.
\label{poles235ii}
\ee
The contributions of the poles \eqref{poles235i} and \eqref{poles235ii} are precisely
the contributions of non-abelian flat connections to the full non-perturbative Chern-Simons partition function~\eqref{Z235}!

Indeed, on the Poincar\'e sphere $M_3=\Sigma (2,3,5)$, there are a total of three $SL(2,\C)$ flat connections,
all of which can be conjugated to the $SU(2)$ subgroup and, therefore, have real values of the classical Chern-Simons action.
In gauge theory literature, these flat connections are usually denoted by $\theta$, $\alpha_1$, and $\alpha_2$.
Here, we denote them by $\alpha_0$, $\alpha_1$, and $\alpha_2$, respectively, since $\theta$ is already used to
denote the angle in the Borel plane.

Since $M_3=\Sigma (2,3,5)$ is an integral homology sphere, that is $H_1 (M_3; \Z) =0$,
there is only one abelian flat connection,\footnote{In particular, $a \in H_1 (M_3) / \Z_2$ that labels $Z_a$ and $\hat Z_a$ in this case takes only one value.}
namely the trivial flat connection $\alpha_0$ (that in gauge theory literature is called $\theta$).
In our analysis here, the $\frac{1}{k}$-expansion of \eqref{Z235} is precisely the perturbative expansion around
the trivial flat connection $\bba_0$:
\be
Z_{\text{pert}} (M_3) \; = \; Z_{\text{pert}}^{\alpha_0} (M_3)\,.
\ee
Clearly, the Chern-Simons functional of $\alpha_0$ is zero, whereas two {\it non-abelian} flat connections $\alpha_1$ and $\alpha_2$ have
\be
\text{CS} (\alpha_1) \; = \; -\frac{49}{120}\mod 1
\qquad , \qquad
\text{CS} (\alpha_2) \; = \; -\frac{1}{120}\mod 1\,.
\label{CSaa12235}
\ee
These are precisely the values of the classical Chern-Simons action \eqref{CSnp}
coming from the poles of the Borel transform $\tilde{BZ}_{\text{pert}} (\xi)$.
Moreover, \eqref{poles235i} and \eqref{poles235ii} lead to a rather striking conclusion that the Borel resummation
of $Z_{\text{pert}}$ with contour shown in Fig. \ref{fig:Psi-contour-thimbles} gives a function
that coincides with the full Chern-Simons partition function \eqref{Z235} at integer and non-integer values of $k$
or, equivalently, gives the unique homological block of $M_3=\Sigma (2,3,5)$.

We also confirmed that Borel transform of trivial flat connection indeed has structure (\ref{Borel-general-pole}) with the following values of the monodromy coefficients in the Stokes phenomenon:
\begin{equation}
	m^{(\alpha_0,0)}_\bbb=\left\{
	\begin{array}{rl}
		1, & \bbb=(\alpha_1,-n^2/120),\; n=1,11,19,29\mod 60\,, \\
		-1, & \bbb=(\alpha_1,-n^2/120),\; n=31,41,49,59\mod 60\,, \\
		1, & \bbb=(\alpha_2,-n^2/120),\; n=7,13,17,23\mod 60\,, \\
		-1, & \bbb=(\alpha_2,-n^2/120),\; n=37,43,47,53\mod 60\,, \\
		0, & \text{otherwise},
	\end{array}
	\right.
\end{equation}
\begin{equation}
	m_\bbb^\bba=0,\qquad \forall\bbb,\bba,\text{ s.t.}\alpha\neq\alpha_0\,,
\end{equation}
where, as before, we use following notation for lifts of flat connections to the universal cover: $\bba=(\alpha,S_\bba)$, where $S_\bba-\text{CS}(\alpha)\in\Z$. As was already mentioned in (\ref{Borel-half-sum}) this gives us the following transseries for analytically continued CS partition function for $\theta=0$:
\begin{equation}
	Z_{\text{CS}} (\Sigma (2,3,5))=Z^{\alpha_0}_\text{pert}+\frac{1}{2}\sum_\bbb m^{(\alpha_0,0)}_\bbb e^{2\pi ikS_\bbb}Z^{\beta}_\text{pert}
\end{equation}
which, as we just checked, reduces (in the sense of (\ref{ZCS-int-k-trans})) to
\be
Z_{\text{CS}} (\Sigma (2,3,5)) \; = \;
\sum_{\alpha=\alpha_0,\alpha_1,\alpha_2}n_{\alpha} e^{2 \pi i k \text{CS} (\alpha)}Z_{\text{pert}}^{\alpha}
\label{ZCSaaa235}
\ee
when $k\in \Z_+$ with the reduced transseries parameters
\be
n_{\alpha_i} \; = \; 1\,, \qquad i=1,2,3\,.
\ee

Thanks to the significant body of the groundwork \cite{Hikami,hikami2006quantum,hikami2011decomposition,MR2191375},
one can perform the resurgent analysis just as concretely for other Seifert manifolds.
In the next subsection we present one more example that illustrates an interesting new phenomenon.

Before we proceed, though, let us note that
\begin{equation}
	\sum_{\bbb\;(\text{fixed }\beta=\alpha_1)}m^{(\alpha_0,0)}_{\bbb}\,{\tilde q}^{-S_\bbb}\;=\;\tilde\Psi^{(1)+(11)+(19)+(29)}_{30}(\tilde q)\,,
\end{equation}
\begin{equation}
	\sum_{\bbb\;(\text{fixed }\beta=\alpha_2)}m^{(\alpha_0,0)}_{\bbb}\,{\tilde q}^{-S_\bbb}\;=\;\tilde\Psi^{(7)+(13)+(17)+(23)}_{30}(\tilde q)
\end{equation}
are the mock modular forms that appear if we perform mock modular transform of (\ref{Z235}) according to (\ref{Psi-modular}),
and $\tilde q =e^{-2\pi i k}$.
For general $M_3$, not necessarily Seifert, the following $\tilde{q}$-series with integer coefficients series appear
in the Stokes phenomenon jump of $Z_\alpha^\text{pert}$ as an overall factor in front of $Z_\beta^\text{pert}$:
\begin{equation}
	m_\beta^\alpha(\tilde q) \equiv \sum_{\bbb\;(\text{fixed }\beta)}m^{(\alpha,\text{CS}(\alpha))}_\bbb \,\tilde{q}^{\text{CS}(\alpha)-S_\bbb}\qquad \in\; \tilde{q}^{\text{CS}(\alpha)-\text{CS}(\beta)}\Z[[\tilde q]]\,.
\end{equation}
It would be interesting to see if, for general $M_3$, they have some mock modular properties.


\subsection{Our first complex flat connection}

In principle, starting with \cite{dglz}, one can systematically compute the exact perturbative expansion $Z_{\text{pert}}$
around any complex flat connection. The classical Chern-Simons action of such $SL(2,\C)$ flat connections is, in general,
a complex number, so we should see them as singularities of the Borel transform $\tilde{BZ}_{\text{pert}} (\xi)$ at generic complex values of $\xi$.
Such complex flat connections, though, should not contribute to the asymptotic expansion of Feynman path integral of the $SU(2)$ theory,
unless the parameters are analytically continued away from their allowed valued.
This is precisely what happens in the volume conjecture \cite{MR1434238,MR1828373}
where the highest weight of $SU(2)_k$ representation is analytically continued,
or in its generalized version \cite{Apol} where both the level and the highest weight are analytically continued to complex values;
the latter also leads to a reformulation of complex Chern-Simons theory as a theory on the ``spectral curve''
that will be relevant to us later.

Relegating a more detailed account of complex flat connections to later sections, here,
as an ``appetizer'', we wish to consider the resurgence for a connection which is
``minimally complex'' in a sense that it can not be conjugated inside $SU(2) \subset SL(2,\C)$,
but nevertheless can be conjugated to a $SL(2,\R)$ flat connection.
Still, since it is not a classical solution to field equations in $SU(2)$ Chern-Simons theory,
we expect the corresponding transseries parameter to be zero (at least for integer values of $k$).

The simplest example that illustrates this behavior is the Brieskorn integer homology sphere:
\be
M_3 \; = \; \Sigma (2,3,7)\,.
\label{M237first}
\ee
In the next section, we describe in more detail flat connections on general Brieskorn spheres and the corresponding
values of the Chern-Simons functional; for now, we only need to know that $M_3 \; = \; \Sigma (2,3,7)$ admits
a total of four $SL(2,\C)$ flat connections, $\alpha_i$, $i=0, 1, 2, 3$.
One of them is the trivial flat connection. Following the notations of the previous example, we denote it by $\alpha_0$.
(In gauge theory literature, it is usually called $\theta$.)
There are two {\it non-abelian} flat connections that can be conjugated inside $SU(2)$; let's call them $\alpha_1$ and $\alpha_2$,
also as in the previous example of the Poincar\'e sphere.
Finally, the remaining flat connection, $\alpha_3$, is also non-abelian, but can only be conjugated inside $SL(2,\R)$,
see \cite{KitanoYamaguchi}.
On each of these connections, the classical Chern-Simons functional takes the following values (mod 1):
\be
\text{CS} (\alpha_0) \; = \; 0
\,, \qquad
\text{CS} (\alpha_1) \; = \; -\frac{25}{168}
\,, \qquad
\text{CS} (\alpha_2) \; = \; - \frac{121}{168}
\,, \qquad
\text{CS} (\alpha_3) \; = \; -\frac{1}{168}\,.
\label{SU2237flat}
\ee

Just as in the previous example, there is only one abelian flat connection on $M_3 \; = \; \Sigma (2,3,7)$
and, hence, there is only one choice of the perturbative series $Z_{\text{pert}}$ around the abelian flat connection $\alpha_0$.
As we shall see in a moment, its Borel resummation with the same simple prescription we used before --- namely, adding half the contribution
of all the poles along the positive imaginary axis in the Borel $\xi$-plane, as in Figure~\ref{fig:Psi-contour-thimbles}  --- gives (for $k\in \Z_+$):
\be
Z_{\text{CS}} (M_3) \; = \; \sum_{\bba \, \in \, \CM_{\text{flat}} (G_{\C},M_3) } \; n_{\alpha} \; e^{2 \pi i k \text{CS} (\alpha)} Z_{\text{pert}}^{\alpha}
\ee
with the transseries parameters
\be
n_{\alpha} \; = \;
\begin{cases}
1 \,, & \text{if}~ \alpha \; \in \; \CM_{\text{flat}} (SU(2),M_3) \\
0 \,, & \text{if}~ \alpha \; \not \in \; \CM_{\text{flat}} (SU(2),M_3) \,.
\end{cases}
\label{transparGGC}
\ee
In particular, $n_{\alpha_3} = 0$.
It is quite remarkable that resurgence of the perturbative series around $\alpha_0$ secretely knows
that $\alpha_3$ should be distinguished from the other flat connections!

In order to see this in practice, we follow the same steps as before.
Namely, the full partition function of $SU(2)$ Chern-Simons theory on $M_3 \; = \; \Sigma (2,3,7)$
can be written as a linear combination of vector-valued mock modular forms \eqref{Psis}:
\be
Z_{\text{CS}} (\Sigma (2,3,7)) \; = \; \frac{1}{i \sqrt{8k} q^{\frac{83}{168}}}
\tilde\Psi^{(1) - (13) - (29) + (41)}_{42} (q)
\label{ZCS237}
\ee
where we use the shorthand notation \eqref{linearPsitilde}.
Its $\frac{1}{k}$- or $\hbar$-expansion has the perturbative part \eqref{Zpertofkhalf} and the rest (``non-perturbative'' or ``instanton'' part);
we are interested in the Borel transform of the perturbative part.
Since the perturbative coefficients of $\tilde\Psi^{(a)}_{p} (q)$ are packaged by the generating function \eqref{sinhgen},
which is additive under \eqref{linearPsitilde} and essentially gives the Borel transform \eqref{exactBorelk},
a simple calculation shows that, in the case of $\tilde\Psi^{(1) - (13) - (29) + (41)}_{42} (q)$,
the structure of poles in the Borel plane is governed by a simple function\footnote{To be more precise,
in order to produce $\tilde{BZ}_{\text{pert}} (\xi)$ we need to change variables via \eqref{zviaxi}
and introduce an extra factor of $\sqrt{\xi}$ in the denominator, as in \eqref{exactBorelk}.
But, since these transformations do not affect the pole structure, it is convenient avoid clutter and stay on the $z$-plane, at least for the time being.}
\be
\frac{2 \sinh (6z) \sinh (14 z)}{\cosh (21 z)}\,.
\ee
There are three groups of poles at $z = \frac{n \pi i}{p}$, with $n$ counted modulo $2p=84$:
\begin{itemize}

\item
$n=1$, 13, 29, 41, 43, 55, 71, 83  have $\text{CS} = -\frac{1}{168}$ and residues $\{ +1,-1,-1,+1,-1,+1,+1,-1 \}$, respectively, up to an overall factor $\frac{i \sin (\pi /7)}{7 \sqrt{3}}$;

\item
$n=5$, 19, 23, 37, 47, 61, 65, 79 have $\text{CS} =- \frac{25}{168}$ and residues $\{ -1,-1,-1,-1,+1,+1,+1,+1 \}$, respectively, up to an overall factor $\frac{i \cos (3\pi /14)}{7 \sqrt{3}}$;

\item
$n=11$, 17, 25, 31, 53, 59, 67, 73 have $\text{CS} = -\frac{121}{168}$ and residues $\{ -1,-1,-1,-1,+1,+1,+1,+1\}$, respectively, up to an overall factor $\frac{i \cos (\pi /14)}{7 \sqrt{3}}$.

\end{itemize}
The position of poles on the Borel $\xi$-plane determines the classical ``instanton'' action,
which is given by the familiar formula \eqref{CSnp} with $p=42$ that follows from the relation \eqref{zviaxi} between variables $z$ and $\xi$.
In particular, we see that the three groups of poles here correspond to the three non-abelian flat connections: $\alpha_1$, $\alpha_2$, and $\alpha_3$.

If we use the same prescription as before and sum over all poles with $1/2$ factor --- which corresponds to the integral
in Figure~\ref{fig:Psi-contour-thimbles} --- then the regularized sum over the residues \eqref{ressumregap} recovers the exact function \eqref{ZCS237}
as the following transseries (with $\theta=0$):
\be
Z_{\text{CS}} (M_3)
\; = \; \frac{1}{2} \Big[ \, S_{\frac{\pi}{2}-\epsilon} Z_{\text{pert}} (k) \; + \; S_{\frac{\pi}{2}+\epsilon} Z_{\text{pert}}^{\alpha_0} (k) \, \Big]
\; = \; \sum_{\bba } \; n_{\bba,0} \; e^{2 \pi i k S_\bba} Z_{\text{pert}}^{\alpha}
\ee
where, as before, the transseries coefficients are related to the Stokes monodromy coefficients as follows:
\begin{equation}
	n_{\bba,0}=\left\{
	\begin{array}{cl}
		1, & \bba=(\alpha_0,0)\\
		\frac{1}{2}m^{(\alpha_0,0)}_\bba,& \text{otherwise}
	\end{array}
	\right.
\end{equation}
which equal to
\begin{equation}
	m^{(\alpha_0,0)}_\bbb=\left\{
	\begin{array}{rl}
		1, & \bbb=(\alpha_1,-n^2/168),\; n=13,29,43,55\mod 84\,,\\
		-1, & \bbb=(\alpha_1,-n^2/168),\; n=1,41,55,71\mod 84\,, \\
		1, & \bbb=(\alpha_2,-n^2/168),\; n=5,19,23,37\mod 84\,, \\
		-1, & \bbb=(\alpha_2,-n^2/168),\; n=47,61,65,79\mod 84\,, \\
		1, & \bbb=(\alpha_3,-n^2/168),\; n=11,17,25,31\mod 84\,, \\
		-1, & \bbb=(\alpha_3,-n^2/168),\; n=53,59,67,73\mod 84\,, \\
		0, & \text{otherwise}.
	\end{array}
	\right.
\end{equation}
While the reduced transseries parameters for $k\in \Z_+$ are indeed as in \eqref{transparGGC}, that is
\be
n_{\alpha}\equiv \sum_{\bba\,\text{(fixed $\alpha$)}} n_{\bba,0} =\left\{
\begin{array}{cl}
	1,& \alpha=\alpha_0,\alpha_1,\alpha_2, \\
	0,& \alpha=\alpha_3.
\end{array}
\right.
\ee
In particular, the residues in the first group of poles magically add up to zero:
\be
n_{\alpha_3} \; = \;
\frac{1}{2} \Big[ \left( 1- \tfrac{1}{42} \right) - \left( 1- \tfrac{13}{42} \right) - \left( 1- \tfrac{29}{42} \right) + \left( 1- \tfrac{41}{42} \right) \Big]
\; = \; 0\,.
\ee
Note that however, in general $n_{\bba}\neq 0$ with $\bba$ being a lift of $\alpha_3$ (i.e. $\alpha=\alpha_3$).


\section{Relation to Yang-Mills instantons on $M_4 = \R \times M_3$}

The starting point of the instanton Floer homology \cite{MR956166}
is a beautiful fact that, in the space of $SU(2)$ gauge connections on $M_3$, the steepest descent (a.k.a. the gradient flow)
trajectories with respect to the circle-valued Morse function $\text{CS} (A)$ are precisely the instantons on $M_4 = \R \times M_3$.
In other words, a family of 3d gauge connections $A(t)$ that obey the flow equation $\frac{\partial A}{\partial t} = - *_3 F_A$
automatically solve the anti-self-duality equation
\be
*_4 F \; = \; *_3 \frac{\partial A}{\partial t} + dt \wedge *_3 F_A \; = \; - F
\label{FFASD}
\ee
where $F = dt \wedge \frac{\partial A}{\partial t} + F_A$ is the curvature of a 4d gauge connection
on $M_4 = \R \times M_3$.

Therefore, when the level is ``pure imaginary'', $k \in i \R$, the anti-self-dual connections on $M_4 = \R \times M_3$
that interpolate between limiting $SU(2)$ flat connections on $M_3$, $\alpha$ and $\beta$, correspond to
the ``broken flow trajectories'' in our resurgent analysis that pass through critical points $\alpha$ and $\beta$.
Put differently, instantons in the sense of resurgent analysis then correspond to instantons in
gauge theory on $M_4 = \R \times M_3$, thereby allowing us to make contact between the two subjects.

\begin{figure}[h]
\centering
\includegraphics[width=3in]{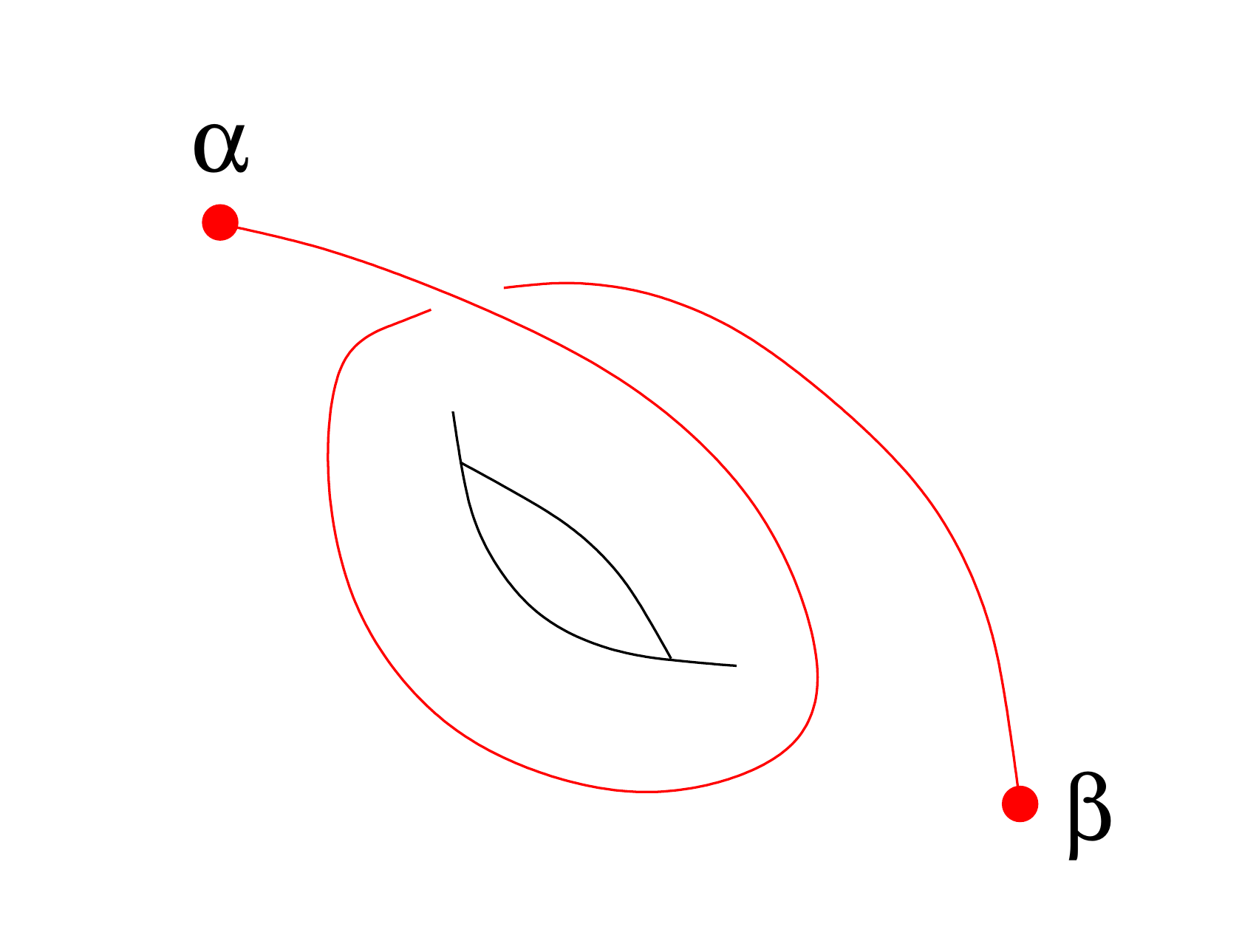}
\caption{The space of gauge connections on $M_3$ modulo gauge transformations is not simply-connected.
As a result, the lift of $\text{CS} (\beta) - \text{CS} (\alpha) \in \R / \Z$ to $\ell \in \R$ depends on
the homotopy class of the path connecting critical points $\alpha$ and $\beta$.
In $SU(2)$ theory, the spectral flow of a relative Morse index around a closed loop in this space
is equal to 8, which, in part, is why the standard instanton Floer homology is only $\Z_8$-graded.}
\label{fig:res6}
\end{figure}

In fact, even the questions that we face in the resurgent analysis of complex Chern-Simons theory are closely
related to questions studied in the traditional gauge theory. Thus, one of the main questions in
the Floer theory is the study of moduli spaces
\be
\CM_{\text{inst}} (M_3 \times \R, \alpha, \beta) \; := \;
\{ A \; \vert \; *_4 F = - F \,, \lim_{t \to + \infty} A = \alpha \,, \lim_{t \to - \infty} A = \beta \} / \text{gauge}\,.
\label{Minstabdef}
\ee
For each choice of the limiting flat connections $\alpha$ and $\beta$, the Yang-Mills action (equivalently, the instanton charge)
is given by
\be
\ell \; = \; \frac{1}{8\pi^2} \int_{M_3 \times \R} \Tr F \wedge F\,.
\label{instnumdef}
\ee
Note, since $M_4 = \R \times M_3$ has two cylindrical ends, the value of this integral does not need to be integer.
In fact, its fractional part is given by the difference of Chern-Simons functional for $\alpha$ and $\beta$,
\be
\ell \; =  \; \text{CS} (\beta) - \text{CS} (\alpha) \quad \text{mod}~\Z\,.
\label{CSaCSb}
\ee
The integer part of $\ell$, on the other hand, depends on the homotopy class of the path connecting $\alpha$ and $\beta$
in the space of fields (modulo gauge transformations), as illustrated in Figure~\ref{fig:res6}.
Indeed, even when $M_3$ is a homology sphere,
the space of all connections in a principal $G$-bundle over $M_3$ is an affine space,
but the group of gauge transformations ({\it i.e.} automorphisms of this bundle)
has $\pi_0 = \Z$, so that the quotient is not simply-connected\footnote{The fundamental group
can be even larger when $M_3$ is not a homology sphere.}
\be
\pi_1 \left( \frac{\text{gauge connections}}{\text{gauge transformations}} \right) \; \cong \; \Z\,.
\label{pioneM3gauge}
\ee
As a result, the moduli space \eqref{Minstabdef} of instantons on $M_4 = \R \times M_3$
is a disjoint union of infinitely many components labeled by the degree of the lift of
$\text{CS} (\beta) - \text{CS} (\alpha)$ from $\R / \Z$ to $\R$,
\be
\CM_{\text{inst}} (M_3 \times \R, \alpha, \beta) \; = \; \bigsqcup_{\ell} \, \CM^{( \ell )}_{\text{inst}} (M_3 \times \R, \alpha, \beta)\,.
\label{Minstabdecomp}
\ee
The expected dimension of each component can be computed from the index formula:
\be
\dim \CM^{( \ell )}_{\text{inst}} (M_3 \times \R, \alpha, \beta) \; = \;
8 \ell - \frac{1}{2} (h_{\alpha} + h_{\beta} + \rho (\alpha) - \rho (\beta))
\label{Minstdim}
\ee
where $h_{\alpha} = \sum_{i=0,1} \dim_{\R} H^i (M_3 , \text{ad} \alpha)$,
and
\be
\rho (\alpha) = \eta_{\alpha} (0) - 3 \eta_{\theta} (0)
\ee
is the Atiyah-Patodi-Singer rho-invariant.
Modulo 8, we also have
\be
\dim \CM_{\text{inst}} (M_3 \times \R, \alpha, \beta) \; = \;
\mu (\alpha) - \mu (\beta) - \dim \text{Stab} (\beta) \qquad \text{mod}~8
\label{Minstdimmod8}
\ee
where $\mu (\alpha)$ is the Floer index of a critical point $\alpha$.
The latter is only defined modulo 8 because going around a closed loop in the space of fields (modulo gauge transformations)
changes its value by 8, {\it cf.} Figure~\ref{fig:res6}.
Indeed, changing $\ell$ by $+1$ increases the value of \eqref{Minstdim} by $+8$.

{}From the resurgence viewpoint, we wish to know the spectrum of values of $\ell$ for which the moduli
spaces $\CM^{( \ell )}_{\text{inst}} (M_3 \times \R, \alpha, \beta)$ are non-empty.
These values of $\ell$ determine the values of the instanton action and, therefore, possible positions of singularities on the imaginary axis of the Borel plane, which are images of $SU(2)$ flat connections. In particular, they can tell us which Stokes monodromy coefficients, $m^\bbb_\bba$, which appear in (\ref{Borel-general-pole}) are possibly non-zero (that is with $S_\bbb-S_\bba=\ell$).
Namely, in a Borel resummation of the perturbative series $Z_{\text{pert}}^{\beta} (M_3)$
around a flat connection $\bbb$, we will see contributions of other flat connections $\bba$
whose Lefschetz anti-thimbles (= stable manifolds, in the language of Morse theory)
meet the Lefschetz thimble (= unstable manifold) of $\bbb$.
Since the flow trajectories are precisely the solutions to the anti-self-duality equation \eqref{FFASD},
we conclude that --- in this infinite-dimensional model based on 4d gauge theory ---
the intersection of the Lefschetz thimble $\Gamma_{\bbb,\frac{\pi}{2}}$ with the Lefschetz anti-thimble $\Gamma_{\bba,-\frac{\pi}{2}}$ intersected with the universal cover of the space $SU(2)$ connections
is precisely the Floer moduli space \eqref{Minstabdecomp}:
\begin{equation}
	\CM_{\text{inst}}^{(\ell)}(M_3 \times \R, \alpha, \beta) =\Gamma_{\bbb,\frac{\pi}{2}}\cap
	\Gamma_{\bba,-\frac{\pi}{2}} \cap \tilde\CA_{SU(2)}
	  \qquad(\text{with }S_\bbb-S_\bba=\ell)\,.
\end{equation}
It would be interesting to find if there is more quantitative relation between
topology of instanton moduli spaces and values of the Stokes monodromy coefficients\footnote{A naive guess would be
\be
m^\bbb_\bba \; = \; \chi \left( \CM^{( \ell )}_{\text{inst}} (M_3 \times \R, \alpha, \beta) \right)\,.
\; 
\ee
One can see, however, that it fails already in a simple case of lens spaces,
where the left hand side is zero, while the right hand side is not.}:
\begin{equation}
	\CM_{\text{inst}}^{(\ell)} (M_3 \times \R, \alpha, \beta)\qquad \stackrel{?}{\rightsquigarrow}\qquad m_\bba^\bbb\qquad (\text{with }S_\bba-S_\bbb=\ell)\,.
\end{equation}
Now, let us illustrate this more concretely for the Brieskorn homology 3-spheres,
\be
M_3 \; = \; \Sigma (p_1, p_2, p_3)
\ee
which were also our examples in the previous section.
Since flat connections on $M_3$ are uniquely characterized by their holonomies,
each irreducible flat connection $\alpha$ corresponds to a representation
\be
\alpha : \quad \pi_1 (\Sigma (p_1, p_2, p_3)) \to SU(2)
\ee
which, in turn, is characterized by a triplet of the so-called ``rotation numbers'' $(\ell_1, \ell_2, \ell_3)$
that satisfy $0 < \ell_i < p_i$ and certain additional conditions \cite{MR1051101}.
For example, in this language, the familiar flat connections $\alpha_i$, $i=0,1,2$, on the Poincar\'e sphere $\Sigma (2,3,5)$
that we encountered in the previous section are characterized by the following data:

\begin{table}[htb]
\centering
\renewcommand{\arraystretch}{1.3}
\begin{tabular}{|@{\quad}c@{\quad}|@{\quad}c@{\quad}|@{\quad}c@{\quad}|@{\quad}c@{\quad}|@{\quad}c@{\quad}|@{\quad}c@{\quad}|}
\hline $\alpha$ & $(\ell_1, \ell_2, \ell_3)$ & $\text{CS} (\alpha)$ & $\mu (\alpha)$ & $h_{\alpha}$ & $\rho (\alpha)$
\\
\hline
\hline $\alpha_0$ & $(0,0,0)$ & $0$ & $-3$ & 3 & 0 \\
\hline $\alpha_1$ & $(1,2,2)$ & $- \frac{49}{120}$ & $5$ & $0$ & $- \frac{97}{15}$ \\
\hline $\alpha_2$ & $(1,2,4)$ & $- \frac{1}{120}$ & $1$ & $0$ & $- \frac{73}{15}$ \\
\hline
\end{tabular}
\caption{Invariants of flat connections on $M_3 = \Sigma (2,3,5)$.}
\label{tab:Poincareconn}
\end{table}

In terms of the triplet $(\ell_1, \ell_2, \ell_3)$, the Chern-Simons functional of the flat connection is
\be
\text{CS}_{\vec \ell}  \; = \; - \frac{p_1 p_2 p_3}{4} \left( \frac{\ell_1}{p_1} + \frac{\ell_2}{p_2} + \frac{\ell_3}{p_3} \right)^2 \qquad \text{mod}~\Z\,.
\ee
For example, for $(p_1,p_2,p_3) = (2,3,5)$ we get, {\it cf.} \eqref{CSaa12235}:
\be
\text{CS}_{\vec \ell = (1,2,2)} \; = \; - \frac{49}{120}
\qquad , \qquad
\text{CS}_{\vec \ell = (1,2,4)} \; = \; -\frac{1}{120}\,.
\ee
Using the data in Table~\ref{tab:Poincareconn}, it is easy to compute the expected dimension of moduli spaces~\eqref{Minstdim}.
As an illustration, let us compute the dimension of the moduli space of anti-self-dual flow trajectories from $\alpha_1$
to $\alpha_2$ with $\ell = \frac{2}{5}$:
\be
\dim \CM^{( \ell )}_{\text{inst}} (M_3 \times \R, \alpha_1, \alpha_2)
\; = \; 8 \cdot \frac{2}{5} - \frac{1}{2} \left( - \frac{97}{15} + \frac{73}{15} \right)
\qquad (= 5 - 1 - 0 ~~\text{mod}~8)\,.
\ee
Note, the result is an integer number, which, modulo 8, agrees with \eqref{Minstdimmod8}. However from the analysis of Borel transform we have $m_{\bba_1}^{\bba_2}=m_{\bba_2}^{\bba_1}=0$ for any lifts $\bba_i$ of $\alpha_i$. Therefore we expect non-emptiness of instanton moduli space to only be a necessary condition for non-vanishing of Stokes monodromy coefficents, but not a sufficient.
A few other examples are summarized in Table~\ref{tab:Poincareinst}.


\begin{table}[htb]
\centering
\renewcommand{\arraystretch}{1.3}
\begin{tabular}{|@{\quad}c@{\quad}|@{\quad}c@{\quad}|@{\quad}c@{\quad}|}
\hline $(\alpha, \beta)$ & $\text{CS} (\beta) - \text{CS} (\alpha)$ &
$\dim \CM^{( \ell )}_{\text{inst}} (\alpha, \beta)$
\\
\hline
\hline $(\alpha_1,\alpha_0)$ & $\frac{49}{120}$ & ~~~$5$~~~(for $\ell=\frac{49}{120}$) \\

\hline $(\alpha_2,\alpha_0)$ & $\frac{1}{120}$ & ~~~$1$~~~(for $\ell=\frac{1}{120}$) \\

\hline $(\alpha_1, \alpha_2)$ & $\frac{2}{5}$ & ~~$4$~~~~(for $\ell=\frac{2}{5}$) \\
\hline
\end{tabular}
\caption{Floer moduli spaces on $M_3 \times \R$ for the Poincar\'e sphere $M_3 = \Sigma (2,3,5)$.}
\label{tab:Poincareinst}
\end{table}

The results of \cite{anvari2013equivariant} also tell us that a necessary condition for
\begin{equation}
\CM^{( \ell )}_{\text{inst}} (M_3 \times \R, \alpha_i, \alpha_0)\neq\varnothing ,
\qquad  i=1,2
\end{equation}
is that $\ell$ is of the form $n^2/4p$ (with $p=30$ for Poincar\'e sphere),
which is in agreement with the analytic structure of the Borel transform.


\subsection{Instantons as solitons in a 2d Landau-Ginzburg model}

Now we formulate a {\it finite-dimensional} model for the problem of instanton counting,
which, among other things, can considerably simplify the study of the moduli spaces \eqref{Minstabdecomp}.
Let us first introduce the model and then explain where it comes from.

The model is very simple: it is a 2d Landau-Ginzburg (LG) model with $\C^*$-valued variables $z_i$
and a potential function $\tilde W (z)$ that depends on the choice of the 3-manifold $M_3$.
The explicit form of $\tilde W (z)$ for many closed 3-manifolds can be found in \cite{Chung:2014qpa},
but we won't need these explicit calculations here (their origin will be explained below)
and focus primarily on the properties.

\begin{figure}[h]
\centering
\includegraphics[width=5in]{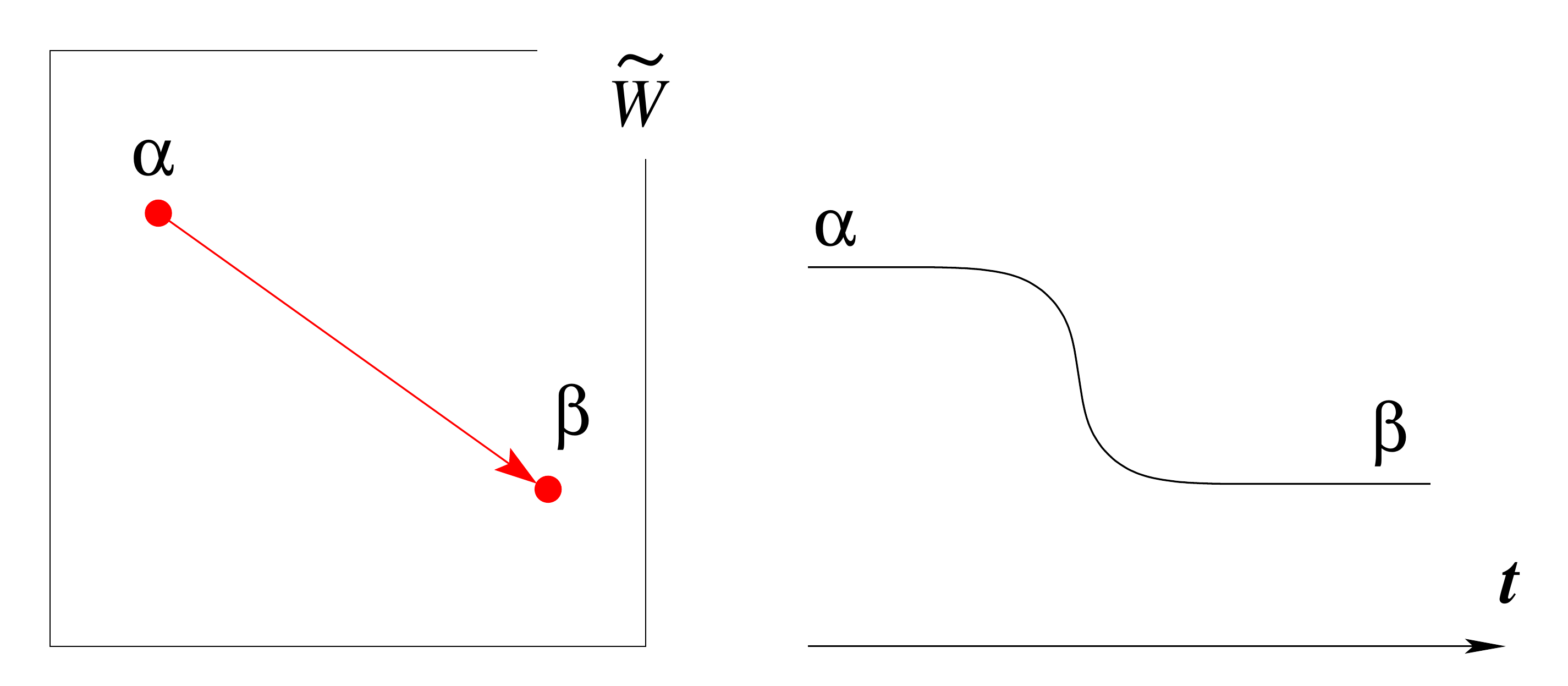}
\caption{The Borel plane of the complex Chern-Simons theory
can be identified with the $\tilde W$-plane of the Landau-Ginzburg model.
Solitons (shown on the right) that tunnel from a vacuum $\bba$ to a vacuum $\bbb$
project to straight lines in the $\tilde W$-plane (shown on the left).}
\label{fig:res7}
\end{figure}

The Borel plane of the complex Chern-Simons theory is the $\tilde W$-plane of the Landau-Ginzburg model,
{\it cf.} Figure~\ref{fig:res7}.
Critical points of $\tilde W (z)$, that is ``LG vacua'', are complex flat connections on $M_3$,
{\it i.e.} critical points of the Chern-Simons functional $\text{CS} (\CA)$.
Moreover, the value of the Chern-Simons functional on a lift of flat connection $\bba$
is equal to the value of the potential function $\tilde W (z)$ at the corresponding critical point:
\be
z^{(\bba)}: \qquad \exp \left( z \frac{\partial \tilde W}{\partial z} \right)_{z = z^{(\bba)}} = 1
\,, \qquad
\tilde W (z^{(\bba)}) = S_\bba\;\;(=\text{CS}(\alpha)\mod 1)\,.
\label{Wcrit}
\ee
Note, what makes this class of 2d Landau-Ginzburg models interesting (and slightly unconventional)
is the fact that $\tilde W (z)$ is a multi-valued function and the space of fields $z_i$ is not simply-connected.
The former corresponds to a similar property of the Chern-Simons functional,
while the latter is the analogue of \eqref{pioneM3gauge} in the infinite-dimensional model.

Continuing with the dictionary, BPS solitons in a LG model are field configurations $z_i (t)$
that depend on one of the 2d coordinates $t$ and solve the flow equation \cite{Cecotti:1992rm,Hori:2000ck}:
\be
\frac{d \bar z_j}{dt} = \frac{e^{i \theta}}{2} \frac{\partial \tilde W}{\partial z_j}
\label{LGBPS}
\ee
which is a finite-dimensional version of the flow equation $\frac{\partial A}{\partial t} = - *_3 F_A$
in gauge theory on $M_4 = \R \times M_3$ that we saw earlier.\footnote{Roughly speaking,
passing from the infinite-dimensional version to the finite-dimensional one involves integrating over $M_3$.}
Moreover, it can be shown that solutions to these equations project onto straight lines in the $\tilde W$-plane
and, in particular, solutions that interpolate between critical points $z^{(\bba)}$ and $z^{(\bbb)}$
must have angle $\theta$ such that \cite{Cecotti:1992rm,Hori:2000ck}:
\be
\theta = \arg \ell_{\bba \bbb} \qquad \text{where} \qquad \ell_{\bba \bbb} = \tilde W (z^{(\bba)}) - \tilde W (z^{(\bbb)})\,.
\label{phaseWW}
\ee
This is precisely the property of the steepest descent trajectories in complex Chern-Simons theory
(or, in fact, in {\it any} theory since it merely follows from the definition of the ``steepest descent'')
and, therefore, is consistent with our identification of the $\tilde W$-plane with the Borel plane in complex Chern-Simons theory.

\begin{table}[htb]
\centering
\renewcommand{\arraystretch}{1.3}
\begin{tabular}{|@{\quad}c@{\quad}|@{\quad}c@{\quad}|}
\hline Landau-Ginzburg model & complex Chern-Simons
\\
\hline
\hline $\tilde W$-plane & Borel plane \\

\hline critical points & flat connections \\

\hline solitons & instantons \\
\hline
\end{tabular}
\caption{A dictionary between complex Chern-Simons theory on $M_3$ and a Landau-Ginzburg model
with the potential $\tilde W_{M_3}$.}
\label{tab:LGdictionary}
\end{table}

In particular, solitons that interpolate between critical points $\bba$ and $\bbb$ are the steepest descent trajectories
for $\bba$ that meet the steepest {\it ascent} trajectories for $\bbb$.
In our finite-dimensional model, these are the ordinary Lefschetz thimbles $\Gamma_{\bba,\theta}$
and $\Gamma_{\bbb,\theta + \pi}$, whose intersection is the moduli space
of BPS solitons ({\it i.e.} flows) from $\bba$ to $\bbb$:
\be
\Gamma_{\bba,\theta} \; \cap \; \Gamma_{\bbb,\theta + \pi}
\; = \; \CM_{\text{soliton}} (\tilde W, \bba, \bbb)
\ee
In the special case $\theta = \frac{\pi}{2}$, we already identified it with the instanton moduli space
in the infinite-dimensional model (= gauge theory on $M_3 \times \R$), so as our last entry in the dictionary
between the two models we can write
\be
\CM_{\text{soliton}} (\tilde W, \bba, \bbb) \; = \; \CM_{\text{inst}} (M_3 \times \R , \bba, \bbb; SL(2,\C)))
\ee
when $\bba$ and $\bbb$ are lifts of flat connections on $M_3$.

Now, let us say a few words about the origin of this finite-dimensional model, first introduced in \cite{DGH}.
The two-dimensional Landau-Ginzburg model presented here is the 3d $\CN=2$ theory $T[M_3]$
compactified on a circle, {\it i.e.} on 3d space-time $\R^2 \times S^1$.
One can derive all of the properties mentioned here by constructing this theory as a 6d five-brane theory
on $\R^2 \times S^1 \times M_3$ or, after compactification on a circle, as a 5d super-Yang-Mills
on D4-brane world-volume
\be
\lefteqn{\overbrace{\phantom{ \R ~~\times~ \R~ \; }}^{\text{LG model}}} \R~~ \times ~\underbrace{\R ~\times~ M_3 }_{\text{YM instantons}}\,.
\label{D4wvol}
\ee
The 2d Landau-Ginzburg model described here lives on $\R^2 = \R \times \R$, parametrized by $x^0$ and $x^1$.
If we identify one of these coordinates with $t$ in \eqref{FFASD}, then the corresponding factor of $\R$ times $M_3$
compose the 4-manifold $M_4$ on which gauge theory instantons live.\footnote{Since we are considering
field configurations translation invariant along one of the $\R$ factors,
one can effectively reduce theory of D4 branes to $\CN=4$ SYM on $\R\times M_3$, which is topologically twisted along $M_3$,
and for which it is well known that the equations of motions are flow equations of complexified CS functional
(see {\it e.g.} \cite{MR2898655,haydys2010fukaya,Witten:2011zz,Gadde:2013sca}).}
Since gauge theory instantons have finite action, given by \eqref{instnumdef}, to a 2d observer on $\R^2$
they appear as finite energy solitons (or, domain walls) localized in one direction and translation invariant along the other,
see Figure~\ref{fig:res7}.

Moreover, if the vortex partition function of $T[M_3]$ on $\R^2 \times S^1$ is expressed as
a contour integral over variables $z_i$, which in $T[M_3]$ are usually associated with 3d $\CN=2$ gauge symmetries,
then the saddle point approximation to the integral looks like
\be
Z \; = \; \int_{\Gamma} \frac{dz}{2\pi i z} \; e^{\frac{1}{\hbar} \tilde W (z) + \ldots}
\label{Zfd}
\ee
where $\tilde W (z)$ is the twisted superpotential of the 3d $\CN=2$ theory $T[M_3]$ on a circle.
As explained in \cite{DGH}, this integral can be interpreted as
a finite-dimensional version of the Chern-Simons path integral on $M_3$,
where one has integrated over all but finitely many modes of the complex gauge field $\CA$ on $M_3$.
Then, the above contour integral is the integral over the remaining modes of the Chern-Simons gauge field on $M_3$,
and the function $\tilde W (z)$ returns the value of the Chern-Simons functional as a function of these modes $z_i$.
Its critical points are in one-to-one correspondence with $G_{\C}$ flat connections on $M_3$
and its values at the critical points are precisely the values of the Chern-Simons functional evaluated
on those flat connections \eqref{Wcrit}.

If the integration contour $\Gamma$ is the steepest descent path through the {\it abelian} critical point $\mathbb{a}$,
then the vortex partition function \eqref{Zfd} should give the homological block $Z_{a} (q)$.
It would be illuminating to verify this by a direct calculation.


\subsection{A-model interpretation}

Here we present yet another model for our problem that involves branes in the Hitchin moduli space.
It was introduced in \cite{Gukov:2007ck} (see also \cite[sec.6]{Gadde:2013wq})
as a ``complexification'' of the standard setting in the Atiyah-Floer conjecture,
where the moduli space of flat $G$-connections on a Riemann surface $\Sigma$ is replaced
by the moduli space of flat $G_{\C}$ connections:
\be
\CM_H (G,\Sigma) \; \cong \; \CM_{\text{flat}} (G_{\C}, \Sigma)
\label{MHtarget}
\ee
This will be the target space of our A-model, which, as usual, deals with holomorphic maps from the world-sheet
to the target space. The world-sheet, on the other hand, is a strip $\R \times I$, where we parametrize $\R$
by the same variable $t$ that in the previous flow equations \eqref{FFASD} and \eqref{LGBPS}
described ``time'' evolution from a critical point $\alpha$ to a critical point $\beta$.

\begin{figure}[h]
\centering
\includegraphics[width=5in]{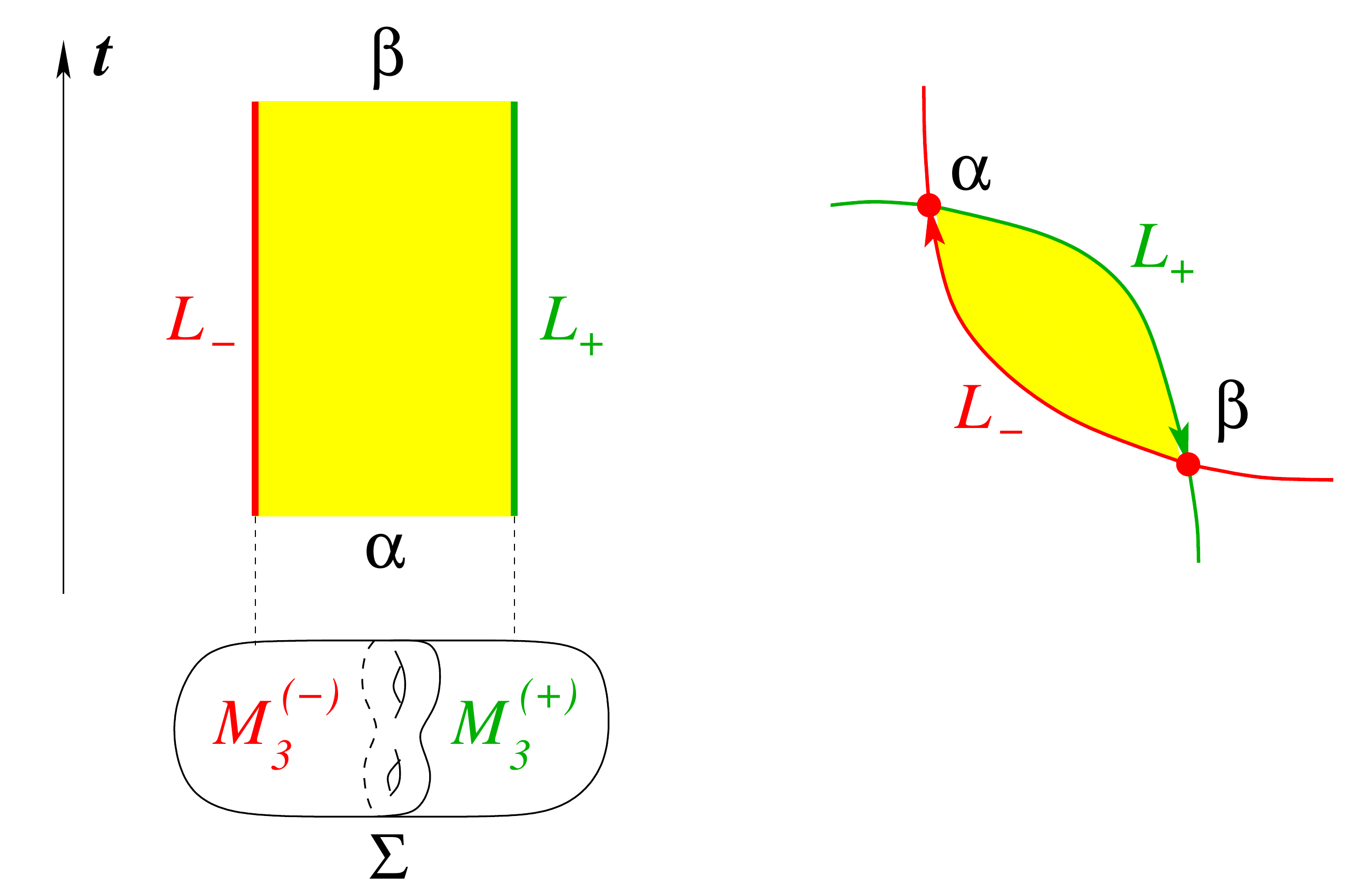}
\caption{The world-sheet of the A-model (shown on the left) and its image in the target space (shown on the right).
The boundary conditions $L_{\pm}$ represent (the image of) the moduli spaces of $G_{\C}$ flat connections on $M_3^{(\pm)}$
in the moduli space of $G_{\C}$ flat connections on $\Sigma$.}
\label{fig:res8}
\end{figure}

Here, the Riemann surface $\Sigma$ is the cross-section of our 3-manifold along a ``neck''
that divides $M_3$ into two pieces that we call $M_3^{(+)}$ and $M_3^{(-)}$:
\be
M_3 \; = \; M_3^{(+)} \cup_{\Sigma} M_3^{(-)}\,.
\label{MMMdecomp}
\ee
In the neck, where $M_3$ looks like $\Sigma \times I$, the part of the D4-brane world-volume that
in \eqref{D4wvol} we called $M_4 = \R \times M_3$ now has the form $\R \times I \times \Sigma$.
Dimensional reduction of the D4-brane world-volume theory down to $\R \times I$ --- which,
in particular, involves integrating over $\Sigma$ --- gives a two-dimensional sigma-model with
the hyper-K\"ahler target space \eqref{MHtarget}.

The reason the world-sheet of this sigma-model is $\R \times I$ (and not $\R^2$) is that the cylindrical
region $\Sigma \times \R$ does not extend to infinity, but rather is ``capped off'' on both ends with
$M_3^{(+)}$ and $M_3^{(-)}$, respectively.
This defines the boundary conditions $L_+$ and $L_-$ at the two boundaries of $\R \times I$, {\it cf.} Figure~\ref{fig:res8}.
Namely, the boundary conditions $L_{\pm}$ are holomorphic Lagrangian submanifolds (more generally, objects in
the Fukaya category of $\CM_H (G,\Sigma)$) which represent those complex flat connections on \eqref{MHtarget}
that can be extended to $M_3^{(\pm)}$:
\be
L_{\pm} \; = \; \CM_{\text{flat}} (G_{\C} , M_3^{(\pm)}) \; \subset \; \CM_{\text{flat}} (G_{\C}, \Sigma)\,.
\ee
The submanifolds $L_{\pm}$ are holomorphic in one of complex structures on \eqref{MHtarget} --- which
in the literature on Higgs bundles is usually denoted by $J$ --- and Lagrangian with respect to
two different symplectic structures, related to each other by $J$.
We denote the corresponding holomorphic symplectic form by $\omega$.
Since $L_{\pm}$ are Lagrangian with respect to $\omega$, by definition, $\omega \vert_{L_{\pm}} = 0$
and it follows that on each $L_+$ and $L_-$ we can define a 1-form $\lambda = d^{-1} \omega$.
Integrating $\lambda$ along a path $\gamma$ in $L_{\pm}$ gives a (multivalued) function $\tilde W_{\pm}$,
such that each $L_{\pm}$ is (locally) a graph of $\partial \tilde W_{\pm}$.

\begin{figure}[h]
\centering
\includegraphics[width=6in]{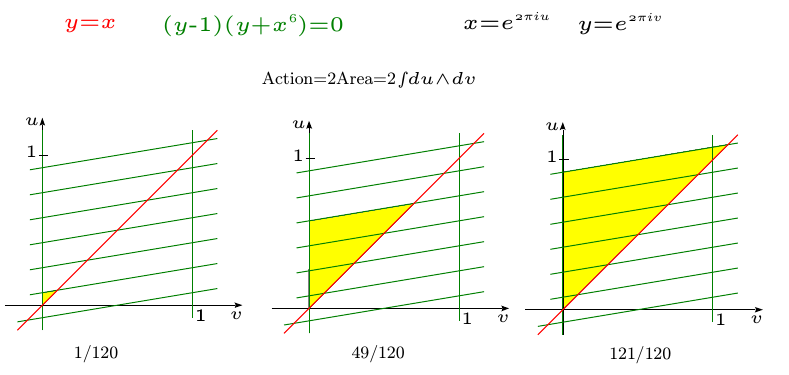}
\caption{Disk instantons for the Poincar\'e sphere $M_3 = \Sigma (2,3,5)$.}
\label{fig:E8disks}
\end{figure}

This setup was used in \cite{Gukov:2007ck} to construct certain homological invariants and it is conceivable
that doubly-graded homology groups of $M_3$ mentioned in \eqref{Zaqt} can be formulated in this language as well.
However, we will not attempt it here and focus primarily on aspects related to the resurgent analysis,
in particular, on the role of instantons, which brings us closer to the general framework of \cite{Kontsevich}.
Instantons in our A-model are holomorphic maps
\be
\phi : \R \times I \; \to \; \CM_H (G, \Sigma)
\ee
such that, in the simplest situation, the two boundaries of the interval $I$ map to
holomorphic Lagrangian submanifolds $L_{\pm}$ in the target space $\CM_H (G, \Sigma)$.
Note, critical points of the function
\be
\tilde W \; = \; \tilde W_{+} - \tilde W_{-}
\label{WWWpm}
\ee
are precisely the intersection points of $L_+$ and $L_-$.
Moreover, the ``complex area'' of a disk instanton can be expressed as an integral of the one-form
$\lambda = d^{-1} \omega$ from one such intersection point $\alpha$ to another intersection point $\beta$
along $L_+$, and then from $\beta$ to $\alpha$ along $L_-$:
\be
\ell_{\alpha \beta}
\; = \; \int_{\alpha}^{\beta} \lambda \vert_{L_+} - \int_{\alpha}^{\beta} \lambda \vert_{L_-} = \int_{\text{disk}} \omega\,.
\label{diskareaviapath}
\ee
Note, this expression coincides with \eqref{phaseWW} if we identify the function \eqref{WWWpm}
with the potential in our finite-dimensional model considered earlier.
In the next section, we will encounter a particular instance of this setup where one of the ``halves'', say $M_+$,
is a knot complement and the other, $M_-$, is a solid toris, so that \eqref{MMMdecomp} is a surgery on a knot in $S^3$.
Figure~\ref{fig:E8disks} illustrates Lagrangian submanifolds $L_{\pm}$ and disks instantons for such
a decomposition of the Poincar\'e sphere $M_3 = \Sigma (2,3,5)$.

\section{Resurgence and surgeries}

There are many ways of constructing 3-manifolds and, correspondingly, many ways of calculating their quantum invariants.
In this section, we apply resurgent analysis to 3-manifolds constructed from surgeries on knots.
Transseries associated with complex flat connections (which are unavoidable on general 3-manifolds)
will now make their full appearance.

\subsection{Exceptional surgeries and the conformal window of $T[M_3]$}

Given a knot $K$ in a 3-sphere, Dehn surgery with a rational\footnote{As usual, $p$ and $r$ are assumed to be coprime. Our peculiar choice of notations with
an extra minus sign here will be justified later.} slope coefficient $-\frac{p}{r}$ ecscavates a tubular neighborhood of $K$
and then glues it back in with a non-trivial diffeomorphism $\varphi$ of the boundary torus
that sends a curve in homotopy class $(p,r)$ to the meridian of the knot complement:
\be
S^3_{-p/r} (K) \; := \; (S^3 \setminus K) \cup_{\varphi} (S^1 \times D^2)\,.
\label{knotsurgery}
\ee
In particular, this is a special case of the gluing construction \eqref{MMMdecomp}
where $M_3^{(+)} = S^3 \setminus K$ is a knot complement and the solid torus
$M_3^{(-)} = S^1 \times D^2$ can be thoughout of as the unknot complement.
For $p \ne 0$, the resulting 3-manifold is a rational homology sphere ($\mathbb{Q}$HS) with
\be
H_1 (S^3_{- p/r} (K)) \; = \; \Z_p\,.
\label{H1Zp}
\ee
A surgery with the slope coefficient $\infty$ is defined to have $-\frac{p}{r} = \frac{1}{0}$
and gives back the original 3-sphere, while $0$-surgery has $H_1 (S^3_{0} (K)) = \Z$.
We will be mostly interested in values of $\frac{p}{r} \in \Q$ other than $0$ or $\infty$.

The construction \eqref{knotsurgery} provides a large supply of interesting 3-manifolds for the resurgent analysis.
In general, they admit all kinds of flat connections: reducible and irreducible, real and complex.
Here, we first present a simple technique to enlist all flat connections on a given surgery manifold
$M_3 = S^3_{-p/r} (K)$ and then explain how to produce the spectrum of values of the Chern-Simons
functional and the transseries coefficients.

First, we note that from \eqref{H1Zp} it immediately follows that, for $G=SU(2)$,
abelian flat connections\footnote{and, therefore, homological blocks}
are labeled by $a \in \Z_p / \Z_2$ for $G=SU(2)$.
(More generally, for $G=U(N)$, they are labeled by $a \in (\Z_p)^N / S_N$.)
At the other extreme, in general we also have irreducible $SL(2,\C)$ flat connections on $M_3$.
A particularly nice class of such flat connections --- sometimes called ``geometric'' --- comes from hyperbolic 3-manifolds,
because a hyperbolic structure on $M_3$ defines an $SL(2,\C)$ flat connection.\footnote{In the language
of Euclidean 3d gravity with negative cosmological constant, it is a complex flat connection $\CA = w + i e$
whose real part is the spin connection and the imaginary part is the vielbein on $M_3$.}
Such geometric $SL(2,\C)$ structures are fairly generic.
Indeed, according to Thurston \cite{MR1435975}, unless $K$ is a torus knot or a satellite knot,
a Dehn surgery on $K$ will be hyperbolic for all but finitely many $p/r$.
Such knots are called {\it hyperbolic}, and the surgeries that produce non-hyperbolic $M_3  = S^3_{-p/r} (K)$
are called {\it exceptional}. It is natural to ask: What is the set of exceptional surgeries for a given knot $K$?

\begin{figure}[h]
\centering
\includegraphics[width=4in]{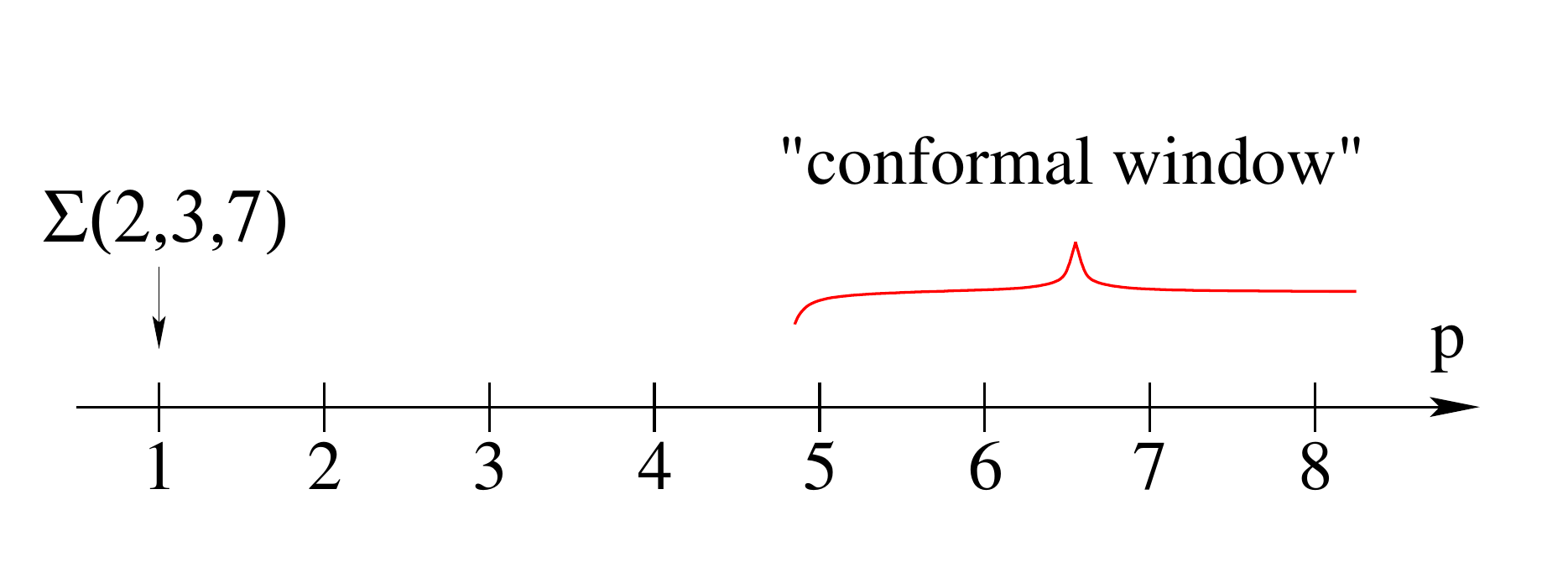}
\caption{The conformal window for 3d $\CN=2$ theories $T[M_3]$, where $M_3 = S^3_p ({\bf 4_1})$
is the integral $p$-surgery on the figure-8 knot ${\bf 4_1}$. The analogue of the central charge function
for these theories qualitatively changes its behavior as we cross the boundary of the conformal window;
in particular, theories for non-hyperbolic surgeries (called exceptional surgeries) have much fewer degrees of freedom.}
\label{fig:res5}
\end{figure}

The answer to this interesting question is not known, but the number of exceptional surgeries is surprisingly small.
Thus, an improved version of the argument by Thurston and Gromov gives a uniform bound $\le 24$ on the number of exceptional
surgeries (for all $K$). In practice, however, the number is even smaller and, according to the Gordon conjecture,
the figure-8 knot $K= {\bf 4_1}$ has the largest number of exceptional surgeries.
Surgeries on the figure-8 knot were also first studied by Thurston; they yield non-Haken\footnote{This condition
has to do with incompressible surfaces in $M_3$ which, when translated to the physics language \cite{Gadde:2013wq},
can be formulated as a condition on the spectrum of local operators in 3d $\CN=2$ theory $T[M_3]$.} hyperbolic 3-manifolds,
except for ten values of $p/r$ (including the slope $1/0$):
\be
\frac{p}{r}
\; = \;
\left\{ \; \frac{1}{0} \,, \; \frac{0}{1} \,, \; \pm \frac{1}{1} \,, \; \pm \frac{2}{1} \,, \; \pm \frac{3}{1} \,, \; \pm \frac{4}{1} \; \right\}\,.
\label{fig8exc}
\ee
In particular, $\pm 1$ surgeries produce the Brieskorn sphere \eqref{M237first}:
\be
S^3_{\pm 1} ({\bf 4_1}) \; = \; \pm \Sigma (2,3,7)
\ee
with four flat connections \eqref{SU2237flat} that we saw earlier.
These surgeries with $\frac{p}{r} = \pm 1$ are examples of the so-called {\it small Seifert fibered spaces},
which are Seifert fibered spaces with genus-0 base and at most three exceptional fibers or, equivalently,
3-manifolds obtained from a product of a circle with a ``pair of pants'' by suitably performing Dehn filling three times.
For the figure-8 knot, the other surgeries that produce small Seifert fibered spaces are $\frac{p}{r} = \pm 2$ and $\pm 3$.
The surgeries with $\frac{p}{r} = 0$ and $\pm 4$ are {\it toroidal}.
Hence, a more refined form of the list \eqref{fig8exc} looks like (see also Figure~\ref{fig:res5}):
\be
{\bf T_{-4}} \;, \quad
{\bf S_{-3}} \;, \quad
{\bf S_{-2}} \;, \quad
{\bf S_{-1}} \;, \quad
{\bf T_{0}} \;, \quad
{\bf S_{+1}} \;, \quad
{\bf S_{+2}} \;, \quad
{\bf S_{+3}} \;, \quad
{\bf T_{+4}}
\ee
where ``${\bf S_{-p/r}}$'' (resp. ``${\bf T_{-p/r}}$'')
indicates small Seifert fibered (resp. toroidal) exceptional surgery with slope $- \frac{p}{r}$.
In general, a closed 3-manifold is not hyperbolic if and only if it is reducible, toroidal, or small Seifert fibered space.
It is believed that exceptional surgeries on a hyperbolic knot $K$ are consecutive and either integral or half-integral,
such that integral toroidal slopes appear at the boundary of this range, except for the figure-8 knot.\footnote{In this sense,
a better representative of a ``typical'' situation could be {\it e.g.} $(-2,3,7)$-pretzel, whose exceptional surgeries are
$$
{\bf T_{16}} \;, \quad
{\bf S_{17}} \;, \quad
{\bf L_{18}} \;, \quad
{\bf T_{37/2}} \;, \quad
{\bf L_{19}} \;, \quad
{\bf T_{20}}
$$
where in addition to the previous notations we introduced ${\bf L_{-p/r}}$ for the Lens space surgery.}

Besides the exceptional values of $(p,r)$ in the set \eqref{fig8exc}, all other Dehn surgeries on
the figure-8 knot supply us with infinitely many examples of hyperbolic 3-manifolds $S^3_{-p/r} ({\bf 4_1})$
whose volumes converge to an accumulation point: the volume $V(\fig8)$ of the cusped hyperbolic 3-manifold, see {\it e.g.} \cite{Apol}.
Moreover, as described in \cite{Apol}, the corresponding critical points of the $SL(2,\C)$ Chern-Simons functional
can be easily deduced by considering the intersection points
\be
\left\{ y \, x^{-\frac{p}{r}} = 1 \right\} \cap \CC_K
\label{curve-intersection}
\ee
with the zero locus of the A-polynomial
\be
\CC_K: \qquad A(x,y)=0 \qquad,\qquad (x,y) \; \in \; \frac{\C^* \times \C^*}{\Z_2}
\label{Acurve}
\ee
which plays the role of a  spectral curve\footnote{In fact, many ideas and techniques from
the theory of hyperbolic (a.k.a. trigonometric) integrable systems apply to this planar algebraic curve.}
in $SL(2,\C)$ Chern-Simons theory.

The counting of intersection points \eqref{curve-intersection} gives a simple practical way to account for {\it all} flat $SL(2,\C)$
connections on the knot surgery \eqref{knotsurgery} that we illustrate below in several concrete examples.
Indeed, if we think of $M_3  = S^3_{-p/r} (K)$ as glued from two pieces, {\it cf.} \eqref{MMMdecomp},
then the spectral curve \eqref{Acurve} simply describes the image of the moduli space of flat
connections on $M_3^{(+)} = S^3 \setminus K$ into the moduli space of flat connections on the boundary torus.
The curve $y \, x^{-\frac{p}{r}} = 1$ in \eqref{curve-intersection} plays a similar role for $M_3^{(-)} = S^1 \times D^2$,
with diffeomorphism $\varphi$ in \eqref{knotsurgery} taken into account in order to write the result in terms
of the same $\C^*$-valued eigenvalues, $x$ and $y$, of $SL(2,\C)$ holonomies along A- and B-cycles of the boundary torus,
$T^2 = \partial (S^3 \setminus K) = \partial ( S^1 \times D^2)$, see \cite{Apol} for details.

In particular, the intersection points \eqref{curve-intersection} which corresponds to $SU(2)$ flat connections
have $x$ and $y$ on the unit circle ({\it i.e.} in the maximal torus of $SU(2)$):
\be
SU(2): \qquad |x|=|y|=1
\ee
whereas intersection points \eqref{curve-intersection} that corresponds to $SL(2,\R)$ flat connections have
$x$ and $y$ either on the real line or unit circle:
\be
SL(2,\R): \qquad x \,, y \; \in \; \R^*\;\;\text{or}\;\; |x|=1,\;\;|y|=1\,.
\ee
Sometimes, it is convenient to work on the covering space, {\it cf.} \cite{Apol}:
\be
x=e^{2\pi i u}
\qquad , \qquad
y = e^{2 \pi i v}
\label{uvviaxy}
\ee
parametrized by $u$ and $v$. Note, in these coordinates, $SU(2)$ flat connections have real values of $u$ and $v$
The $\Z_2$ Weyl group acs on these variables simply as $(u,v) \mapsto (-u,-v)$.

Let us illustrate how this works in practice.
Thus, our main examples so far, the Poincar\'e sphere $\Sigma (2,3,5)$ and the Brieskorn sphere $\Sigma (2,3,7)$,
both can be constructed as surgeries on the trefoil knot $K = {\bf 3_1}$, whose A-polynomial consists of two factors:
\be
A(x,y) = (y-1)(y+x^6)\,.
\label{Apoltrefoil}
\ee
Correspondingly, the spectral curve \eqref{Acurve} has two irreducible components, one of which (namely, $y=1$)
is present for every knot and corresponds to abelian flat connections on the knot complement, while the form of
the other component ($y+x^6=0$) is not universal and depends on the knot $K$ in a non-trivial way.
The two component meet at three points (or, six points before we quotient by $\Z_2$ in \eqref{Acurve}):
\be
x \; = \; \pm i \,, \; \pm e^{\pi i/6} ,\; \pm e^{5 \pi i/6}\,.
\label{trefintbranches}
\ee
Note, in the space of $u$ and $v$, the spectral curve \eqref{Acurve} of the trefoil knot is a linear subspace,
whose real slice is shown in Figure~\ref{fig:E8disks}.
In particular, one can see very clearly in this Figure the intersection points \eqref{trefintbranches} all of
which have real values of $u$ and $v$.
In general, the $SU(2)$ bifurcation points where components of the representation variety
associated to reducible and irreducible flat connections come together are the solutions to
\be
\Delta_K (x^2) \; = \; 0
\label{AlexApol}
\ee
where $\Delta_K (q)$ is the Alexander polynomial of $K$.
Thus, for the trefoil knot $\Delta_{{\bf 3_1}} (q) = q^2 - q + 1$, so that $x= \pm e^{\pi i/6}$
and $x = \pm e^{5 \pi i/6}$ are precisely the $SU(2)$ bifurcation points that we found in \eqref{trefintbranches}.
See \cite{MR1008696,FGSS} for details.

However, our main interest here is not in the A-polynomial curve $\CC_K$ itself, but rather in its intersection
with  $y^r = x^p$ that tells us about flat connections on the knot surgery $M_3  = S^3_{-p/r} (K)$.
For example, in the case of the trefoil knot, from the explicit form of the A-polynomial \eqref{Apoltrefoil}
it is easy to see that the intersection points \eqref{curve-intersection} with $p/r = +1$ perfectly reproduce
the three flat connections (three critical points) on the Poincar\'e sphere $S^3_{-1} ({\bf 3_1}) = \Sigma (2,3,5)$:
\be
\{ y = x^{p/r} \} \,\cap\, \{ A(x,y)=0 \} \; = \;
\begin{cases}
(x,y) = (1,1) & \text{trivial}, \\
x=y = -e^{\frac{2 \pi i a}{5}}, & a=1,2
\end{cases}
\label{trefPoinconns}
\ee
that in our previous analysis we called $\alpha_0$, $\alpha_1$, and $\alpha_2$, see \eqref{ZCSaaa235}
and Table~\ref{tab:Poincareconn}.
This is a good place to point out that, since \eqref{curve-intersection} is the intersection of character varieties,
it may ``over-count'' flat connections on $M_3  = S^3_{-p/r} (K)$, {\it i.e.} produce intersection points
that actually do not lift to representations $\alpha: \pi_1 (M_3) \to SL(2,\C)$.
For this reason, as explained in \cite{Chung:2014qpa}, the intersection point $(x,y)=(-1,-1)$
is excluded from our list of flat connections \eqref{trefPoinconns} on $S^3_{-1} ({\bf 3_1}) = \Sigma (2,3,5)$.
Near each critical point, the curve $\CC_K$ looks like:
\ben
~~~~~~\alpha_0: & \quad & \tilde y=1   \\
\alpha_1 ~\text{and}~\alpha_2: & \quad & \tilde y=6 \, \tilde x \nonumber
\een
where we introduced local coordinates $\tilde x = x - x_{\alpha_*}$ and $\tilde y = y - y_{\alpha_*}$
in the neighborhood of a given critical point~$\alpha_*$.

Surgeries on other knots can be treated in a similar way.
For example, as we mentioned earlier, the Brieskorn sphere $\Sigma (2,3,7)$
can be also produced by a surgery on the figure-8 knot $K=\fig8$, whose A-polynomial looks like
\be
A(x,y) \; = \; (y-1) (x^4 - (1 - x^2 - 2x^4 - x^6 + x^8)y + x^4 y^2)
\label{Afigure8}
\ee
so that the corresponding algebraic curve $\CC_K$
is a three-fold branched cover of the $x$-plane $\C_x$.
The ``reducible'' branch $y=1$ meets the other two branches at the following 6 points:
\begin{equation}
x=\,\pm i,\;\frac{\pm 1\pm \sqrt{5}}{2}
\end{equation}
while the two ``non-abelian'' branches meet each other at the 12 points:
\begin{equation}
x=\,\pm i,\;\frac{\pm 1\pm \sqrt{5}}{2},\;\pm 1,\;\pm e^{\pm \pi i/3}\,.
\end{equation}
The intersection points $(x,y) = \left( \frac{\pm 1\pm \sqrt{5}}{2} , 1 \right)$
between abelian and non-abelian branches also follow from \eqref{AlexApol} since the Alexander polynomial
of the figure-8 knot is $\Delta_{\fig8} (q) = -q^{-1} + 3 - q$.
The A-polynomial curve $\CC_K$ has nodal singularities precisely at these points.

{}From the explicit form of the A-polynomial \eqref{Afigure8}, it is easy to find the intersection points \eqref{curve-intersection}
and produce a list of critical points of the $SL(2,\C)$ Chern-Simons functional on the surgery manifold $M_3=S_{-p/r}(\fig8)$.
For $r=1$ and $|p| \le 4$, there are only $SU(2)$ and $SL(2,\R)$ flat connections which manifests in the fact that all points
of \eqref{curve-intersection} lie either on the real line or on the unit circle in $\C_x$.
In other cases, there are $SL(2,\C)$ flat connections which lie neither in $SU(2)$ nor in $SL(2,\R)$ subgroup
and correspond to generic values of $x$ and $y$, as illustrated in Figure~\ref{fig:fig8-x}.


\subsection{Instantons and ``complex geodesics'' on the A-polynomial curve}

\begin{figure}[h]
\centering
\includegraphics[width=3in]{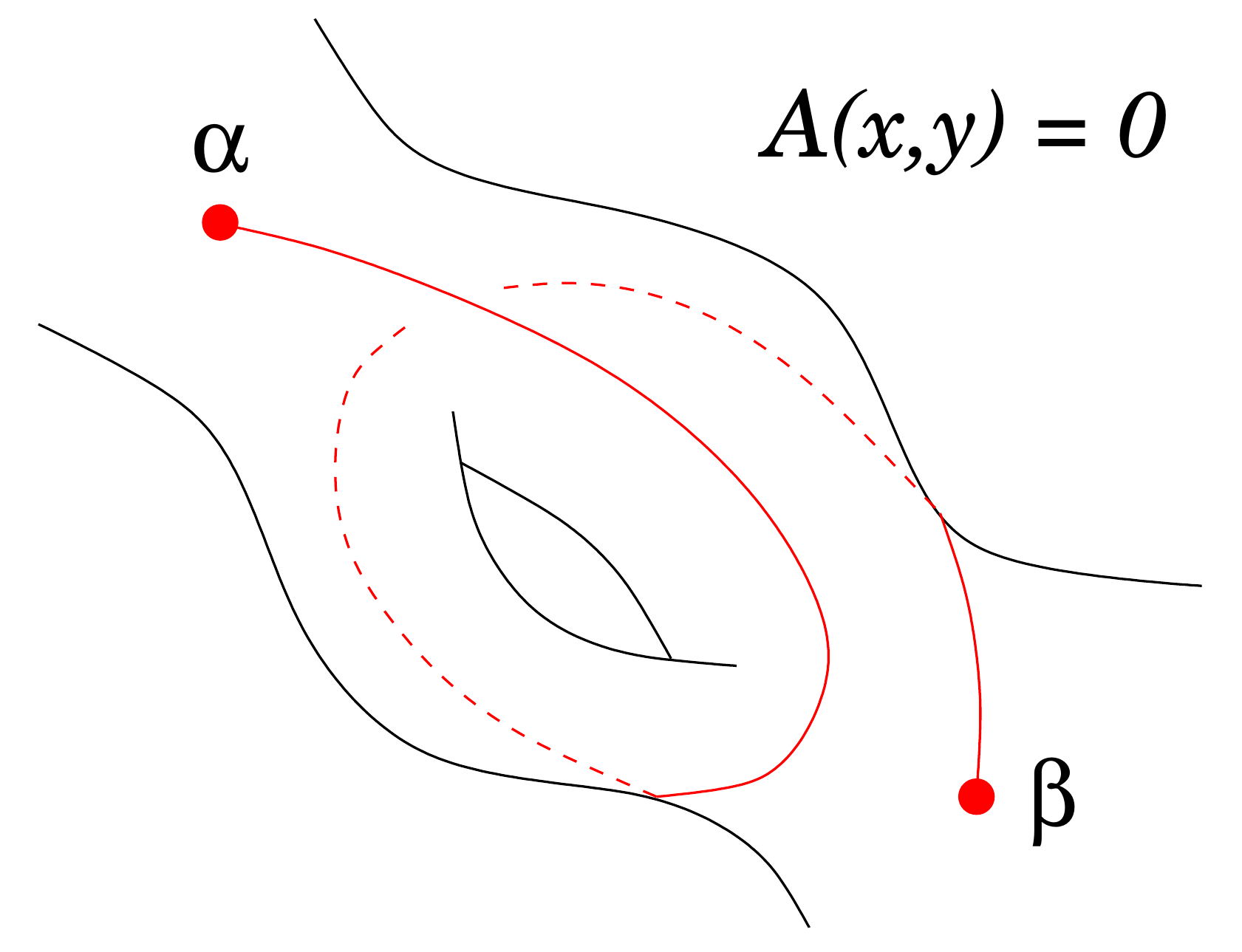}
\caption{A path $\gamma$ on the A-polynomial curve.}
\label{fig:res6a}
\end{figure}

For a knot $K$ (or, more generally, for a 3-manifold with a toral boundary) one can study
lengths spectra of ``complex geodesics'' on the A-polynomial curve\footnote{The choice of lifts $\bba$, $\bbb$ corresponds to a choice of contour.}~\eqref{Acurve}:
\be
\ell_{\bba \bbb} \; := \; \int_{\alpha}^{\beta} \lambda
\label{labint}
\ee
where $\lambda$ is a differential on $\CC_K$ given by
\be
\lambda \; = \; \frac{1}{2\pi^2}  \log  y \, \frac{dx}{x}\,.
\label{Apol1form}
\ee
As explained in \cite{Apol}, this problem naturally arises in quantization of complex Chern-Simons theory and,
in particular, such length spectra describe variation of the Chern-Simons functional, {\it cf.} \eqref{CSaCSb}:
\be
\ell_{\bba \bbb} \; =  \; \text{CS} (\beta) - \text{CS} (\alpha) \quad \text{mod}~\Z
\ee
for $SL(2,\C)$ flat connections $\alpha$ and $\beta$ on the knot complement $S^3 \setminus K$.
Of course, the value of the integral \eqref{labint} depends on the homotopy class of the path $\gamma$
that connects the two points, $\alpha = (x_{\alpha}, y_{\alpha})$ and $\beta = (x_{\beta}, y_{\beta})$,
as illustrated in Figure~\ref{fig:res6a} (or, equivalently, on the choice of lifts $\bba$ and $\bbb$ in the universal cover).
In particular, note that the integral vanishes on the ``abelian branch'', $y=1$. In a related context,
a version of this problem that involves counting length spectra of {\it closed} geodesics on the curve $\CC_K$
was studied in \cite{Gadde:2013wq} and, for more general curves (not directly related\footnote{However,
finding bridges between these different problems is, of course, not a bad idea.} to the A-polynomial),
similar problems often appear in the study of BPS spectra in string theory, see {\it e.g.} \cite{Klemm:1996bj}. The relation between instantons and periods of the complex curves has also appeared, for example in \cite{balian1978discrepancies,Drukker:2011zy}.

In our present context, the length spectra \eqref{labint} can be useful for finding values of
the $SL(2,\C)$ Chern-Simons functional on a knot surgery \eqref{knotsurgery},
which is composed of two 3-manifolds with a toral boundary.
Correspondingly, the Chern-Simons functional consists of two pieces, both of the form \eqref{labint},
one obtained by integrating $\lambda$ on the A-polynomial curve of the knot complement
and the other given by a similar integral on the curve $y^r=x^p$ that describes
$SL(2,\C)$ flat connections on the ``solid torus'' part of \eqref{knotsurgery}:
\be
\ell_* \; = \; \frac{1}{2\pi^2} \int\limits_{\gamma_*\subset \CC_K}\left(\log y-\frac{p}{r}\log x\right)\frac{dx}{x}\,.
\label{S-contour-int}
\ee
Here, $\gamma_*$ can be any path connecting critical points $\alpha_0$ and $\alpha_*$
that, as we learned in \eqref{curve-intersection}, correspond to the intersection points\footnote{These
intersection points, that is critical points of the $SL(2,\C)$ Chern-Simons functional on $M_3=S_{-p/r}(K)$,
are precisely the zeroes of the differential 1-form on the right-hand side of \eqref{S-contour-int}:
\be
\lambda_{p/r} \; := \; \left( \log y-\frac{p}{r}\log x\right)\frac{dx}{x}
\ee
}
of $\CC_K$ with the curve $y^r=x^p$.
It is convenient to choose $\alpha_0$ to be some reference point, {\it e.g.} the point $(x,y)=(1,1)$
associated with the trivial flat connection, and $\alpha_*$ to be our point of interest.
Then, an instanton on $M_3\times \R$ interpolating between the trivial flat connection on $M_3$ at $t = -\infty$
and an interesting $SL(2,\C)$ flat connection associated with one of the intersection points (\ref{curve-intersection})
at $t = +\infty$ has the classical action \eqref{S-contour-int}.

Not coincidentally, the integral \eqref{S-contour-int} coincides with \eqref{phaseWW}
and also with \eqref{diskareaviapath} if we think of $(\C^* \times \C^*) / \Z_2$
as the target space of our A-model and for the Lagrangian submanifolds $L_+$ and $L_-$
choose the curves $\CC_K$ and $y^r=x^p$, respectively.
Indeed, both of these curves are holomorphic Lagrangian with respect
to the symplectic form\footnote{Note, the symplectic structure $\omega$ is flat in the logarithmic variables $u$ and $v$
introduced in \eqref{uvviaxy}. The differential \eqref{Apol1form} also takes a very simple form in this variables, $\lambda = 2v \,du$.}
\be
\omega \; = \; \frac{1}{2\pi^2} \, \frac{dy}{y} \wedge \frac{dx}{x}
\ee
whose primitive 1-form $\lambda = d^{-1} \omega$ is precisely \eqref{Apol1form}.
In particular, Figure~\ref{fig:E8disks} illustrates the computation of the instanton action \eqref{S-contour-int}
for the Poincar\'e sphere represented as a $-1$ surgery on the trefoil knot.
In this case, $L_+ = \CC_K$ is the zero locus of the A-polynomial \eqref{Apoltrefoil} and $L_- = \{ y=x \}$.

In the Borel plane, the singularity nearest to the origin is associated with the instanton that has the smallest absolute value of $\ell_*$.
For example, for the $(-\frac{p}{r})$-surgeries on the figure-8 knot, this leading instanton could be found as follows.
In the case $|p/r|<4$, the position $x_*$ of the corresponding flat connection in the $x$-plane
can be chosen to lie in the interval $\left( 0 , \frac{\sqrt{5}-1}{2} \right)$. In the case $|p/r|>4$,
we can choose $x_*$ to be inside the upper-right quadrant of the unit disk, $|x|<1$.
The flat connection we are interested in has the smallest value of $x_*$ compared to the other flat connections.

\begin{figure}[ht]
\centering
\begin{tabular}{cc}
 \includegraphics[scale=0.7]{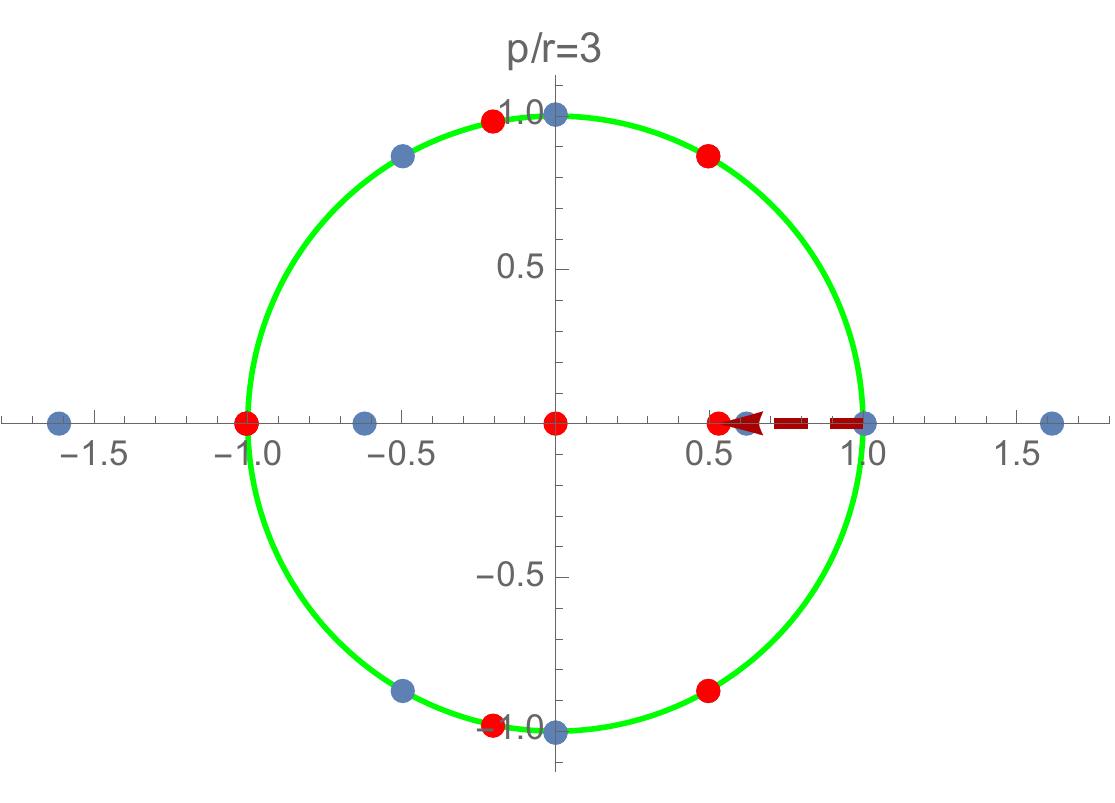} &  \includegraphics[scale=0.7]{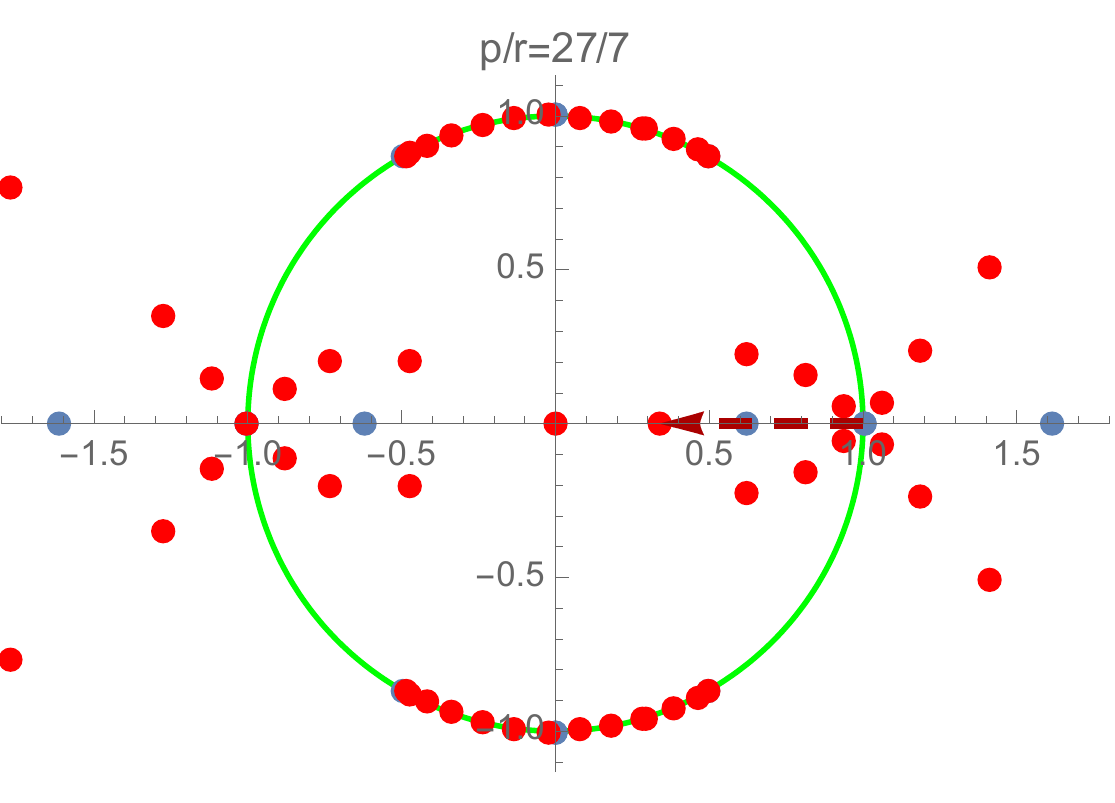} \\
 \includegraphics[scale=0.7]{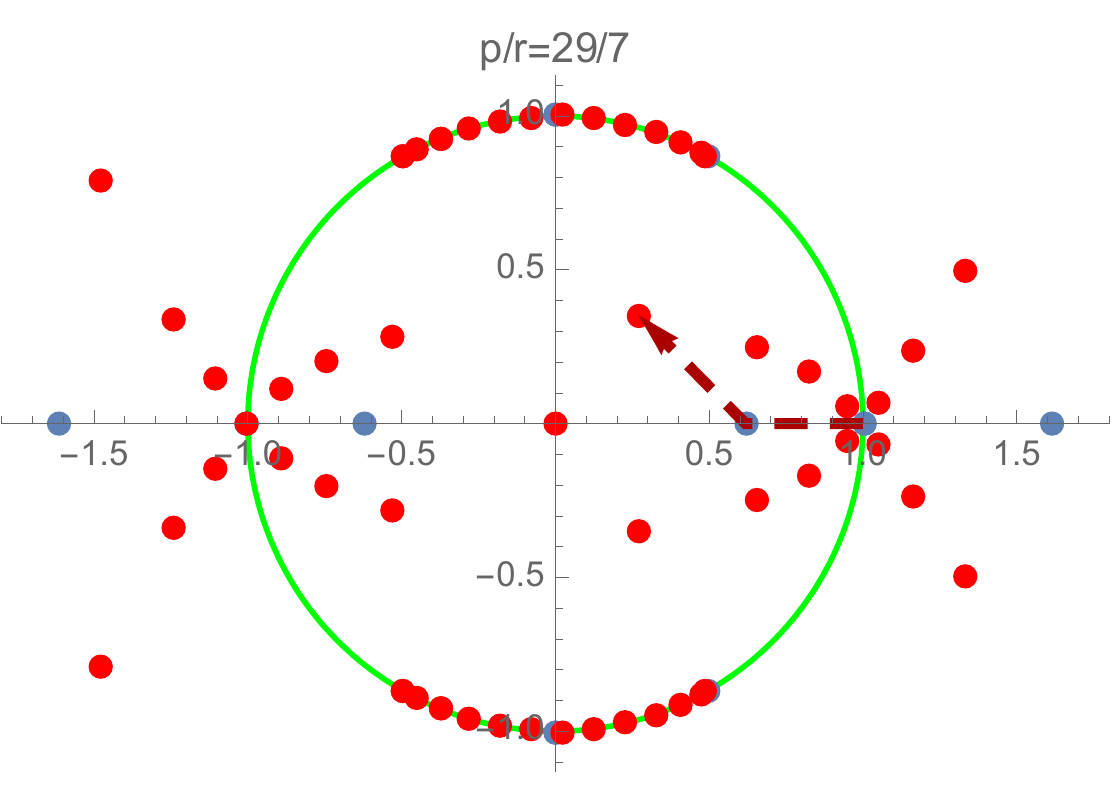} &  \includegraphics[scale=0.7]{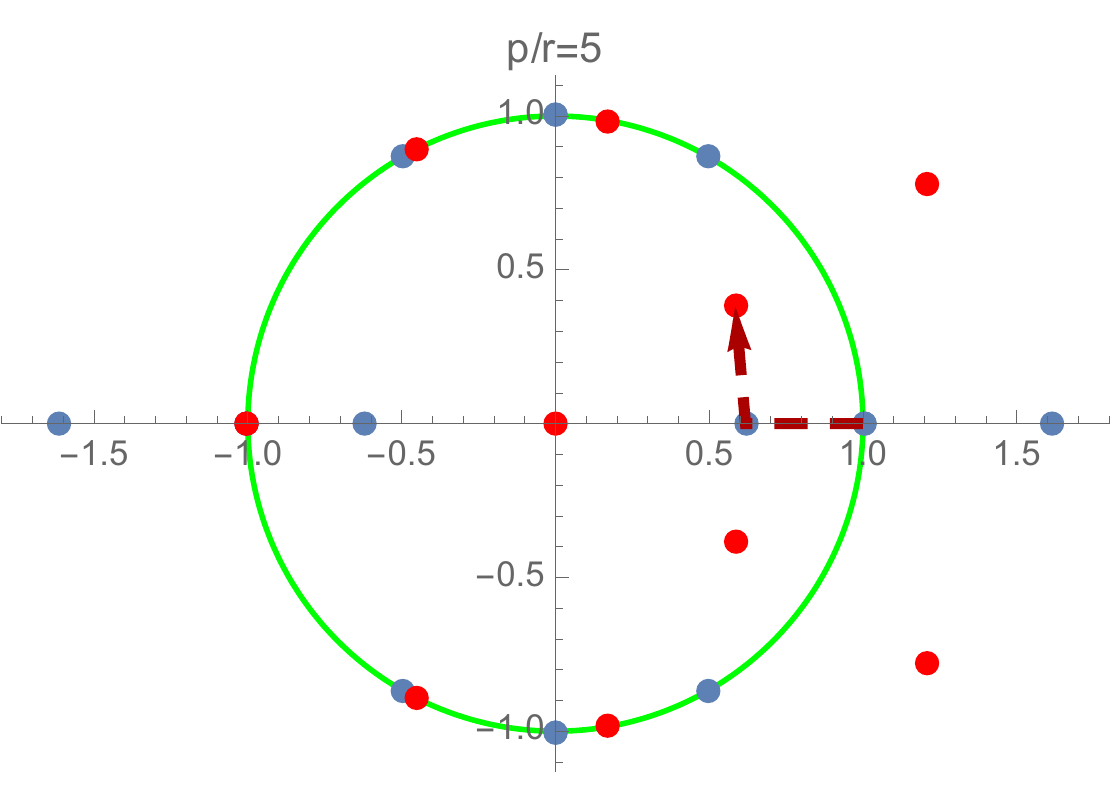}
\end{tabular}
\caption{The $x$-plane for various values of the surgery coefficient $-\frac{p}{r}$.
Blue dots are branch points of the solutions to $A(y,x)=0$ w.r.t. $y$.
Red points are the intersection points $\CC_K\cap \{y^r=x^p\}$.
Dashed arrow shows the path $\gamma_*\subset\CC_K$ connecting the point
associated with the trivial flat connection (at $x=1$)
to the point associated with the irreducible flat connection with the smallest instanton action (at $x=x_*$).
Unlike some other critical points (which can appear and disappear as $r$ changes),
this critical point is always present and $x_*$ depends continuously on the value of $p/r$
(and approaches a puncture at $x=0$ as $|\tfrac{p}{r}|\rightarrow 4$ from either side).}
\label{fig:fig8-x}
\end{figure}

The contour $\gamma_*\subset \CC_K$ can be chosen as follows. It first connects $x=1$ with $x= \frac{\sqrt{5}-1}{2}$
on the abelian branch $y=1$, then passes to the ``irreducible branch'' and reaches $x_*$ along a straight line (see Figure~\ref{fig:fig8-x}). The contour integral (\ref{S-contour-int}) then naturally splits into two pieces:
\be
2\pi^2 \ell_*=-\frac{p}{2r}(\log x)^2|_{x=(\sqrt{5}-1)/2}+\int\limits_{((\sqrt{5}-1)/2,1)}^{(x_*,x_*^{p/r})}\left(\log y-\frac{p}{r}\log x\right)\frac{dx}{x}\,.
\label{S-integral}
\ee
Note, the action  $\ell_*$ is normalized such that the corresponding contribution to
the path integral is weighted by $e^{2\pi ik \ell_*}$.

\subsection{From cyclotomic expansion to the asymptotic one}

Let $J_n[K]$ be the colored Jones polynomial of a knot $K$ normalized to $1$ on the unknot.
Consider Habiro's cyclotomic expansion~\cite{habiro2002quantum}:
\be
J_n (K) \; = \; \sum_{m=0}^{\infty} C_{m} [K] \cdot (qx^2)_m (q/x^2)_m \,, \qquad (x^2 = q^n)
\label{Jncyclotomic}
\ee
where $C_m [K] \in \Z [q,q^{-1}]$ and we use the standard notation
\be
(x)_n \; \equiv \; (x;q)_n \; := \; (1-x)(1-xq) \ldots (1 - xq^{n-1})\,.
\ee
In writing \eqref{Jncyclotomic}, we replaced $q^n$ by the variable $x^2$ to emphasize that the color-dependence
of $J_n[K]$ is encoded entirely in the $x$-dependence.
In fact, it is the same variable that, via the generalized volume conjecture,
becomes the $x$-variable of the A-polynomial curve~\eqref{Acurve}.

The cyclotomic expansion \eqref{Jncyclotomic} proves to be very helpful in writing explicit surgery formulae
for the quantum invariants of $M_3 = S^3_{-p/r} (K)$ in terms of the coefficients $C_m [K]$.
Thus, according to \eqref{H1Zp}, for $|\tfrac{p}{r}| =1$ the resulting 3-manifold is the integer homology sphere ($\Z$HS)
and the corresponding surgery formulae have a very nice and simple form\footnote{See footnote \ref{WRTfootnote} for relation between $Z_\text{CS}$ and $\tau$}~\cite{habiro2002quantum,BBLlaplace,MR2457479}:
\be
\tau (S^3_{+1} (K)) \; = \; \frac{1}{1-q} \sum_{m=0}^{\infty} (-1)^m q^{-\frac{m(m+3)}{2}} C_m [K] (q^{m+1})_{m+1}
\label{Habiroplusone}
\ee
and
\be
\tau (S^3_{-1} (K)) \; = \; \frac{1}{1-q} \sum_{m=0}^{\infty} C_m [K] (q^{m+1})_{m+1}\,.
\label{Habirominusone}
\ee
For example, in the case of the (right- and left-handed) trefoil knot and the figure-8 knot that we used
as our main examples throughout the paper, the coefficients of the cyclotomic expansion look like:
\be
C_m [ {\bf 3_1^r} ] = q^{-m(m+2)} \,, \qquad
C_m [ {\bf 3_1^\ell} ] = q^m \,, \qquad
C_m [ {\bf 4_1} ] = (-1)^m q^{-\frac{m(m+1)}{2}}\,.
\label{treffig8cycl}
\ee
Correspondingly, the srugery formula \eqref{Habirominusone} gives the familiar expression for
the WRT invariant of the Poincar\'e sphere $\Sigma (2,3,5) = S^3_{-1} ({\bf 3_1^\ell})$
as a surgery on the trefoil knot, {\it cf.} \eqref{Z235}:
\be
\tau (S^3_{-1} ({\bf 3_1^\ell})) \; = \; \frac{1}{1-q} \sum_{m=1}^{\infty} q^{m-1} (q^{m})_{m}\,.
\label{ZPoincaretref}
\ee
This expression, as well as its generalizations to $(-\frac{p}{r})$-surgeries for other knots,
can be conveniently written in terms of a ``Laplace transform'' $\CL_{p/r}$ acting
on the cyclotomic expansion \eqref{Jncyclotomic} presented as a function of $x$.
(This was another reason to write \eqref{Jncyclotomic} in terms of $x$.)
Namely, $\CL_{p/r}$ is a linear (over $\Z[q,q^{-1}]$) operation acting on monomials in $x$
as follows, {\it cf.}~\cite{BBLlaplace,MR2457479}:
\be
\CL_{p/r}: \qquad x^m \mapsto q^{\tfrac{r m^2}{4p}}\,.
\ee
Then, it is easy to see that the WRT invariant \eqref{ZPoincaretref} can be expressed as the simplest
version of the Laplace transform
\be
\tau (S^3_{-1} ({\bf 3_1^\ell})) \; = \; \frac{1}{1-q} \; \CL_1 \left[ \sum_{m=1}^{\infty} q^{m-1} \, (x^2)_m \, (1/x^2)_m \right]
\ee
which sends $x^m \mapsto q^{m^2 / 4p}$ or, equivalently, as
\be
\tau (S^3_{-1} ({\bf 3_1^\ell})) \; = \; -\frac{1}{2 (1-q)} \; \CL_1
\left[ \sum_{m=0}^{\infty} (x^2 - 2 + x^{-2}) \, q^m \, (q x^2)_m \, (q/x^2)_m \right]\,.
\ee

More generally\footnote{A generalization in a different direction would produce surgery formulae for homological blocks:
\be
\hat Z_a(q) \; = \; \CL_p^{(a)} \, [F(q,x)] \; = \; \sum_{m,n = pj + a} N_{m,n} \, q^{m + \tfrac{j^2}{4p}}
\; = \; \int_{|x|=1} \frac{dx}{2\pi i x} \, F(q,x) \, \theta_p^{(a)} (x)
\label{Laplace}
\ee
where $F(q,x) = \sum_{m,n} N_{m,n} \, q^m x^{n}$ denotes the right-hand side of \eqref{cyc-Laplace}
and $\CL_p^{(a)}$ is a ``generalized Laplace transform'' that sends
$x^{n} \mapsto q^{n^2/4p}$ and sums only over $n$ of the form $n = pj + a$ for $j \in \Z$.
Here, we also expressed it as a convolution with a theta-function:
\be
\theta_p^{(a)} (x) = \sum_{n \; = \; a \, \text{mod} \, p} q^{\frac{n^2}{4p}} x^n
\ee
Since for the trivial flat connection labeled by $a=0$, the elements of the S-matrix \cite{Gukov:2016gkn}
are all $S_{ab}=1$, the expression for $Z_a (q)=\sum_{a} S_{ab} \hat Z_b (q)$ obtained by the generalized
Laplace transform \eqref{Laplace} agrees with \eqref{cyc-Laplace}.}
--- and importantly for resurgent analysis! --- acting with $\CL_{p/r}$
on the cyclotomic expansion \eqref{Jncyclotomic} written in terms of $x$ gives the surgery formula
for the asymptotic $\frac{1}{k}$-expansion around the trivial flat connection with arbitrary $p/r$
({\it cf.} \cite{Rozansky:1994qe,Rozansky:1994ba}):
\begin{equation}
\sqrt{2kir}\,q^{\Delta_{p,r}} \, Z_\text{triv} \; = \; \CL_{p/r}
\underbrace{\left[\,(x-x^{-1})(x^{1/r}-x^{-1/r}) \sum_{m=0}^\infty C_m[K]\, (qx^2)_m(q/x^2)_m \right]}_{F(q,x)}
\label{cyc-Laplace}
\end{equation}
where $\Delta_{p,r} \in \Q$ is some knot-independent rational number.
Note, this expression gives us a well defined element in $\Q[[2\pi i/k]]$,
\begin{equation}
 \sqrt{2kir}q^{\Delta_{p,r}}\,Z_\text{triv}= \sum_{m=1}^\infty \frac{a_m}{k^{m}},
\label{Ztrivpert}
\end{equation}
because
\begin{equation}
\CL_{p/r}\left[\,(x-x^{-1})(x^{1/r}-x^{-1/r})
	 C_m[K]\, (qx^2)_m(q/x^2)_m\right]=O\left(\frac{1}{k^{m+1}}\right).
\end{equation}
This makes \eqref{cyc-Laplace} convenient for resurgent analysis and allows to extract
the perturbative coefficients $a_m$ from the coefficients of the cyclotomic expansion.

\begin{figure}[ht]
\centering
\begin{tabular}{cc}
 \includegraphics[scale=0.7]{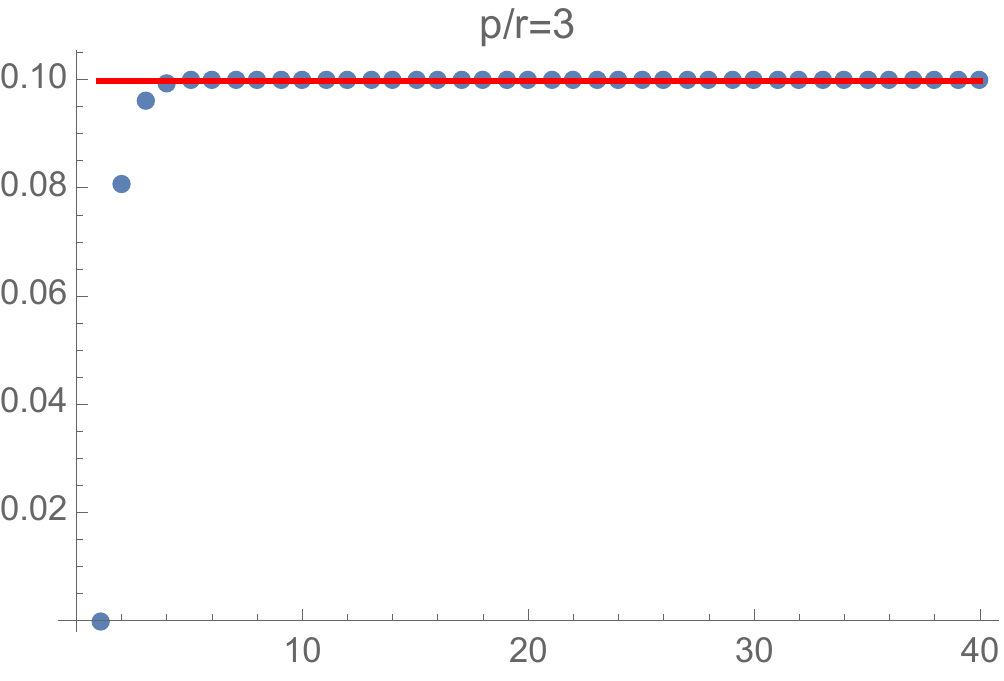} &  \includegraphics[scale=0.7]{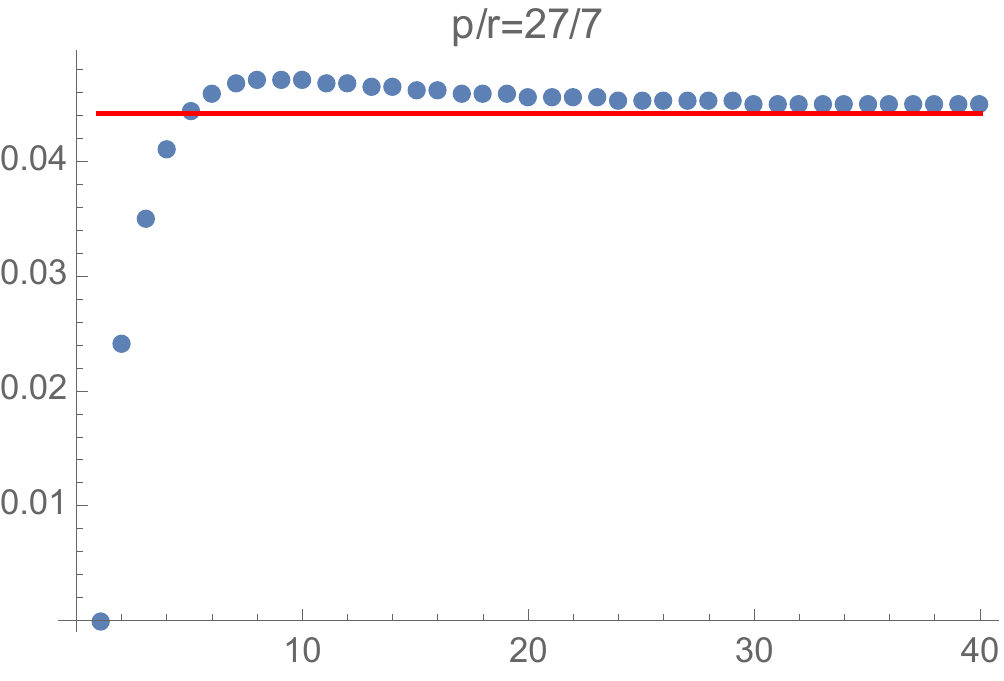} \\
  \includegraphics[scale=0.7]{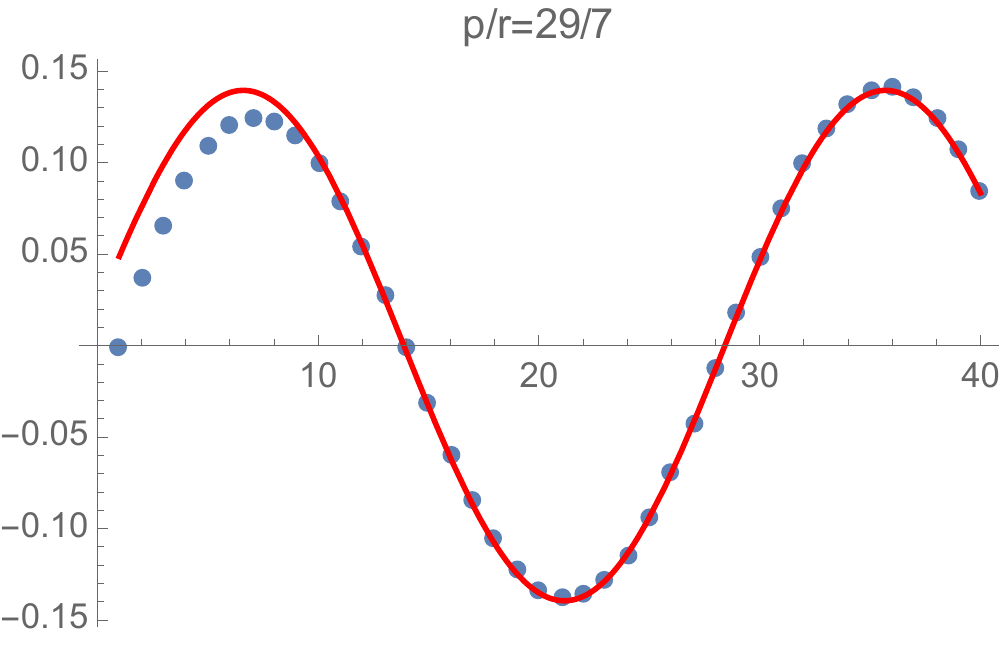} &  \includegraphics[scale=0.7]{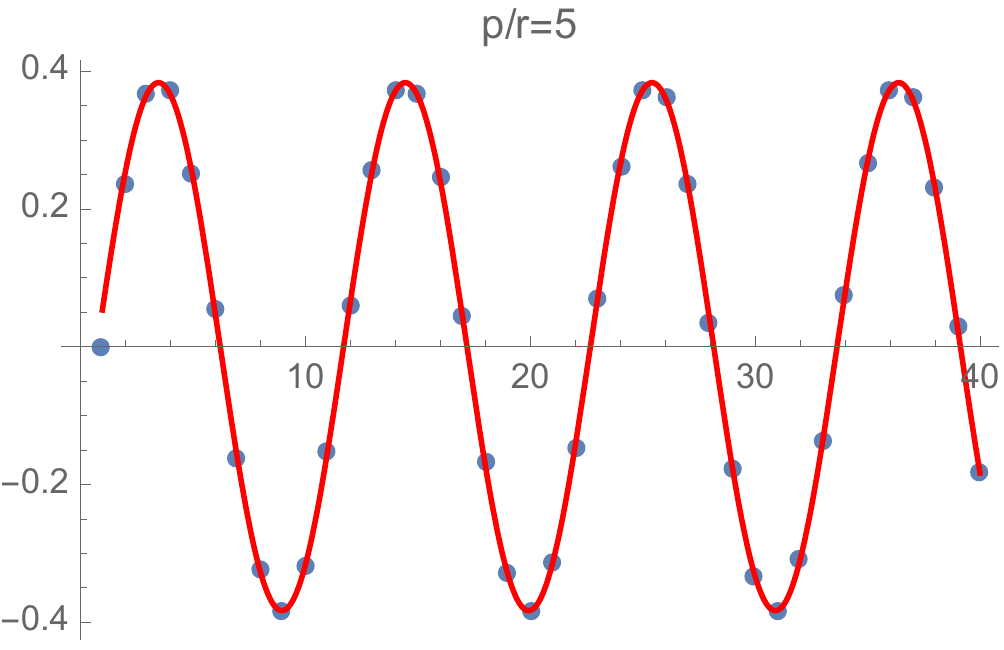}
\end{tabular}
\caption{Blue dots depict values of the sequence (\ref{atilde}) for various values of the surgery coefficient $-\frac{p}{r}$.
Red curves are plots of the functions of $m$ in the r.h.s. of (\ref{atilde-asympt}) with numerically fitted values of $C_*$, that is amplitude and phase of oscillations, in the case of complex $\ell_*$.
In the case of real $\ell_*$ we use a higher Richardson transform of the sequence $\tilde{a}_m$ to determine the value of $C_*$. }
\label{fig:fig8-asympt}
\end{figure}

Indeed, as before, let $\ell_*$ be the instanton action with the smallest absolute value.
When $\ell_*$ is real (which is the case for $SU(2)$ and $SL(2,\R)$ flat connections),
we expect the coefficients of the perturbative expansion \eqref{Ztrivpert} to have the following asymptotics:
\begin{equation}
a_m \; \sim \; C_*\,\frac{\Gamma(m+1/2)}{(2\pi i \ell_*)^{m}} \,, \qquad m\rightarrow \infty
\label{asympt-real}
\end{equation}
which correspond to the statement that singularity of the Borel transform closest to the origin has the following form (cf. (\ref{Borel-general-pole})):
\begin{equation}
	B^\text{triv}(\xi)\sim \frac{C_*}{\xi -2\pi i \ell_*}+\ldots
	\label{Borel-sing-approx}
\end{equation}
in the leading order.

On the other hand, when $\ell_*$ is complex, there is also an instanton with the complex conjugate
value of the action, $\bar{\ell_*}$, which gives the contribution to the asymptotics of the same order.
This results in a different behavior:
\begin{equation}
a_m \; \sim \; |C_*| \,\frac{\Gamma(m+1/2)}{(-2\pi i |\ell_*|)^{m}} \cos(m\arg(- \ell_*)+\arg C_*) \, ,\qquad m\rightarrow \infty
\label{asympt-complex}
\end{equation}
where the coefficient $C_*$ is usually called the {\it Stokes parameter}.
Such asymptotic behaviors can be tested by constructing a new sequence from
the coefficient of the $\frac{1}{k}$-expansion:
\begin{equation}
 \tilde{a}_m=
 \begin{cases}
  \frac{a_m(2\pi i \ell_*)^{m}}{\Gamma(m+1/2)} \,, & \mathrm{Im}\, \ell_*=0\,, \\
   \frac{a_m(-2\pi i |\ell_*|)^{m}}{\Gamma(m+1/2)} \,, & \mathrm{Im}\, \ell_*\neq 0\,.
   \end{cases}
\label{atilde}
\end{equation}
This new sequence has the following asymptotics when $m\rightarrow \infty$:
\begin{equation}
 \tilde{a}_m\sim \left\{
 \begin{array}{ll}
  C_*, & \mathrm{Im}\, \ell_*=0\,, \\
   |C_*| \cos(m\arg(- \ell_*)+\arg C_*) , & \mathrm{Im}\, \ell_*\neq 0\,,
   \end{array}
 \right.
 \label{atilde-asympt}
\end{equation}
for some $C_*$, if and only if the correct $\ell_*$ is chosen in (\ref{atilde}).

In our example of the figure-8 knot, we can
use this to check that the value of $\ell_*$ in (\ref{asympt-real}) or (\ref{asympt-complex})
indeed coincides with the one computed by the integral (\ref{S-integral});
we find an excellent agreement (see Figure~\ref{fig:fig8-asympt}).
The cyclotomic coefficients are given in \eqref{treffig8cycl} and
the Stokes coefficient $C_*$ can be computed by fitting the asymptotics (\ref{atilde-asympt})
to the numerical values of $\tilde{a}_m$.
In the table below we present numerical values of $\ell_*$ and $C_*$ for the examples
shown in Figures \ref{fig:fig8-x} and \ref{fig:fig8-asympt}:
\begin{equation}
\begin{array}{|c|c|c|}
\hline
p/r & C_* & \ell_* = \text{CS} (\alpha_*) \\
\hline\hline
5 & -0.161002+0.347784 i & -0.0385191 + 0.0248584 i \\
29/7 & 0.0196256\, +0.13811 i & -0.043208 + 0.00948317 i \\
27/7 & 0.0441796 &  -0.0340532 \\
3 & 0.0997329 & -0.0208333 \\
\hline
\end{array}
\end{equation}
It would be interesting to check that subleading terms in the asymptotics (\ref{asympt-real}) (or (\ref{asympt-complex})), which correspond to the subleading, less singular terms, in (\ref{Borel-sing-approx}) agree with the structure (\ref{Borel-general-pole}).


\subsection{Borel integrals as localization integrals in 3d $\CN=2$ theory}

According to the generalized volume conjecture,
the asymptotic behavior of the function $F(q,x)$ on the right-hand side of \eqref{cyc-Laplace}
in the limit $q = e^{\hbar} \to 1$ has the form \cite{Apol}:
\be
F(q,x) \; = \; e^{\frac{1}{\hbar} \tilde W_K (x) + \ldots}
\qquad (q \to 1 \,, \; x = \text{fixed})
\ee
where the graph of function $\partial \tilde W_K$ is precisely the A-polynomial curve $\CC_K$
introduced in \eqref{Acurve}:
\be
\CC_K: \qquad \exp \left( \frac{\partial \tilde W_K}{\partial \log x} \right) \; = \; y\,.
\ee
In fact, if we express \eqref{cyc-Laplace} as a contour integral, similar to \eqref{Laplace},
then in the limit $q = e^{\hbar} \to 1$ it will take precisely the form of the vortex partition function \eqref{Zfd}
in a 3d $\CN=2$ theory $T[S_{-p/r} (K)]$ with the twisted superpotential
\be
\tilde W (x) \; = \; \tilde W_K (x) - \frac{p}{2r} (\log x)^2\,.
\ee
Critical points of this multivalued function are precisely the intersection points \eqref{curve-intersection}
that gave us a list of flat connections on the knot surgery $M_3 = S^3_{-p/r} (K)$:
\be
\left\{ y \, x^{-\frac{p}{r}} = 1 \right\} \cap \CC_K
\qquad \Leftrightarrow \qquad
\exp \left( \frac{\partial \tilde W}{\partial \log x} \right) \; = \; 1\,.
\ee
Indeed, from the viewpoint of 3d $\CN=2$ theory, gluing the two pieces in the surgery operation \eqref{knotsurgery}
corresponds to gauging $U(1)_x$ global symmetry of $T[S^3 \setminus K]$ and $T[S^1 \times D^2]$,
which at the level of vortex partition functions means integrating over $x$,
\be
Z_{T[M_3]} \; = \; \int \frac{dx}{2 \pi i x} \; Z_{T[S^3 \setminus K]} (x) \cdot Z_{\varphi \, \circ \, T[S^1 \times D^2]} (x^{-1})
\label{ZZZgluing}
\ee
and at the level of the corresponding superpotentials means extremizing $\tilde W$ with respect to $x$, {\it cf.} \eqref{Wcrit}.

It would be interesting to pursue these ideas further and to explore the connection between Borel integrals
and localization integrals in SUSY gauge theories, in the context of 3d-3d correspondence and beyond.


\section{Triangulations}

There are various other ways of building a 3-manifold $M_3$ out of basic pieces,
which include handle decompositions, triangulations, {\it etc.} Here we consider triangulations.

\begin{figure}[h]
\centering
\includegraphics[width=3in]{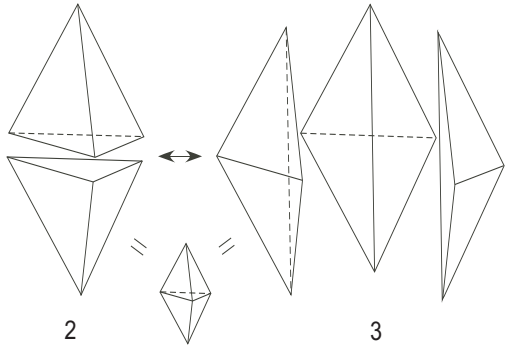}
\caption{The 2-3 Pachner move.}
\label{fig:Pachner}
\end{figure}

In this approach, $Z_{\text{pert}} (M_3)$ is computed by decomposing $M_3$ into tetrahedra
(typically, ideal tetrahedra) and associating to each tetrahedron a function that ensures invariance
under the 2-3 Pachner move (see Figure~\ref{fig:Pachner}).
Various forms of the quantum dilogarithm function are ideally suited for this role,
with the five-term pentagon relation responsible for the invariance under the 2-3 Pachner move.
Beautiful as it is, this idea usually leads to state integral models for the $SL(2,\C)$ Chern-Simons partition function,
including the one in \cite{dglz}, that do not ``see'' abelian flat connections paramount in the resurgent analysis.
Nevertheless, a state integral can successfully produce an all-loop
perturbative expansion $Z_{\text{pert}} (M_3)$ around a {\it non-abelian} critical point that we explore here.

In studying triangulations of hyperbolic 3-manifolds, one usually starts with a complement of
the figure-8 knot as the basic example, obtained by gluing two ideal tetrahedra~\cite{MR1435975}.
Correspondingly, the ``state integral'' model for the $SL(2,\C)$ Chern-Simons partition function \cite{dglz}:
\be
Z (\hbar) \; = \; \frac{1}{\sqrt{2\pi \hbar}} \int \frac{\Phi_{\hbar} (z)}{\Phi_{\hbar} (-z)} \rd z
\label{fig8pert}
\ee
involves two copies of the Faddeev's quantum dilogarithm function \cite{Faddeev:1995nb}
(sometimes called ``noncompact'' quantum dilogarithm),\footnote{The function $\Phi_{\hbar} (z)$ here
is related to $\Phi_\mb (p)$ used {\it e.g.} in \cite{gk} by a simple change of variable
$p=\frac{z}{2 \pi \mb}$ and the parameter $\hbar = 2 \pi i \mb^2 = 2\pi i /k$.
We conform to the former, which is more convenient in state integral models.
Also note, that \cite{dglz} used $q = e^{2 \hbar}$, so that $\hbar_{\text{here}} = 2 \hbar_{\text{DGLZ}}$.}
\be
\Phi_{\hbar} (z) = \exp \left(
\frac{1}{4} \int_{\R + i \epsilon} \frac{dx}{x} \frac{e^{-izx}}{\sinh (\pi x) \sinh (\tfrac{\hbar}{2i} x)} \right)
= \prod_{r=1}^{\infty} \frac{1 + q^{r - \frac{1}{2}} e^z}{1 + {}^Lq^{\frac{1}{2}-r} e^{{}^Lz}}\,.
\label{Faddeevdilog}
\ee
To all orders in perturbative $\hbar$-expansion, one can omit the terms involving
${}^L q = \exp (- \tfrac{4\pi^2}{\hbar})$ and ${}^L z = \frac{2\pi i}{\hbar} z$.
Hence, for the purpose of studying the perturbative power series in $\hbar$,
we can replace $\Phi_{\hbar} (z)$ in \eqref{fig8pert} by the $q$-Pochhammer symbol $(-q^{1/2} e^z ; q)_{\infty}$
defined for $|q|<1$ and all $z \in \C$.
Specifically, if we use the asymptotic expansion of the quantum dilogarithm \cite[sec.3.3]{dglz}:
\be
\log \Phi_{\hbar} (z)  \; \sim \;
\sum_{n\ge 0} \hbar^{n-1} \frac{B_n (1/2)}{n!} {\rm Li}_{2-n}(-\re^z) \,,
\ee
we can write the integral \eqref{fig8pert} as
\be
Z (\hbar) \; = \; \frac{1}{\sqrt{2\pi \hbar}} \int \re^{- k V(z, k)} \rd z,
\label{fig8pertxx}
\ee
where, using $k = \frac{2\pi i}{\hbar}$ ($={1\over \mb^2} $) and the fact that $B_n (1/2) = 0$ for $n$ odd,
we have
\be
\ba
V(z, k)& \; \sim \; -\sum_{n\ge 0} {B_{2n}(1/2) \over (2n)!} \frac{(2 \pi \ri)^{2n-1}}{k^{2n}}
\left({\rm Li}_{2-2n}(-\re^z) -{\rm Li}_{2-2n}(-\re^{-z})  \right)\\
&=V_0 (z) +  \CO \left( \tfrac{1}{k} \right) \,,
\ea
\ee
with
\be
V_0(z) \; = \; {1\over 2 \pi \ri} \left( {\rm Li}_2 (-\re^{-z}) -{\rm Li}_2(\re^{-z}) \right)
\ee
This function has critical points at
\be
\zeta_n^{\pm}=\pm {2 \pi \ri \over 3}  + 2 \pi \ri n , \qquad n \in \IZ.
\ee
If we denote the hyperbolic volume of the figure-8 knot complement by
\be
V (\fig8)=2 {\rm Im} \left( {\rm Li}_2 (\re^{\pi \ri /3}) \right)= 2.0298832...
\ee
then $V_0 \left( \zeta_n^\pm\right)=\mp {V(\fig8) \over 2 \pi}$.

\begin{figure}[tb]
\begin{center}
\begin{tabular}{cc}
\resizebox{70mm}{!}{\includegraphics{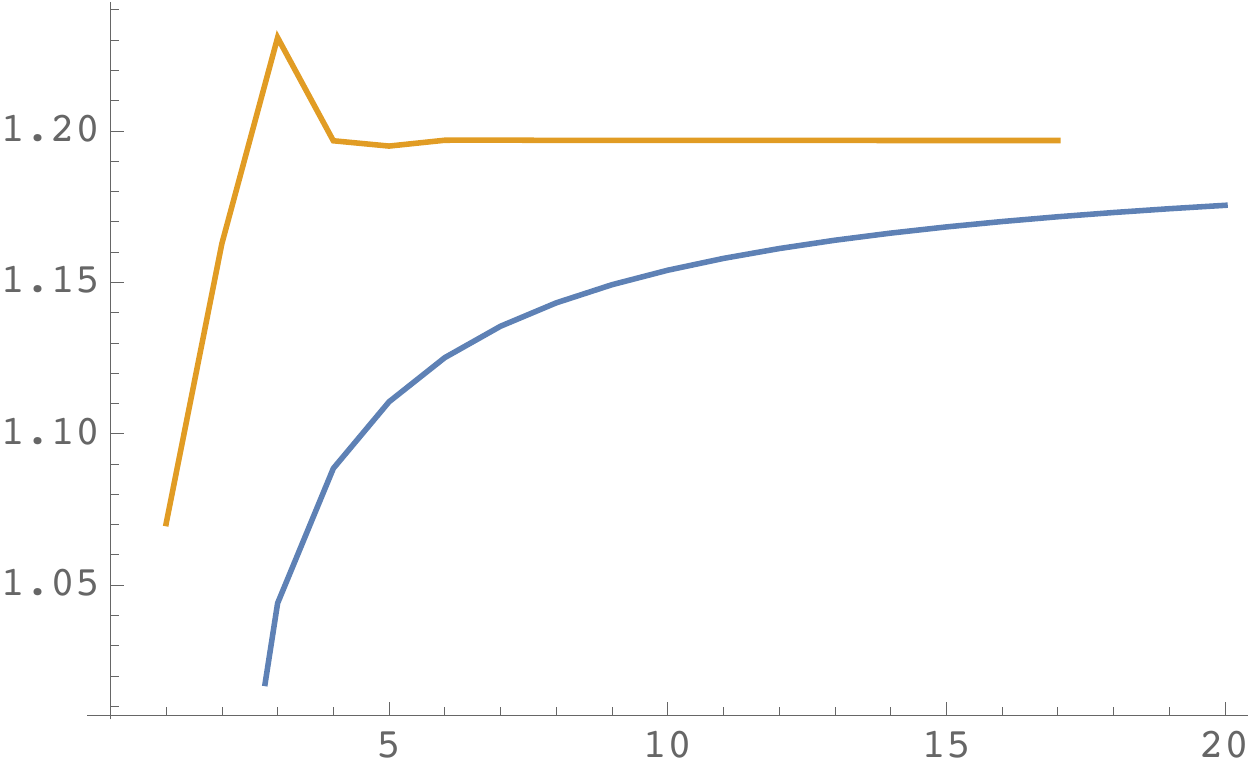}}
\hspace{3mm}
&
\resizebox{65mm}{!}{\includegraphics{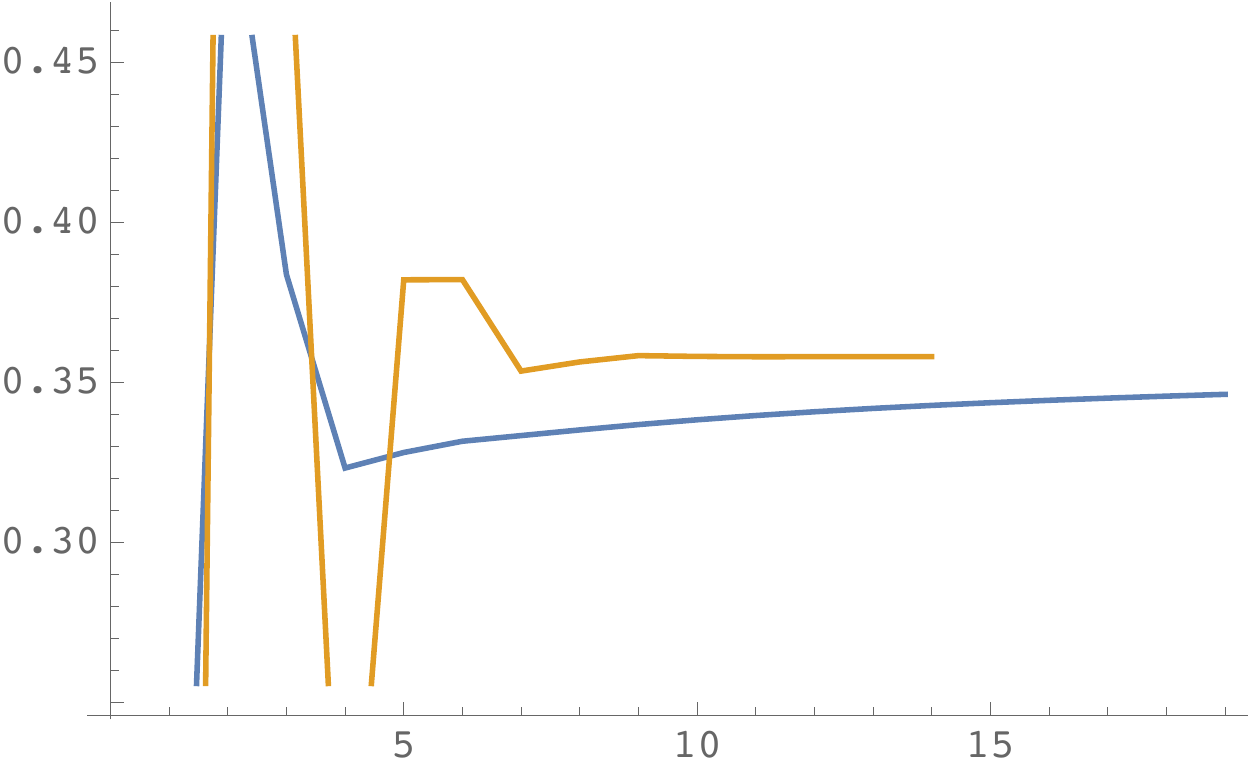}}
\end{tabular}
\end{center}
\caption{(Left) The sequence $b_n$ defined in (\ref{ln-seq}) and its third Richardson transform,
showing the convergence to the Stokes parameter. (Right) The sequence $s_n$ defined in (\ref{sn-seq})
and its fifth Richardson transform, showing the convergence to (\ref{f-corr}). }
\label{sn}
\end{figure}

We want to analyze the integral \eqref{fig8pertxx} by using the tools of resurgence.
As usual, to each critical point we associate a trans-series by doing a formal saddle-point evaluation.
For example, to obtain the trans-series associated to $\zeta_0^+$,
we write
\be
V(z,k) \; = \; V_0 \left( \zeta_0^+\right) +{\sqrt{3} \over 4 \pi} (z-\zeta_0^+)^2
+ \sum_{n,m} c_{n,m} \frac{(z-\zeta_0^+)^n}{k^m} \,,
\ee
and we integrate.\footnote{It is convenient to change variables to $u=z \sqrt{ \frac{k \sqrt{3}}{2 \pi} }$.}
Note, since the expansion of the ``potential function'' starts with a quadratic term,
this critical point is non-degenerate (as we expect, because the corresponding flat connection is irreducible).
Let us denote the trans-series associated to $\zeta_0^+$ by $Z_{\text{pert}}^+$. We find
\be
Z_{\text{pert}}^+ \; = \; \frac{1}{3^{1/4}} \re^{k V(\fig8) \over 2 \pi} \left( 1 -{11 \pi \over 36 k {\sqrt{3}}} + {697 \pi^2 \over 776 k^2} +\cdots\right) \,.
\ee
If we write the perturbative coefficients of this series as
\be
Z_{\text{pert}}^+ \; = \; \frac{1}{3^{1/4}} \re^{k V(\fig8) \over 2 \pi} \; \sum_{n \ge 0} \frac{a_n}{k^n} \,,
\ee
then the series around the saddle point $\zeta_0^-$ looks like
\be
\label{z-minus}
Z_{\text{pert}}^{-} \; = \;
\frac{1}{3^{1/4}} \re^{- \frac{k V(\fig8) }{2 \pi}} \; \sum_{n \ge 0} \, (-1)^n \, \frac{a_n}{k^n} \,.
\ee
In the setting of \cite{dglz}, the series $Z_{\text{pert}}^+$ (resp. $Z_{\text{pert}}^-$) corresponds to
the perturbative expansion around the ``geometric'' (resp. ``conjugate'') flat connection on $M_3 = S^3 \setminus \fig8$.
In our setting, if we regard $Z_{\text{pert}}^+$ as the perturbative series,
then $Z_{\text{pert}}^-$ should be considered as an instanton trans-series.
This is similar to the Airy function, where ${\rm Ai}$ is the perturbative series and ${\rm Bi}$
can be regarded as the instanton trans-series (see the discussion in Example 2.12 in \cite{mm-lectures}).
This has consequences for the large order behavior of the coefficients $a_n$.
One expects that (see for example eq.(2.156) in \cite{mm-lectures}):
\be
\label{asy}
a_n \; \sim \; {S \over 2 \pi \ri } \Gamma(n-\delta) A^{-n+\delta} \left[1+ {\varphi_{1,n} A \over n-\delta-1}+\cdots\right]
\ee
where $\varphi_{1,n}$ are the coefficients of the trans-series, which in our case is given by (\ref{z-minus}), {\it i.e.}
\be
\varphi_{1,n} \; = \; (-1)^n a_n \,.
\ee
The action in (\ref{asy}) should be the difference between the actions of the saddles,
\be
A \; = \; V_0(\zeta_0^+)-V_0(\zeta_0^-) \; = \; -{V(\fig8) \over  \pi} \,.
\ee
In addition, we expect $\delta=0$. Indeed, it can be easily checked numerically that
\be
a_n \; \sim \; A^{-n} (n-1)! \,.
\ee
This corresponds to a singularity in the Borel plane located at $A<0$.
In Figure~\ref{sn}, we show the sequence
\be
\label{ln-seq}
b_n = {a_n \over A^{-n} (n-1)!}
\ee
which should converge to the Stokes parameter $S/(2 \pi \ri)$. Numerically, we find
\be
{S\over 2 \pi \ri} \approx  1.196827
\ee
In addition, we can test the next-to-leading order asymptotics by considering the sequence
\be
\label{sn-seq}
s_n= n^2 \left[ {a_{n+1} \over n a_n A } -1\right],
\ee
which should converge to
\be
\label{f-corr}
-\varphi_{1,1} A = {11 \pi \over 36 {\sqrt{3}}}  {V(\fig8) \over \pi} \approx 0.358097
\ee
By using just twenty terms in the series we find a good match with this value.
For example, after five Richardson transforms, the best approximation to this number gives
\be
0.3580976...
\ee
We show the sequence $s_n$ and its fifth Richardson transform in Figure~\ref{sn}.

The series $Z_{\text{pert}}^+$ is Borel summable,
so we can consider its Borel resummation $SZ_{\text{pert}}^+(k)$.
For example, at $k=1$ we find\footnote{It is curious to compare it with the value of the integral \eqref{fig8pert}
which can be computed in closed form for rational values of $k \in \Q$, and for $k=1$ gives \cite{gk}:
\be
Z(1) 
= {1\over {\sqrt{3}}}\left( \re^{{ V(\fig8) \over 2 \pi}} - \re^{-{ V(\fig8) \over 2 \pi}}\right)
\approx 0.379568  \,.
\ee
Of course, we do not expect to find an agreement since the perturbative
expansion in $\hbar$ or in $1/k$ does not ``see'' the second $q$-Pochhammer symbol
in the denominator of \eqref{Faddeevdilog}.
Rather, the numerical value of the Borel resummation here should be compared with
the version of the integral \eqref{fig8pert} --- which is precisely the vortex partition function of $T[S^3 \setminus \fig8]$ ---
where $\Phi_{\hbar} (z)$ is replaced by $(-q^{1/2} e^z ; q)_{\infty}$.}
\be
SZ_{\text{pert}}^+(1) \approx 0.7242
\ee


\acknowledgments{We would like to thank Miranda Cheng, Mikhail Kapranov, Amir Kashani-Poor, Albrecht Klemm,
Maxim Kontsevich, Cumrun Vafa, Edward Witten, Masahito Yamazaki for useful comments and discussions.
The work of S.G. is funded in part by the DOE Grant DE-SC0011632 and the Walter Burke Institute for Theoretical Physics.
The work of M.M. is supported in part by the Swiss National Science Foundation, subsidies 200020-149226, 200021-156995,
and by the NCCR 51NF40-141869 ``The Mathematics of Physics" (SwissMAP).
P.P. gratefully acknowledges support from the Institute for Advanced Study.
Opinions and conclusions expressed here are those of the authors and do not necessarily reflect the views of funding agencies.
}


\appendix


\section{Notes on Borel resummation and Picard-Lefschetz theory}

\label{app:PL}

The goal of this section is to review the relation between Borel resummation and Picard-Lefschetz theory with
generalization to the case of non-isolated critical points. Some relevant references are (the case of isolated critical points): \cite{Pham,berry1990hyperasymptotics,berry1991hyperasymptotics,howls1997hyperasymptotics,Cherman:2014ofa}.
The case of non-isolated critical points is  discussed in \cite{Witten:2010cx} (however without application to Borel resummation)
and from a slightly different angle in \cite{Kontsevich}.

We are interested in a multidimensional integral of the following type:
\begin{equation}
 I_\Gamma=\int_\Gamma dz_1\ldots dz_M e^{2\pi i k S(z_1,\ldots,z_M)}
 \label{Int-Gamma}
\end{equation}
where $\Gamma$ is a contour of real dimension $M$ in $\C^M$ chosen such that the integral is convergent. The integral $I_\Gamma$ only depends on the class of $\Gamma$ in the relative homology $H_M(\C^M,Y_k)$ where $Y_k \subset \d \C^M$ where $-\mathrm{Re} [2\pi i k S(z_1,\ldots,z_M)]$ becomes very large. Obviously, $Y_k$ depends only on the argument $k$. Note that the story can be easily generalized to the case when a polynomial function is inserted in (\ref{Int-Gamma}) and also to the case of an arbitrary complex manifold instead of $\C^M$.

\begin{figure}[ht]
\centering
 \includegraphics[scale=1.8]{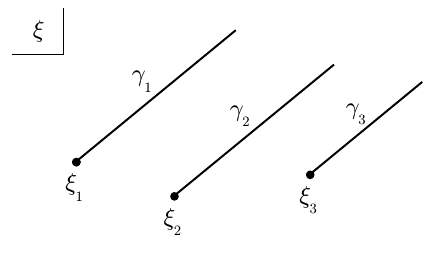}
\caption{Contours $\gamma_\bba$, the images of Lefschetz thimbles under the map $S:\C^M\rightarrow \C_\xi$, for generic value of $k=|k|e^{i\theta}$.}
\label{fig:PL-gammas}
\end{figure}

The function $-2\pi iS(z_1,\ldots, z_M)$ defines a fibration of $\C^M$ over $\C$:
\begin{equation}
\begin{CD}
\C^M\\
@V{-2\pi iS}VV\\
\C_{\xi}
\end{CD}
\end{equation}
where $\xi$, the value of the action, is a complex coordinate parameterizing the base. The fiber over a point with coordinate $\xi$ is an $(M-1)$-dimensional hypersurface
\begin{equation}
X_\xi =\{-2\pi i S(z_1,\ldots,z_M)=\xi\} \subset \C^M
\end{equation}
which becomes singular at critical values of the action $\{\xi_\bba\}_\bba\subset \C_\xi$.
Over a generic point
\begin{equation}
\xi \in \C\setminus \{\xi_\bba\}_\bba
\end{equation}
the fiber $X_\xi$ is the same manifold when considered in smooth category which we denote as $X_*$. The integration contour $\Gamma$ is projected to the contour $\gamma$ on the base. Locally $\Gamma$ is fibered over $\gamma$ as
\begin{equation}
\Gamma \stackrel{\text{locally}}{\approx} \gamma \times \Gamma_* ,\qquad \Gamma_*\subset X_*
\end{equation}
where $\Gamma_*$ is a representative of middle-dimensional homology\footnote{Note that here it is the usual, non the {\it relative} homology.} of the fiber $X_*$:
\begin{equation}
[\Gamma_*] \subset H_{M-1}(X_*,\Z).
\end{equation}
For generic value of $k$ there is a well defined decomposition of $\Gamma$ into Lefschetz thimbles, the generators of $H_M(\C^M,Y_k)$:
\begin{equation}
[\Gamma]=\sum_{\bba,j}n_{\bba,j}[\Gamma_{\bba,j}]
\end{equation}
so that
\begin{equation}
\Gamma_{\bba,j}\stackrel{\text{locally}}{\approx} \gamma_\bba \times \Gamma^{\bba,j}_*
\end{equation}
where
\begin{equation}
\gamma_\bba = \{\xi\in \xi_\bba +\frac{1}{k}\,\R_+\} \subset \C_\xi
\end{equation}
(see Figure~\ref{fig:PL-gammas}) and $\{\Gamma_*^{\bba,j}\}_j$ are the set of middle-dimensional cycles of $X_*$ vanishing at $\xi=\xi_\bba$.

Assume that each critical value $\xi_\bba$ corresponds to a connected submanifold of critical points $\CM_\bba\subset \C^M$. Let us first review the standard case when $\CM_\bba\cong \text{pt}$ is an isolated critical point $z^\bba=(z_1^\bba,\ldots,z_M^\bba)$. Then there is an analytic local change of coordinates $z=f(\tilde z)$ such that $z_\bba=f(0)$ and
\begin{equation}
	-2\pi iS= \xi_\bba+\tilde{z}_1^2+\ldots+ \tilde{z}_M^2
\label{actionA1}
\end{equation}
in the vicinity of the critical point. So that $X_\xi$ locally looks like
\begin{equation}
	\{\tilde{z}_1^2+\ldots+ \tilde{z}_M^2 =\xi-\xi_\bba \}\cong T^*S^{M-1}_{\sqrt{|\xi-\xi_\bba|}}\;,
\end{equation}
the cotangent bundle over a sphere whith radius $\sqrt{|\xi-\xi_\bba|}$. There is a unique cycle $\Gamma_*^{\bba}\cong S^{M-1}_{\sqrt{|\xi-\xi_\bba|}}$ in $X_\xi$ vanishing at $\xi=\xi_\bba$.

Now suppose $\CM_\bba$ is a submanifold of complex dimension $d_\bba$. Then there is local change of coordinates $z=f(\tilde z)$ such that $\tilde{z}_1,\ldots,\tilde{z}_{d_\bba}$ locally parametrize $\CM_\bba$ and in the vicinity of $\CM_\bba$ we have
\begin{equation}
	-2\pi iS= \xi_\bba+\tilde{z}_{d_\bba+1}^2+\ldots+ \tilde{z}_M^2\,.
\end{equation}
Therefore $X_\xi$ locally looks like $T^*S^{M-d_\bba-1}_{\sqrt{|\xi-\xi_\bba|}}$ fibration over $\CM_\bba$. Consider $\Lambda_{\bba,j}$, middle dimensional cycles in $\CM_\bba$ which are representatives of the generators in $H_{d_\bba}(\CM_\bba)$. Then vanishing cycles $\Gamma^{\bba,j}_*$ can be chosen to be $S^{M-d_\bba-1}_{\sqrt{|\xi-\xi_\bba|}}$ fibrations over $\Lambda_{\bba,j}$, that is
\begin{equation}
	\Gamma^{\bba,j}_* \stackrel{\text{locally}}{\approx}  \Lambda_{\bba,j}\times S^{M-d_\bba-1}_{\sqrt{|\xi-\xi_\bba|}}.
	\label{vanishing}
\end{equation}
Note that compared to the case of isolated critical point, cycle $\Gamma_{\bba,j}$ does not shrink to a point when $\xi\rightarrow \xi_\bba$ but rather shrinks to a lesser-dimensional cycle $\Lambda_{\bba,j}$.

As in \cite{Kontsevich,Witten:2010cx}, we are interested in applications where
$H_{d_\bba}(\CM_\bba)$ are one dimensional, so that there is unique $\Lambda_\bba$ and moreover, $\CM_*^\bba\cong T^*\Lambda_\bba$ where $\Lambda_\bba$ is an orbit of the action of a compact Lie group ($\CM_*^\bba$ is its complexification).
Correspondingly, there is a unique vanishing cycle $\Gamma_*^{\bba}$ for each $\xi_\bba$ and we can drop extra index $j$.

The integral (\ref{Int-Gamma}) over a cycle $\Gamma=\Gamma_\bba$ then reads
\begin{equation}
	I_{\Gamma_\bba}=
	\int_{\Gamma_\bba} dz_1\ldots dz_M e^{2\pi i k S(z)}=
	\int_{\gamma_\bba}d\xi e^{-k\xi} B^\bba(\xi)
	\label{Int-thimble}
\end{equation}
where
\begin{equation}
	B^\bba(\xi)=\int_{\Gamma^\bba_*(\xi)} \Omega(\xi)
\end{equation}
and $\Omega(\xi)$ is the reduction of the holomorphic volume form to $X_\xi$. Formally one can write\footnote{
To be precise, one has to make a change of coordinates $z=f(u,\xi)$ such that $u$ parametrizes $X_\xi$. Then
\begin{equation}
	\Omega(\xi) = J_f(u,\xi)\, du_1\ldots du_{M-1}
\end{equation}
where $J_f$ is the Jacobian corresponding to the coordinate transformation $f$.
}
\begin{equation}
	\Omega(\xi)=\frac{dz_1\ldots dz_M}{d\xi}.
\end{equation}

\begin{figure}[ht]
\centering
 \includegraphics[scale=1.9]{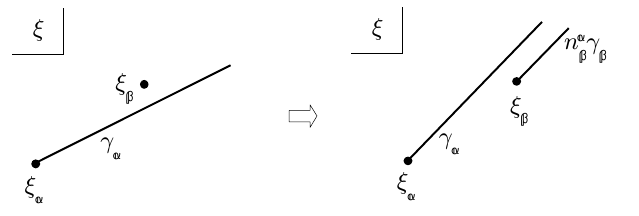}
\caption{Graphical representatation of Stokes phenomenon (\ref{Int-stokes}) which reflects the change $I_{\Gamma_\bba}$ when $\theta$ changes from $\arg(\xi_\bba-\xi_\bbb)-\epsilon$ to $\arg(\xi_\bba-\xi_\bbb)+\epsilon$ for a small epsilon.}
\label{fig:PL-stokes-simple}
\end{figure}

The cycles $[\Gamma^\bba_*(\xi)]\in H_{M-1}(X_\xi,\Z)$ in general have monodromy of the following form
\begin{equation}
	[\Gamma^\bba_*(\xi)]\rightarrow [\Gamma^\bba_*(\xi)]+m^\bba_\bbb[\Gamma^\bbb_*(\xi)],
	\qquad m^{\bba}_{\bbb}\in\Z
	\label{cycle-mon}
\end{equation}
when $\xi$ goes around $\xi_\bbb$. Note that we do not expect the matrix $m^{\bba}_\bbb$ to be symmetric as in the case of the usual Picard-Lefschetz theory when all critical points are isolated. In particular
\begin{equation}
	m^\bba_\bbb \neq \pm \,\#(\Gamma_*^\bba \cap \Gamma_*^\bbb)
	\label{no-PL}
\end{equation}
in general. Locally defined functions $B_\bba(\xi)$ have the same monodromy:
\begin{equation}
	B^\bba(\xi)\rightarrow B^\bba(\xi)+m^\bba_\bbb B^\bbb(\xi).
	\label{B-mon}
\end{equation}
Therefore $B^\bba(\xi)$ is expected to have a branch point at $\xi=\xi_\bbb$. From (\ref{Int-thimble}) and (\ref{B-mon}) it follows that the same integer coefficients appear in the Stokes phenomenon
\begin{equation}
	I_{\Gamma_\bba}\longrightarrow I_{\Gamma_\bba}+m^\bba_\bbb\,I_{\Gamma_\bbb}.
	\label{Int-stokes}
\end{equation}
which happens when $\gamma_a$ passes through $\xi_a$ when rotate $k$ in clockwise direction (see Figure~\ref{fig:PL-stokes-simple}).

From (\ref{vanishing}) it follows that $B^\bbb(\xi)$ has the following expansion (which has finite radius of convergence) when $\xi\approx\xi_\bbb$:
\begin{equation}
	B^\bbb(\xi)= (\xi-\xi_\bbb)^{\frac{M-d_\bbb}{2}-1}\,
	\left(c_0^\bbb+c_1^\bbb\,(\xi-\xi_\bbb)+c_2^\bbb\,(\xi-\xi_\bbb)^2+\ldots\right)\,.
	\label{B-series}
\end{equation}
Its coefficients are simply related to the coefficients of the asymptotic expansion of the integral (\ref{Int-thimble}):
\begin{equation}
	I_{\Gamma_\bbb}=\frac{1}{k^{(M-d_\bbb)/2}}\,e^{- k\xi_\bbb}\left({\Gamma((M-d_\bbb)/2)}\,{c_0^\bbb}+
	\Gamma((M-d_\bbb)/2+1)\,\frac{c_1^\bbb}{k}
	+\ldots\right)\,.
	\label{I-series}
\end{equation}
Equivalently, $B^\bbb(\xi)$ is the Borel transform of the asymptotic series above.

From (\ref{B-mon}) and (\ref{B-series}) it follows that $B^\bba(\xi)$ should have expansion of the following form when $\xi\approx\xi_\bbb$:
\begin{multline}
 B^\bba(\xi)\;=\;\text{regular}\;+\;  \\
 (\xi-\xi_\bbb)^{\frac{M-d_\bbb}{2}-1}\,\left(c_0^\bbb+c_1^\bbb\,(\xi-\xi_\bbb)+c_2^\bbb\,(\xi-\xi_\bbb)^2+\ldots\right)\times
 \left\{\begin{array}{cl}
 \frac{1}{2\pi i}m^\bba_\bbb\log(\xi-\xi_\bbb), & (M-d_\bbb)\;\; \text{even} \\
 -\frac{1}{2} m^\bba_\bbb, & (M-d_\bbb)\;\; \text{odd} \\
 \end{array}\right.
\label{cross-behavior}
\end{multline}

For practical purposes one wants to allow some extra freedom in the definition of $B^\bba(\xi)$. Namely, suppose we want to rescale integrals as\footnote{In principle, one can also consider rescaling by a non-integer power of $k$.}
\begin{equation}
 I'_{\Gamma_\bba}=k^N I_{\Gamma_\bba},\qquad N\in \Z_+
 \label{Int-rescaling}
\end{equation}
This is equivalent to the following redefinition of the Borel transforms:
\begin{equation}
{B'}^\bba(\xi)=\d_\xi^N B^\bba(\xi).
\label{Borel-redef}
\end{equation}
However if $N$ is large enough the relation (\ref{Int-thimble}) might be modified by a finite number of terms arising from integration by parts:
\begin{equation}
	I'_{\Gamma_\bba}=
	\text{non-negative powers of $k$}\,+\,\int_{\gamma_\bba}d\xi e^{-k\xi} B'^\bba(\xi)\,.
	\label{Int-thimble-new}
\end{equation}
The following redefinition is also might be useful for practical purposes, especially in the case of Chern-Simons theory:
\begin{equation}
	I_{\Gamma_\bba} \rightarrow e^{\frac{2\pi i \Delta}{k}}\,I_{\Gamma_\bba}\,,
\end{equation}
\begin{multline}
	B^\bba(\xi) \rightarrow e^{2\pi i\Delta\,\d_\xi^{-1}}\,B^\bba (\xi)=\\
	B^\bba (\xi)+2\pi i\Delta\int_{\xi_\bba}^\xi d\xi_1B^\bba(\xi_1)
	+\frac{(2\pi i\Delta)^2}{2}\int_{\xi_\bba}^\xi d\xi_1\int_{\xi_\bba}^{\xi_1} d\xi_2 B^\bba(\xi_2)+\ldots
	\label{Borel-redefinition}
\end{multline}
Since $e^{\frac{2\pi i \Delta}{k}}$ has expansion in $1/k$ starting with 1 with infintite radius of convergence, this does not qualitatively the change behavior of $B^\bba(\xi)$ around critical points and radius of convergence of series (\ref{B-series}). In particular, (\ref{B-series}) and (\ref{cross-behavior}) will remain the same but with different values of subleading coefficients. The integrals in (\ref{Borel-redefinition}) are performed along the path of analytic continuation of $B^\bba(\xi)$.


\subsection{A simple example}

Let us consider the following simple example which illustrates well general features described above:
\begin{equation}
 I_\Gamma=\int_\Gamma dz_1dz_2dz_3dz_4 e^{- k(z_1^2+z_2^2+z_3^2+z_4^2-(z_1^2+z_2^2)(z_3^2+z_4^2))}\,.
 \label{example-integral}
\end{equation}
The action has $U(1)\times U(1)$ symmetry. There are two critical values:
\begin{equation}
\xi_0=0
\end{equation}
corresponding to an isolated critical point
\begin{equation}
\CM_0=\{z_1=z_2=z_3=z_4=0\}\cong \text{pt},
\end{equation}
and
\begin{equation}
\xi_1=1
\end{equation}
corresponding a submanifold
\begin{equation}
\CM_1=\{z_1^2+z_2^2=z_3^2+z_4^2=1\}\cong T^*T^2
\end{equation}
of complex dimension 2 which is a nontrivial orbit of $U(1)\times U(1)$ action.

The action defines a fibration over the Borel plane $\C_\xi$ with the following fiber:
\begin{equation}
X_\xi=\{z_1^2+z_2^2+z_3^2+z_4^2-(z_1^2+z_2^2)(z_3^2+z_4^2)=\xi\}\,.
\end{equation}
Equivalently, one can define it by the following system of equations:
\begin{equation}
X_\xi=\left\{
\begin{array}{c}
(U-1)(V-1)=1-\xi\\
z_1^2+z_2^2=U\\
z_3^2+z_4^2=V
\end{array}
\right\}
\end{equation}
which realizes $X_\xi$ as a fibration over the rational curve
\begin{equation}
\CC_\xi=\{(U-1)(V-1)=1-\xi\}
\label{example-curve}
\end{equation}
with fiber
\begin{equation}
\left\{
\begin{array}{c}
z_1^2+z_2^2=U\\
z_3^2+z_4^2=V
\end{array}
\right\}\cong T^*S^1_{\sqrt{|U|}}\times T^*S^1_{\sqrt{|V|}}\cong
T^*T^2.
\label{example-T2}
\end{equation}
\begin{figure}[ht]
\centering
 \includegraphics[scale=2.3]{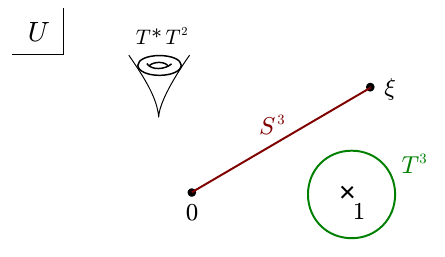}
\caption{The fiber $X_\xi$ presented as a $T^*T^2$ fibration over the $U$-plane. Red interval and green circle depict non-trivial cycles in middle-dimensional homology as $T^2$ fibrations.}
\label{fig:PL-example-1}
\end{figure}
The rational curve (\ref{example-curve}) can be parametrized by the $U$ plane with a puncture at $U=1$ (see Figure~\ref{fig:PL-example-1}). The points $U=0$ and $U=\xi$ are where one of two cycles of $T^2$ in (\ref{example-T2}) shrinks. Hypersurface $X_\xi$ has two middle dimensional cycles, $\Gamma^0_*\cong S^3$ and $\Gamma^1_*\cong T^3$. The first is realized as a $T^2$ fibration over an interval connecting $U=0$ and $U=\xi$ in the $U$ plane in the usual way. The second is realized as a product of a circle around the puncture at $U=1$ and the same $T^2$.

When $\xi$ goes around $1$ it is clear that the interval ``absorbs'' the circle, which correspond to the following monodromy:
\begin{equation}
[S^3] ~ \rightarrow ~  [S^3] \; +\; [T^3]\,.
\end{equation}
However there is no monodromy when $\xi$ goes around $0$. Therefore the only non-zero element of the matrix of monodromy coeffients in (\ref{cycle-mon}) is $m^0_1=1$, that is
\begin{equation}
m=\left(\begin{array}{cc}
0 & 1 \\
0 & 0
\end{array}
\right)\,.
\label{example-nmatr}
\end{equation}
Note that $\#(\Gamma_*^0 \cap \Gamma_*^1)=0$, which confirms (\ref{no-PL}).

Let us compare this with behavior of the Borel transform. Using Gaussian integration one can bring (\ref{example-integral}) to the following form\footnote{If the original contour is $\Gamma=\R^4$, the contour $\gamma$ goes along $\R_+$ and ``dodges'' the singularity at $\xi=1$ from below.}:
\begin{equation}
I_\Gamma =\frac{\pi^2}{k}\int_\gamma d\xi\,\frac{1}{1-\xi}\,e^{-k\xi}=
-{\pi^2}\int_\gamma d\xi\,\log(1-\xi)\,e^{-k\xi}
\end{equation}
so that
\begin{equation}
\begin{array}{l}
B^0=-\pi^2\log(1-\xi)\,, \\
B^1=-2\pi^3 i
\end{array}
\end{equation}
which is in agreement with the monodromy matrix (\ref{example-nmatr}).


\subsection{$SU(2)$ Chern-Simons}

Rescaling (\ref{Int-rescaling}) becomes especially useful when one consider path integral, that is when $M\rightarrow\infty$. Consider for simplicity the case when all critical points are isolated if we quotient over the total gauge symmetry. That is, the space
\begin{equation}
 \text{Hom}(\pi_1(M_3),SL(2,\C))\,/SL(2,\C)
\end{equation}
is discrete. This is the case for example for Seifert fibrations over a sphere with 3 exceptional fibers. Then if we quotient over the total gauge symmetry except the $SL(2,\C)$ symmetry at a refernce point of $\pi_1(M_3)$, the submanifolds of
\begin{equation}
 \text{Hom}(\pi_1(M_3),SL(2,\C))
 \label{Hom-space}
\end{equation}
 corresponding to central, abelian and irreducible flat connections have complex dimensions 0, 2 and 3 respectively. We want to normalize the $SU(2)$ Chern-Simons partition function so that
\begin{equation}
e^{2\pi ikS_\bba}Z_\bba \equiv  I'_{\Gamma_\bba} \sim \left\{
\begin{array}{cl}
 k^{-3/2} \,,&~~~\bba \,= \, \text{central}\\
 \frac{1}{\sqrt{k}}e^{-k \xi_\bba} \,,&~~~\bba \, \in \, \text{abelian}\\
 e^{-k \xi_\bba} \,,&~~~\bba \, \in \, \text{irreducible}\\
\end{array}\right.
\label{CS-pert-scaling}
\end{equation}
in the leading order, where $\bba$ denotes a lift of the connected component of (\ref{Hom-space}) to the universal cover of the space of gauge connections (\text{cf.} beginning of section \ref{sec:BorelCS}). The values of the CS action on flat connections are related to the critical values of $\xi$ as $\xi_\bba=-2\pi i S_\bba$. Note that $Z_\bba$ itself does not depend on the choice of the lift, so one can denote it as $Z_\alpha$, where $\alpha$ labels connected component in (\ref{Hom-space}), without a lift. We will do the same to some other quantities that actually depend only on $\alpha$. Normalization (\ref{CS-pert-scaling}) corresponds to the standard ``physical'' normalization for which $S^3$ parition function is $Z (S^3) = \sqrt{\frac{2}{k}} \sin(\pi/k)$. Such behavior is in agreement with dependence on dimensions $d_\bba$ in (\ref{I-series}).

\begin{figure}[ht]
\centering
 \includegraphics[scale=2.0]{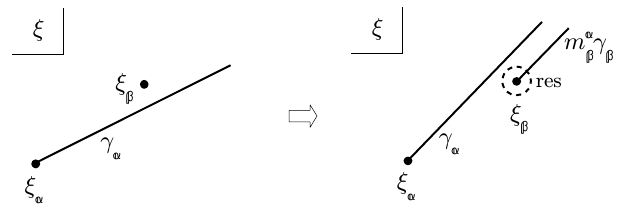}
\caption{Graphical representatation of a Stokes phenomenon in the Borel plane. The dashed circle depicts contribution of non-integral terms in (\ref{Borel-stokes}) or, more generally, in (\ref{Borel-stokes-multi}) which are given by residue of $B'^\bba(\xi)e^{-k\xi}$ at $\xi=\xi_\bbb$.}
\label{fig:PL-stokes}
\end{figure}

The corresponding (redefined as in (\ref{Borel-redef})) Borel transforms therefore behave as follows when $\xi\approx\xi_\bbb$:
\begin{equation}
 B'^\bbb(\xi)
 =\left\{
 \begin{array}{cl}
  \xi^{1/2}(c^\beta_0+c^\beta_1\xi+c^\beta_2\xi^2+\ldots) \,,&~~\bbb \, = \, \text{trivial},\\
(\xi-\xi_\bbb)^{-1/2}(c^\beta_0+c^\beta_1(\xi-\xi_\bbb)+c^\beta_2(\xi-\xi_\bbb)^2+\ldots) \,,&~~\bbb \, \in \, \text{abelian},\\
  (c^\beta_0+c^\beta_1(\xi-\xi_\bbb)+c^\beta_2(\xi-\xi_\bbb)^2+\ldots) \,,&~~\bbb \, \in \, \text{irreducible}.\\
 \end{array}\right.
\end{equation}
However, as was pointed out in (\ref{Int-thimble-new}), we now have the following relation:
\begin{equation}
Z_\beta =c_{-1}^\beta+\,\int_{\gamma_\bbb}d\xi e^{-k(\xi-\xi_\bbb)} B'^\bbb(\xi),\qquad
\bbb \in\text{irreducible}
\end{equation}
Redefinition $B^\bba\rightarrow B'^\bba$ will also result in a modification of (\ref{cross-behavior}). The behavior of $B'^\bba(\xi)$ near $\xi=\xi_\bbb$ now has the following form:
\begin{multline}
B'^\bba(\xi)\;=\;\text{regular}\;+\;\\
 m^\bba_\bbb\,\left[\frac{1}{2\pi i}\,\frac{c_{-1}^\beta}{\xi-\xi_\bbb}+\frac{\log(\xi-\xi_\bbb)}{2\pi i}
 \left(c_0^\beta+c_1^\beta\,(\xi-\xi_\bbb)+c_2^\beta\,(\xi-\xi_\bbb)^2+\ldots\right)\right],\qquad \bbb \in\text{irreducible} .
\end{multline}
In particular, the pole at $\xi=\xi_\bbb$ has appeared from action of $(-\d_\xi)^N$ on the logarithm. Of course, this is in agreement with the Stokes phenomenon
\begin{multline}
 Z_\alpha = \int_{\gamma_\bba}d\xi e^{-k(\xi-\xi_\bba)} B'^\bba(\xi)\longrightarrow\\
Z_\alpha +m^\bba_\bbb e^{- k(\xi_\bbb-\xi_\bba)}c^\beta_{-1}+\,m^\bba_\bbb\int_{\gamma_\bbb}d\xi e^{-k(\xi-\xi_\bba)} B'^\bbb(\xi)=Z_\alpha+m^\bba_\bbb e^{- k(\xi_\bbb-\xi_\bba)}\,Z_\beta,\\
\bba\notin\text{irreducible} ,\;\;\bbb \in\text{irreducible}
\label{Borel-stokes}
\end{multline}
which happens when $\gamma_\bba$ passes through $\xi_\bbb$ and schematically depicted in Figure~\ref{fig:PL-stokes}.

One can easily generalize this to the case when there is a connected component of the moduli space of irreducible flat connections
\begin{equation}
 \CM_\beta \subset \text{Hom}(\pi_1(M_3),SL(2,\C))\,/SL(2,\C)
\end{equation}
of dimension $d_\beta\equiv \text{dim}_\C \CM_\beta>0$. Suppose for simplicity that $d_\beta$ is even. We then have the following behaviors:
\begin{equation}
 B'^\bbb(\xi)
 =
  c^\beta_0+c^\beta_1(\xi-\xi_\bbb)+c^\beta_2(\xi-\xi_\bbb)^2+\ldots,\qquad \xi\approx\xi_\bbb,
\end{equation}
\begin{multline}
B'^\bba(\xi)\;=\;\text{regular}\;+\;
 m^\bba_\bbb\,\left[\frac{1}{2\pi i}\sum_{n=1}^{d_\beta/2+1}\frac{c_{-n}^\beta}{(\xi-\xi_\bbb)^n}+\right. \\
\left.\frac{\log(\xi-\xi_\bbb)}{2\pi i}
 \left(c_0^\beta+c_1^\beta\,(\xi-\xi_\bbb)+c_2^\beta\,(\xi-\xi_\bbb)^2+\ldots\right)\right],\qquad \xi\approx\xi_\bbb,.
\label{Borel-stokes-multi}
\end{multline}
while
\begin{equation}
Z_\beta =\sum_{n=1}^{d_\beta/2+1}(-k)^{n-1}\,\frac{c_{-n}^\beta}{(n-1)!}+\,\int_{\gamma_\bbb}d\xi e^{-k(\xi-\xi_\bbb)} B'^\bbb(\xi).
\end{equation}




\bibliographystyle{JHEP_TD}
\bibliography{classH}

\end{document}